\documentclass[twocolumn, table, x11names]{aastex631}
\usepackage{courier}
\usepackage{xurl}
\usepackage{multirow}
\usepackage{xcolor}

\shorttitle{}
\shortauthors{Claytor et al.}

\newcommand{\nspoc}{4,127}\newcommand{\ntasoc}{3,545}\newcommand{\nboth}{427}\newcommand{\nperiods}{7,245}\newcommand{\nnorot}{35,664}\newcommand{\nspec}{2,461}\newcommand{\ngold}{1,149}\newcommand{\ncool}{585}
\newcommand{\nplatinum}{567}
\newcommand{\nrossby}{1,023}
\newcommand{\nkdetected}{266}
\newcommand{\nknondet}{209}

\begin{document}

\title{TESS Stellar Rotation up to 80 days in the Southern Continuous Viewing Zone}

\correspondingauthor{Zachary R. Claytor}
\email{zclaytor@ufl.edu}

\author[0000-0002-9879-3904]{Zachary R. Claytor}
\affiliation{Department of Astronomy, University of Florida, 211 Bryant Space Science Center, Gainesville, FL 32611, USA}
\affiliation{Institute for Astronomy, University of Hawai‘i at Mānoa, 2680 Woodlawn Drive, Honolulu, HI 96822, USA}

\author[0000-0002-4284-8638]{Jennifer L. van Saders}
\affiliation{Institute for Astronomy, University of Hawai‘i at Mānoa, 2680 Woodlawn Drive, Honolulu, HI 96822, USA}

\author[0000-0002-8849-9816]{Lyra Cao}
\affiliation{Department of Physics and Astronomy, Vanderbilt University, Nashville, TN 37212, USA}
\affiliation{Department of Astronomy, The Ohio State University, Columbus, OH 43210, USA}

\author[0000-0002-7549-7766]{Marc H. Pinsonneault}
\affiliation{Department of Astronomy, The Ohio State University, Columbus, OH 43210, USA}

\author[0009-0008-2801-5040]{Johanna Teske}
\affiliation{Earth \& Planets Laboratory, Carnegie Institution for Science, 5241 Broad Branch Road, NW, Washington, DC 20015, USA}

\author[0000-0002-1691-8217]{Rachael L. Beaton}
\affiliation{Space Telescope Science Institute, Baltimore, MD, 21218, USA}
\affiliation{Department of Physics and Astronomy, Johns Hopkins University, Baltimore, MD 21218, USA}

\accepted{to \apj\ December 11, 2023}

\begin{abstract}
     The TESS mission delivers time-series photometry for millions of stars across the sky, offering a probe into stellar astrophysics, including rotation, on a population scale. However, light curve systematics related to the satellite's 13.7-day orbit have prevented stellar rotation searches for periods longer than 13 days, putting the majority of stars beyond reach. Machine learning methods have the ability to identify systematics and recover robust signals, enabling us to recover rotation periods up to 35 days for GK dwarfs and 80 days for M dwarfs. We present a catalog of \nperiods~rotation periods for cool dwarfs in the Southern Continuous Viewing Zone, estimated using convolutional neural networks. We find evidence for structure in the period distribution consistent with prior \textit{Kepler} and K2 results, including a gap in 10--20-day cool star periods thought to arise from a change in stellar spin-down or activity. Using a combination of spectroscopic and gyrochronologic constraints, we fit stellar evolution models to estimate masses and ages for stars with rotation periods. We find strong correlations between the detectability of rotation in TESS and the effective temperature, age, and metallicity of the stars. Finally, we investigate the relationships between rotation and newly obtained spot filling fractions estimated from APOGEE spectra. Field star spot filling fractions are elevated in the same temperature and period regime where open clusters' magnetic braking stalls, lending support to an internal shear mechanism that can produce both phenomena. \end{abstract}


\section{Introduction}
Rotation, activity, and magnetism are all deeply connected to the structure and evolution of stars. 
{In stars with convective envelopes (on the main sequence, similar to and less massive than the Sun), rotation powers magnetism \citep[e.g.,][]{Parker1955, Spiegel1992}}, which influences stellar winds and causes flares. Magnetized winds create torque on stars, causing them to spin down over time \citep{Weber1967}; this allows us to infer stellar ages from rotation periods using gyrochronology \citep{Skumanich1972, Barnes2003}. Stellar magnetism is the source of space weather, which directly affects life on Earth as well as the habitability of planets around other stars. 
Because of the inextricable links to rotation, a complete picture of stellar activity and magnetism demands a grasp of rotation across all types of stars.

The \textit{Kepler} mission \citep{Borucki2010} enabled rotation period estimates for more than 50,000 stars in a single 100-square-degree patch of sky \citep{McQuillan2014, Santos2019, Santos2021}, revolutionizing our understanding of stellar rotation. \textit{Kepler}'s rotation periods enabled precise age estimates for field stars \citep[e.g.,][]{Claytor2020, Lu2021} and investigations of changing stellar activity with time \citep{Mathur2023}. The mission also revealed departures from expected rotational behavior, such as a gap in the period distribution of cool stars \citep{McQuillan2014} and a halt of magnetic braking in middle-aged Solar-like stars \citep{Angus2015, vanSaders2016, David2022}. Stellar evolution and population synthesis models failed to predict these behaviors \citep{vanSaders2019}, highlighting the need for updated theory as well as more period measurements.

{With the loss of \textit{Kepler}'s second reaction wheel, the second \textit{Kepler} mission K2 \citep{Howell2014} repurposed the telescope to observe stars along the ecliptic plane. K2 delivered 30,000 field star rotation periods \citep{Reinhold2020, Gordon2021}, which featured the same rotational anomalies seen in the \textit{Kepler} field.}

As successful as \textit{Kepler} {K2 were} in measuring rotation periods, the survey designs imposed tight limitations on the kinds of stars that could be studied in rotation. The \textit{Kepler} mission's goal of finding Earth-like planets around Sun-like stars resulted in a complex selection function that biased observed stellar samples in comparison to the underlying population \citep[e.g.,][]{Borucki2010, Wolniewicz2021}. For example, the choice to avoid young stellar populations biases the \textit{Kepler} sample toward less active, slowly rotating stars intrinsically more difficult to detect in rotation. Any \textit{Kepler} study preserves these biases, and the selection function is difficult to correct for. {K2's periods provided a look at other fields, but the short observing baselines limited the lengths of periods that could be measured.} Furthermore, the small observing footprints means that any new rotational physics inferred from \textit{Kepler} {and K2} must be tested against other samples across the sky. The solution is an untargeted, all-sky survey.

The \textit{Transiting Exoplanet Survey Satellite} \citep[TESS][]{Ricker2015} stares at millions of stars in its search for transiting planets, surveying the entire sky in 27-day sectors. In addition to short-cadence light curves for pre-selected targets, TESS delivers full-frame images (FFIs), enabling high-precision photometry for any source brighter than $\sim$15th magnitude. Importantly, TESS does not rely only on postage stamps for selected targets as \textit{Kepler} {and K2} did. {Rather,} the FFIs permit investigators to design their own surveys. While primarily a planet-finding mission, the mission's short cadence and long temporal baseline also make it suitable for studying stellar variability due to oscillations, pulsations, and rotational spot modulation. While studies of stellar oscillations and pulsations have achieved some success \citep[e.g.,][]{SilvaAguirre2020, Mackereth2021, Chontos2021, Hon2021, Hon2022, Stello2022}, systematics related to TESS's observing strategy and data processing have slowed the quest for {measuring long} rotation periods \citep{Oelkers2018b, CantoMartins2020, Avallone2022, Kounkel2022, Fetherolf2023}. It is worth noting that the \textit{Kepler} mission faced similar challenges; the seminal stellar rotation paper \citep{McQuillan2014} was published 5 years after the satellite was launched.

TESS's unique 2:1 resonance orbit of the Earth-Moon system subjects the detectors to earthshine and moonlight on the timescale of the orbit, 13.7 days \citep{TESSHandbook}. The earthshine itself has time-varying signals within it, such as a 1-day modulation from the rotation of the Earth \citep{Luger2019}. Besides earthshine, TESS encounters systematics related to angular momentum dumps, detector heating, data downlinks, and more, all on timescales that interfere with astrophysical signals. The telescope's large field of view (24$^\circ$ by 96$^\circ$ total) makes the background spatially non-uniform as well. Because of these effects, throughout a sector the TESS detectors encounter systematics on different pixels at different times and with varying intensity, making them difficult to correct. 

Attempts to remove or correct spurious instrumental signals may also attenuate astrophysical signals, particularly those on the timescales of the telescope's orbital period (13.7 days) and longer. Rapid rotators, which also have larger spot modulation amplitudes \citep[e.g.,][]{Santos2021}, are affected less, and conventional rotation searches with TESS have been largely successful at measuring periods shorter than 13 days {(see, for example, \citealt{CantoMartins2020} with 131 periods ranging from 0.3 to 13.2 d, \citealt{Avallone2022} with 169 from 0.9 to 10.5 d, \citealt{Holcomb2022} with 13,504 from 0.4 to 14 d, \citealt{Kounkel2022} with $\sim$100,000 from 0.1 to 12 d, and \citealt{Fetherolf2023} with 84,000 from 0.01 to 13 d, but includes all types of variability)}. However, the same searches have struggled to recover longer periods, instead catching periods associated with the TESS systematics. So far, only \citet{Lu2020} have claimed to recover long periods in TESS, but they relied heavily on priors from the observed \textit{Kepler} period distribution.

The efforts to correct TESS systematics have yielded broadly useful public pipelines and tools like \textit{eleanor} \citep{Feinstein2019}, TESS-SIP \citep{Hedges2020}, \textit{Unpopular} \citep{Hattori2022}, and T'DA \citep{Handberg2021, Lund2021}. While each pipeline makes different but well-motivated choices to handle the systematics, each decision runs the risk of accidentally removing stellar signals \citep{Hattori2022, Kounkel2022}. Rather than trying to remove the systematics at the risk of removing astrophysical signals, we adopt deep machine learning methods that see the periodicity with the noise and disentangle them.

Deep learning methods are now widely used in stellar astrophysics. \citet{Breton2021} used random forests to classify and detect rotation signals in \textit{Kepler} light curves, while \citet{Lu2020} used random forests to draw connections between stellar parameters to estimate TESS rotation periods. \citet{Feinstein2020} employed convolutional neural networks (CNNs) to identify stellar flares in light curves, while \citet{Hon2021} applied similar techniques to detect oscillations in red giant stars. CNNs are particularly powerful when working with images or image-like data, which are ubiquitous in astronomy. CNNs can be trained to identify images of many different classes despite contaminating features, making them particularly attractive for our problem with TESS systematics.

In a pilot study of 21 objects, \citet{Claytor2022} demonstrated that long periods can be recovered from TESS data using CNNs trained on simulated data. In this work we apply the \citet{Claytor2022} approach to a greatly expanded sample to infer rotation periods with uncertainties for cool, main-sequence stars in the TESS Southern Continuous Viewing Zone (SCVZ). We employ new training sets tailored to specific period ranges and the specific light curves in which we search for periods. We present the periods, their associated uncertainties, and model-inferred stellar parameters in the first catalog probing long stellar rotation periods with TESS.

The paper is outlined as follows. In Section~\ref{sec:sample} we describe our data and sample selection. In Section~\ref{sec:deeplearning} we outline our deep learning framework, including the training sets and model architectures. Section~\ref{sec:modeling} details our method to fit stellar evolutionary models to stars with reliable rotation periods. In Section~\ref{sec:periods} we present the rotation periods and analyze the TESS SCVZ period distribution, comparing and contrasting with \textit{Kepler} and K2. In Section~\ref{sec:detectability} we explore the detectability of periods as a function of temperature, metallicity, age, and convection zone properties to understand the effects of detection limits on the period distribution. In Section~\ref{sec:spots} we use new spot filling fraction measurements from infrared spectroscopy to examine the effects of spottedness on the detectability of rotation, and we finally conclude in Section~\ref{sec:conclusion}.

\section{Data and Sample Selection} \label{sec:sample}

\begin{deluxetable*}{cccr}
    \label{tab:sample}
    \centering
    \tablecaption{Sample Selection}
    \tablehead{\colhead{Designation} & \colhead{Description} & \colhead{Criteria} & \colhead{\# targets}}
    \startdata
        A1 & TESS-SPOC Dwarfs & \texttt{eclat} $\leq -78^\circ$ \& \texttt{Tmag} $\leq 15$ \& \texttt{Teff} $\leq 10,000$ K & 38,215 \\
        A2 & TASOC Dwarfs & \texttt{eclat} $\leq -78^\circ$ \& \texttt{Tmag} $\leq 15$ \& \texttt{Teff} $\leq 5,000$ K & 29,609 \\
        \hline
        B1 & APOGEE--TESS-SPOC & A1 \& APOGEE DR17 & 16,545 \\
        B2 & APOGEE--TASOC & A2 \& APOGEE DR17 & 3,156 \\
        \hline
        C1 & Rotators & Either A1 or A2 \& reliable period & \nperiods \\
        C2 & Non-Rotators & Either A1 or A2 \& no reliable period & \nnorot \\
        \hline
        D & APOGEE Rotators & Either B1 or B2 \& C1 & \nspec \\
        \hline
        Gold & Binary-Cleaned Rotators & D \& \texttt{STAR\_BAD} $=$ \texttt{SNR\_BAD} $=0$ \&  \texttt{RUWE} $< 1.2$ \& \texttt{contratio} $< 0.1$ & \ngold \\
        \hline
        Platinum & Single Cool Dwarfs & Gold \& $M_G > 4.4$ \& $G_{BP}-G_{RP} > 1$ \& $|\Delta M_G| < 0.4$ & \nplatinum \\
    \enddata
    \tablecomments{Our sample selections using TIC version 8.2 \citep{Stassun2019, Paegert2021}, APOGEE DR17 \citep{Abdurro'uf2022}, and \textit{Gaia} DR3 \citep{Gaia2023}. Selection criteria are given as table column names where convenient and include ecliptic latitude (TIC \texttt{eclat}), TESS magnitude (TIC \texttt{Tmag}), effective temperature (TIC \texttt{Teff}), ASPCAP \citep{GarciaPerez2016} spectral fit flags \texttt{STAR\_BAD} and \texttt{SNR\_BAD}, absolute $G$-magnitude (\textit{Gaia} $M_G$), color index (\textit{Gaia} $G_{BP} - G_{RP}$), and \textit{Gaia} photometric excess above the main sequence ($|\Delta M_G|$). The lettered samples are described in Section~\ref{sec:sample}, while the Gold and Platinum science samples are detailed in Sections~\ref{sec:mresults} and \ref{sec:spots} respectively.}
\end{deluxetable*}

For the period search, we targeted stars cool enough to maintain surface convection zones and dynamos capable of producing surface spots. The steps of our full sample selection our outlined in Table~\ref{tab:sample}. We excluded evolved red giants, of which the large majority are slow rotators \citep{Ceillier2017}. The 1--2\% that rotate rapidly are typically the products of binary star interactions \citep{Carlberg2011}, and not reliable age tracers. We selected relatively bright, cool dwarf and subgiant stars in the TESS SCVZ, a 450 square degree field centered around the southern ecliptic pole. TESS observed the SCVZ continuously for 350 days in its first year, taking FFIs every 30 minutes. The long baseline ensures sufficient coverage for the most slowly-rotating stars we might hope to detect. For example, an M-dwarf rotating once every 100 days will complete 3.5 rotations under observation in the CVZs. In the same interval, an old K-dwarf rotating at 45 days will rotate nearly 8 times, and a G-dwarf at 30 days will rotate more than 10. 

We selected stars from the TESS Input Catalog \citep[TIC,][]{Stassun2019, Paegert2021} with effective temperature $\leq 10,000$ K, TESS magnitude $\leq 15$, and ecliptic latitude $\leq -78^\circ$ to target the SCVZ. There are 398,977 such stars in the TIC, but requiring public photometry narrowed the sample considerably. 38,215 targets had public FFI photometry from the TESS Science Processing Operations Center \citep[TESS-SPOC,][]{Jenkins2016, Caldwell2020}. We also used FFI data products from the TESS Asteroseismic Science Operations Center \citep[TASOC,][]{Handberg2021, Lund2021}, but we selected only the 29,609 targets with TIC $T_\mathrm{eff} < 5,000$~K to prioritize the most likely rotation detections. We motivate the choice to use both TESS-SPOC and TASOC products in Section~\ref{sec:photometry}, and we detail each pipeline's target selections in Sections~\ref{sec:spoc} and \ref{sec:tasoc}.

\subsection{APOGEE Spectroscopy}
While the TESS Input Catalog has metallicities and surface gravities for all the stars in our sample, the sources are a heterogeneous combination of photometry and spectroscopy, observations and models. Furthermore, the TIC has no information on detailed abundances, which are useful when investigating changing Galactic chemistry with time, and which are important to the connection between rotation and magnetism \citep[e.g.,][]{Claytor2020}. We therefore supplement TESS photometric rotation periods with spectroscopic parameters from the Apache Point Observatory Galactic Evolution Experiment \citep[APOGEE]{Majewski2017}. 

APOGEE collects high-resolution ($R \sim 22,500$), near-infrared (1.51--1.70 $\mathrm{\mu}$m) spectra and provides calibrated, model-dependent estimates of effective temperature, surface gravity, metallicity, and detailed abundances for hundreds of thousands of stars across the entire sky. The TESS/APOGEE survey within APOGEE-2S \citep[Section 5.8 of][]{Santana2021} targeted 38,000 stars in the TESS SCVZ with 2MASS color and magnitude ranges $7 < H < 11$ and $J-K > 0.3$, and about 9,000 other SCVZ stars were observed for other programs. We cross-matched the TIC SCVZ cool dwarfs with APOGEE Data Release 17 \citep{Abdurro'uf2022}, {matching the TIC \texttt{TWOMASS} column with the \texttt{APOGEE\_ID} columns in the \texttt{apogeeStar} and \texttt{aspcapStar} tables.} This cross-match returned spectroscopic parameters for 47,142 stars. Of those, 16,545 have TESS-SPOC data products, and 3,156 have data products in our TASOC subsample. These combine to yield 17,796 unique targets with APOGEE spectroscopy and either TESS-SPOC or TASOC photometry.

We adopted calibrated effective temperatures, metallicities, and $\alpha$-element abundances estimated by the APOGEE Stellar Parameters and Abundances Pipeline \citep[ASPCAP,][]{GarciaPerez2016}. Comparisons between APOGEE stellar parameters and high-fidelity measurements have demonstrated the ASPCAP-derived uncertainties to be underestimated for giants \citep{Serenelli2017}, and dwarfs alike \citep{Birky2020, Sarmento2021}. Pinsonneault et al. (in prep.) find temperature errors of 30 K in giants, larger for dwarfs, and scatter in clusters of 0.05 dex in metallicity and 0.03 dex in $\alpha$ enhancement. We therefore set minimum uncertainty floors of 50 K for $T_\mathrm{eff}$, 0.05 dex for [M/H], and 0.03 dex for [$\alpha$/M]. While these likely still underestimate the error in the ASPCAP measurements, they were large enough for our fitting routines to find self-consistent models that successfully predicted other stellar parameters, e.g., luminosity or surface gravity.

\vfill
\subsection{Gaia}

We supplemented our sample with data from \textit{Gaia} DR3 \citep{Gaia2023,10.26131/IRSA544}, including $G$, $G_{BP}$, and $G_{RP}$ magnitudes, parallaxes, and Renormalized Unit Weight Error (RUWE). {The TIC contains \textit{Gaia} DR2 identifiers, and most stars' identifiers did not change from DR2 to DR3, so no further cross-matching was needed.} \textit{Gaia} data were available for all our targets. Computing the absolute magnitude $M_G$ from $G$ and parallax, we use a photometric excess and RUWE to identify and remove likely binaries before population analysis.

\subsection{Photometry} \label{sec:photometry}
There are several publicly available light curve sets, pipelines, and tools designed and optimized for TESS data. We review some of the most widely used in Appendix~\ref{app:lightcurves}. After trying several systematics removal pipelines and data products, we found that all pipelines were too aggressive and removed stellar signal. Instead, we used the apertures from two public pipelines and performed our own minimal corrections. Due to data availability and lightweight data products, we determined the apertures from the TESS-SPOC \citep{Jenkins2016, Caldwell2020} and TASOC \citep{Handberg2021, Lund2021} to be the best available for a rotation search at the time of writing. 

TESS-SPOC provides data products for fewer stars over a longer baseline, while TASOC provides products for a larger sample, but over a shorter baseline in TESS year 1. The two pipelines feature different target and aperture selections, providing two slightly overlapping stellar samples so that we can maximize the number of rotation periods while testing for robustness of periods against the pipelines' different apertures. We summarize the pipelines' key differences in the next two sections; we then describe our custom photometry using the pipeline apertures in Section~\ref{sec:lightcurves}. Both pipelines' target pixel file (TPF) and aperture data are publicly available on MAST, \citet{10.17909/t9-wpz1-8s54} for TESS-SPOC data and \citet{10.17909/t9-4smn-dx89} for TASOC data.

\subsubsection{TESS-SPOC} \label{sec:spoc}
The SPOC pipeline \citep{Jenkins2016} was initially used to calibrate the TESS FFIs and generate TPFs and light curves for all two-minute cadence targets. \citet{Caldwell2020} more recently used the SPOC pipeline to create TPFs and light curves for FFI targets, providing the TESS-SPOC light curves on MAST.

\citet{Caldwell2020} selected a maximum of ten thousand targets per CCD from the TIC for a maximum of 40,000 stars in the SCVZ. For each CCD, the selection order was (1) all two-minute cadence targets; (2) potentially high-value planet host candidates with $H$ magnitude $\leq 10$ or distance $\leq 100$ pc, flux contamination $\leq 50\%$, and TESS magnitude $Tmag \leq 16$; (3) field star targets brighter than $Tmag \leq 13.5$, log surface gravity $\geq 3.5$ (CGS units), and flux contamination $\leq 20\%$. The depth $Tmag \leq 13.5$ was chosen to ensure sufficient signal-to-noise. We estimated the 6-hour {combined differential photometric precision \citep[CDPP,][]{Christiansen2012}} of our custom TESS-SPOC light curves to be about {400 ppm at $Tmag = 13.5$. At this faint limit, a 5$\sigma$ detection should vary at the 0.2\% level. About 25\% of \textit{Kepler} rotators varied at or above this level \citep{Santos2019, Santos2021}.}

TESS-SPOC computed photometric apertures using the same module as was used for \textit{Kepler}. Briefly, the module uses a synthetic FFI produced from the input catalog and the real pixel response function to compute the optimal aperture for each target. \citet{Caldwell2020} detail the full FFI target selection, \citet{Jenkins2016} describe the SPOC pipeline, and \citet{Smith2016} outline the aperture selection. The TESS-SPOC pipeline has produced TPFs, which include target apertures, for all sectors in year 1. We queried all TPFs available for our sample, yielding time-series images and photometric apertures for 38,215 targets. 

\subsubsection{TASOC} \label{sec:tasoc}
TASOC has performed photometry for all stars brighter than TESS magnitude $\leq 15$ for use in asteroseismology \citep{Handberg2021, Lund2021}. To date, only sectors 1--6 from the first year have been processed, yielding time-series FFI photometry with a 160-day baseline and 30-minute cadence. While fewer sectors of data are available from TASOC, limiting us to shorter rotation periods than TESS-SPOC, TASOC's fainter magnitude limit and lack of number cap (i.e., TESS-SPOC processed not more than 10,000 stars per CCD, but TASOC has no such limit) complements the TESS-SPOC data. To compute light curves, we downloaded the TASOC apertures and applied them to cutouts from the calibrated FFIs.

The TASOC pipeline computed apertures for all TIC targets brighter than $Tmag \leq 15$. Aperture selection is fully described by \citet{Handberg2021}, but uses the clustering algorithm DBSCAN \citep{Ester1996} to find clusters of pixels associated with TIC targets. The watershed image segmentation routine from \texttt{scikit-image} \citep{vanderWalt2014} is then used to segment apertures containing more than one target. In general, the apertures created by the TASOC pipeline are larger than those created by TESS-SPOC, resulting in light curves with higher photometric precision. We estimated our custom TASOC light curves to have 6-hour CDPP of {300 ppm at $Tmag = 13.5$. A 5$\sigma$ detection at this magnitude will vary at the 0.15\% level. In \textit{Kepler}, about 33\% of rotating stars varied at or above this level \citep{Santos2019, Santos2021}.}

TASOC data products are also available on MAST. To obtain the likeliest targets for detecting rotation, we queried data for TIC dwarf stars cooler than 5,000 K, yielding FFI cutouts for 29,609 targets spanning the first 6 sectors of the TESS mission.

\subsubsection{Custom Light Curves and Wavelet Transform} \label{sec:lightcurves}
For both datasets, we began with the publicly available TPF cutouts from calibrated FFIs. The FFI calibrations include traditional bias, dark, and flat field corrections, cosmic ray removal, corrections for variations in pixel sensitivity, and removal of smear signals resulting from the cameras' lack of shutters \citep{Jenkins2016}. After FFI calibration, both the TESS-SPOC and TASOC pipelines perform background subtraction and systematics correction to produce light curves; we opt not to use this next level of data correction, as they can have the unintended consequence of removing or attenuating the stellar signals. To mitigate the removal of stellar rotation signals, we performed custom photometry using the apertures supplied by the pipelines. For each available TPF, we computed light curves as follows:
\begin{enumerate}
    \item reject cadences with bad quality flags, which are usually associated with cosmic rays, data downlinks, or angular momentum dumps
    \item compute a raw light curve using simple aperture photometry, adding all pixels within the aperture
    \item remove the first three principal components of the time series using \texttt{lightkurve.RegressionCorrector}
    \item reject 5$\sigma$ outliers from light curve.
\end{enumerate}
Although neural networks can perform regression in spite of systematics to some extent, \emph{some} systematics removal is necessary. 
We sought to perform as little systematics correction as possible in order to preserve the underlying stellar signals. Removing the first three principal components corrected the largest TESS systematics---Earthshine and angular momentum dumps---while leaving smaller systematics and stellar signals mostly intact. To determine the optimal number $n_\mathrm{pca}$ of principal components to remove, we removed 1, 2, 3, 4, and 5 components from a set of 10 randomly selected light curves. We then visually inspected the resulting light curves to determine for what value of $n_\mathrm{pca}$ the largest systematics were removed. Meanwhile, removing 5$\sigma$ outliers cleaned the light curves from systematic jumps and stellar flares. Next, we median-divided the light curves for each target and stitched them together, linearly interpolating to fill any gaps. Finally, we computed Morlet wavelet transforms following \citet{Claytor2022} and binned them to $64\times64$ pixels to be used as input to the convolutional neural network.

\subsubsection{Variability Amplitudes}
We computed the photometric variability amplitudes $R_\mathrm{per}$ and $S_\mathrm{ph}$ for all our stars with estimated periods. Like \citet{McQuillan2014}, to compute $R_\mathrm{per}$ we measured the interval between the 5th and 95th percentile of normalized flux in each period bin, then took the median of those values. We computed $S_\mathrm{ph}$ as in \citet{Mathur2023}, by partitioning the light curve into segments of duration 5$P_\mathrm{rot}$, then taking the standard deviation of the light curve flux over each segment. This creates a time series of standard deviations; $S_\mathrm{ph}$ is taken to be the median value. For different analyses we use either $R_\mathrm{per}$ or $S_\mathrm{ph}$, but in theory the two metrics are related by $S_\mathrm{ph} \approx 0.35 R_\mathrm{per}$\footnote{This approximation holds true for perfect sinusoids observed for an integer number $N$ of cycles or in the limit of large $N$. However, our measured $S_\mathrm{ph}$ and $R_\mathrm{per}$ follow this relation remarkably well.}. We verified this relation in our {period sample}, so for this work we consider the two metrics to be interchangeable.

We emphasize that due to sector-to-sector stitching and detector sensitivity changes during momentum dumps, variability amplitudes for periods longer than about 13 days and especially 27 days will inevitably be suppressed. To attempt to account for this, we ran a series of noiseless, sinusoidal light curve simulations through a renormalization and stitching algorithm and compared the measured amplitudes to the true input amplitudes. For perfect sinusoids, we found that the amplitude suppression factor decays exponentially with period past 27 days. However, applying a correction to our measured amplitudes did not affect our results; to avoid artificially injecting period--amplitude biases we leave our reported amplitudes uncorrected.

Finally, we also measured the 6-hour CDPP, which quantifies the photometric noise, for each of our light curves. Since the CDPP is measured on timescales shorter than the typical TESS systematics, it should be unaffected by momentum dumps and sector-to-sector stitching.

\section{Deep Learning Framework} \label{sec:deeplearning}
To infer rotation periods from TESS light curves, we applied the method of \citet{Claytor2022} with a few modifications. Namely, we generated new training sets tailored to both the TESS-SPOC and TASOC samples (mostly to represent the different light curve lengths), and we optimized different neural networks for each data set.

\subsection{Training Set}
In \citet{Claytor2022} we trained a convolutional neural network on a set of synthetic light curves made from physically realistic spot evolution simulations, combined with real TESS noise from SCVZ galaxy light curves. Inactive galaxies do not vary on yearlong timescales or shorter, and thus they are a robust standard sample that can be useful to infer systematics. Other quiescent objects can serve the same role, such as hot stars, which we employ here. 

Two weaknesses of our previous approach were that (1) we were not successful in recovering periods of less than 10 days from our held-out test set, and (2) the neural network overfit within a few (of order 10) iterations over the training set. The first weakness was due to the choice of a loss function that enabled the network to estimate period uncertainty. In the presence of uncertainty, inferred periods are biased toward the mean of the distribution and away from the upper and lower limits. The effect is most pronounced for the top and bottom 10\% of the training set period range, affecting the ranges from 0--18 days and 162--180 days. Since the ability to estimate the period uncertainty is a key strength of our approach, we worked around this problem by using multiple training sets with different period ranges.

\subsubsection{Light Curve Simulations}
We created four separate simulated training sets {of 1 million light curves each} using \texttt{butterpy} \citep{butterpy, Claytor2022} with periods ranging from 0.1 day to 30, 60, 90, and 180 days. Having a shorter upper limit such as 30 days allows us to more successfully probe the short-period range---here only 0--3 days and 27--30 days are severely affected---while having multiple training sets with increasing upper limits gives us multiple period estimates that we can mutually compare for extra tests of reliability.  The distributions of all simulation input parameters besides period were the same as in \citet{Claytor2022} (the simulations for the 180 day upper limit \emph{are} the same as in the previous work \citep{10.17909/davg-m919}), and the same simulations were used for both the TESS-SPOC and TASOC training sets. The only other difference was the source of the light curves combined with the simulations to emulate instrumental noise and systematics. We note that using multiple training realizations yields multiple period estimates for the same star; we discuss the breaking of ties in Section~\ref{sec:periods}.

\subsubsection{Noise and Systematics Templates}
\label{sec:templates}
The second shortcoming was simply due to the small number ($\sim 2,000$) of galaxy light curve examples. If there are too few examples of noise in the training set, the neural network learns the noise quickly and overfits the data. Since there are many more bright stars than galaxies in TESS, we addressed this by combining our simulations with light curves of stars in temperature ranges that should be comparatively quiescent to emulate TESS noise. \citet{McQuillan2014} detected periods in \textit{Kepler} stars hotter than the Sun half as often as in cooler stars. Given TESS's slightly worse photometric precision and redder pass band than \textit{Kepler}, we expect TESS stars hotter than the Sun to be even harder to detect in rotation. This makes stars in the temperature range above $\sim5,800$ K ideal for use as quiescent light curve sources. At first we queried TPFs and computed light curves for TASOC stars in the range 5,800 K $\leq T_\mathrm{eff} \leq$ 6,000 K. We kept light curves with at least 4 sectors to allow for gaps in the data while ensuring that there were data for more than half the time baseline. This yielded a set of 22,090 TASOC noise templates {with $6 \leq Tmag \leq 15$}, an order of magnitude more than the number of galaxy sources used in the previous exercise. The same range of temperatures in TESS-SPOC, requiring that light curves have at least 7 sectors to cover more than half of the time baseline, has only 6,000 targets, so a larger temperature range was required. We used the range 6,000 K $\leq T_\mathrm{eff} \leq$ 8,000 K, which contained 17,637 sources {with $6 \leq Tmag \leq 13$}. {We note that since both noise sets have a bright limit of 6th TESS magnitude, there is little chance of saturation, which may occur at $Tmag \lesssim 6$ \citep{Handberg2021}.} Table~\ref{tab:training} details the noise light curves samples that make up the TESS-SPOC and TASOC training sets.

\begin{table}
    \centering
    \caption{TESS-SPOC and TASOC Noise Light Curves for Training Set}
    \begin{tabular}{l|c|c}
        \hline \hline
        & TESS-SPOC & TASOC \\
        \hline
        Sectors included & 1--13 & 1--6 \\
        Minimum number of sectors & 7 & 4 \\
        Time baseline (days) & 350 & 160 \\
        $T_\mathrm{eff}$ range of sources (K) & 6,000--8,000 & 5,800--6,000 \\
        {$Tmag$ range of sources} & 6--13 & 6--15 \\
        Number of sources & 17,637 & 22,090 \\
        \hline
    \end{tabular}
    \label{tab:training}
\end{table}

{The noise examples are distributed roughly uniformly across the entire SCVZ and therefore probe location-dependent systematics across all four CCDs of TESS Camera 4. Furthermore, the light curves span the same sectors as the target stars, probing the same time-dependent pixel-to-pixel and sector-to-sector systematics. The noise templates have the same distribution of brightness as the target sample, so they should share the same shot noise properties as well. Since shot noise limits the photometric precision, we do not expect temperature-dependent noise differences (e.g., granulation) to be present in the light curves. To be sure, we computed the CDPP distributions of the noise and target light curves and verified that they were the same. A brief discussion of the light curve noise properties is in Appendix~\ref{app:noise}.}

We note that the temperature range for the TESS-SPOC noise light curves overlaps with the $\delta$ Scuti instability strip \citep[e.g.,][]{Murphy2019} and with the $\gamma$ Doradus strip \citep[e.g.,][]{Balona2011, Bradley2015}. Of our TESS-SPOC noise targets, 1,724 ($\sim$10\%) fall within the $\delta$ Scuti strip, and depending on the criterion used, as few as 30\% \citep{Balona2011} and as many as two-thirds \citep{Bradley2015} are within the $\gamma$ Dor strip. Because $\delta$ Scuti stars pulsate with periods on the order of hours and $\gamma$ Dor less than about 3 days, we do not expect significant contamination from pulsation in our training light curves. The TASOC noise sample does not overlap with either instability strip. The presence of contaminants in the training set should make the CNN more robust against contamination (i.e., misidentifying a pulsator as a rotator), but {future work should test this using training sets that include pulsators, eclipsing binaries, and other kinds of periodic variables}.

\subsubsection{Noise-Combined Synthetic Light Curves}
{As in \citet{Claytor2022}, we combined the simulated rotational light curves with the ``pure noise'' curves to create our training sets. For the noise combination, both light curves were divided into thirteen 27-day sectors, then the simulated curve was interpolated to the cadences of the noise light curve and combined point-by-point. The individual sector light curves were then median normalized and stitched together. Finally, as with the ``real'' light curves, we computed Morlet wavelet transforms and binned them to $64\times64$ pixels for use by the CNN.}

{We produced eight training sets in all: four each for the TESS-SPOC and TASOC data, with maximum rotation periods of 30, 60, 90, and 180 days. We partitioned each set of light curves 80\%/10\%/10\% for our training/validation/test samples. Each sample was combined with a proportionate set of noise light curves, and no light curves were shared between the training, validation, or test samples. Because we had only about 20,000 noise light curves for either the TESS-SPOC or TASOC sets, each noise light curve was used roughly 50 times in each training set. While there is still the chance of overfitting due to the limited number of noise examples, this is a factor of 10 improvement over previous work in \citet{Claytor2022}.}

\subsection{Convolutional Neural Network}
\label{sec:network}
We began with the same CNN as in \citet{Claytor2022}, which uses the Adam optimizer \citep{Kingma2014} and negative log-Laplacian loss, enabling the estimation of uncertainty along with the rotation period. The loss function has the form
\begin{equation} \label{laplace}
    \mathcal{L} = \ln\left(2b\right) + \frac{|P_\mathrm{true} - P_\mathrm{pred}|}{b},
\end{equation}
where $b$, the median absolute deviation, is taken to represent the uncertainty. {We added 10\% dropout to each layer, which improved performance overall. Table~\ref{tab:architecture} summarizes the CNN architecture.}

\begin{deluxetable*}{lcccccc}
\tablecaption{Convolutional Neural Network Architecture}
\tablehead{\colhead{Layer Type} & \colhead{Number of Filters} &\colhead{Filter Size} & \colhead{Stride} & \colhead{Activation} & \colhead{Dropout} & \colhead{Output Size}} 
\startdata
Input image &  -      & -             & -         & -         & -     & $64\times64$ \\
Conv2D      & $x$     & $3\times3$    & $1\times1$& ReLU      & -     & $62\times62\times x$\\
MaxPool2D   & 1       & $1\times3$    & $1\times3$& -         & 10\%  & $62\times20\times x$\\
Conv2D      & $y$     & $3\times3$    & $1\times1$& ReLU      & -     & $60\times18\times y$\\
MaxPool2D   & 1       & $1\times3$    & $1\times3$& -         & 10\%  & $60\times6\times y$\\
Conv2D      & $z$     & $3\times3$    & $1\times1$& ReLU      & -     & $58\times4\times z$\\
MaxPool2D   & 1       & $1\times4$    & $1\times4$& -         & 10\%  & $58\times1\times z$\\
Flatten     & -       & -             & -         & -         & -     & $58z$ \\
Dense       & -       & -             & -         & ReLU      & 10\%  & 256 \\
Dense       & -       & -             & -         & ReLU      & 10\%  & 64 \\
Dense       & -       & -             & -         & Softplus  & -     & 2 \\
\enddata
\label{tab:architecture}
\tablecomments{We used three 2D convolution layers with ReLU activation, max-pooling, and 10\% dropout. As in \citet{Claytor2022}, we used 2D max-pooling with a 1-dimensional kernel to pool in the time dimension, imparting translational invariance, but not the frequency dimension, to preserve frequency resolution. We tried four different combinations of filter numbers in the three convolution layers. We denote the number of filters as $x$, $y$, and $z$. The four runs had (A) $x=8$, $y=16$, and $z=32$ filters; (B) 16, 32, 64; (C) 32, 64, 128; (D) 64, 128, 256. The convolution output goes through a series of fully connected, or dense, layers, also with ReLU activation and 10\% dropout. Our architecture uses softplus output and negative log-Laplacian loss to predict the rotation period with its uncertainty.}
\end{deluxetable*} 
We experimented with different architectures to optimize different networks to the TESS-SPOC and TASOC training sets. The original architecture had three convolution layers with (A) 8, 16, and 32 kernels, respectively, but we also tried (B) 16, 32, and 64 kernels; (C) 32, 64, and 128; and (D) 64, 128, and 256. More kernels or filters per layer allow the network to learn more features if they are present in the data, but they may also cause the network to overfit the data faster. {Only the number of filters changed between runs---the activation, loss, dropout, layer sequence, and training sets never changed.} We trained each architecture individually on each training set {until the validation loss stopped decreasing} and chose the architecture with the best overall recovery on a held-out test sample. For the TESS-SPOC set, architecture C performed best overall, but architecture A was optimal for the TASOC set. We discuss the details of architecture optimization in Appendix~\ref{app:architecture}.

\section{Rotational Modeling} \label{sec:modeling}
With newly obtained TESS rotation periods, we will be able to look for trends of rotation detectability and variability across fundamental stellar parameters. Stars spin down and become less active as they age \citep{Skumanich1972}, so we expect both detectability and variability to decrease with age. We also know activity to vary with Rossby number, the ratio of rotation period to the convective overturn timescale \citep[e.g.,][]{Noyes1984, Wright2011}. To validate these relationships and look for potential departures from expected behavior in TESS, we will need to derive ages and Rossby numbers for our sample. We employ the stellar evolution and rotational modeling using \texttt{kiauhoku} \citep{Claytor2020, kiauhoku} to infer ages, masses, convective timescales, and Rossby numbers for our stars with rotation periods and APOGEE spectroscopy.

\begin{deluxetable}{lc}
    \label{tab:physics}
    \centering
    \tabletypesize{\scriptsize}
    \tablecaption{Input Physics to Stellar Evolution Models}
    \tablehead{\colhead{Parameter} & \colhead{Value/Source}}
    \startdata
        Atmosphere & \citet{Castelli2004} \\
        Convective overshoot & False \\
        Diffusion & True \\
        Equation of state & OPAL \citep{Rogers2002}\\
        High-temperature opacities & OP \citep{Mendoza2007} \\
        Low-temperature opacities & \citet{Ferguson2005} \\
        Mixing length $\alpha$ & 1.86 \\
        Mixture and solar $Z/X$ & \citet{Grevesse1998} \\
        Nuclear reaction rates & \citet{Adelberger2011} \\
        Solar $X$ & 0.7089 \\
        Solar $Y$ & 0.2728 \\
        Solar $Z$ & 0.0183 \\
        $\Delta Y / \Delta Z$ & 1.4 \\
        Surface $(Z/X)_\odot$ & 0.02289 \\
        \hline
        Angular momentum evolution & \citet{vanSaders2016} \\
        Initial rotation period & 8.134 d \\
        Critical Rossby number & 2.16 \\
        Critical $\omega$ for saturation & 3.394$\times10^{-5}$ s$^{-1}$ \\
        $f_k$ & 6.575\\
        Disk coupling timescale & 0.281 Myr\\
    \enddata
\end{deluxetable}

The stellar evolution tracks were generated using the non-rotating version of the Yale Rotating Stellar Evolution Code \citep[YREC,][]{Demarque2008}, then global stellar properties were used to calculate angular momentum evolution following the magnetic braking law of \citet{vanSaders2016}. The models are fully described by \citet{Claytor2020}, but we list the input physics and solar calibration here in Table~\ref{tab:physics}. The angular momentum evolution includes weakened magnetic braking beginning about halfway through the main sequence \citep{vanSaders2016}, but does not include core-envelope decoupling \citep[e.g.,][]{Spada2020} or the apparent stalling of spin-down that appears to occur in young, cool stars \citep{Curtis2019}.

Using the Markov-chain Monte Carlo (MCMC) tools in \texttt{kiauhoku}, we interpolated and fit stellar evolution models to observational data. For the MCMC we used a $\chi^2$ log-likelihood of the form
\begin{equation}
    \mathcal{L}_{\chi^2} = -\frac{1}{2} \sum_{i} \frac{\left(x_i - x_i'\right)^2}{\sigma_{x_i}^2}, \nonumber
\end{equation}
where $x_i$ and $\sigma_{x_i}$ are the observational input parameters and uncertainties, respectively, $x_i'$ is the computed value from the model, and $i$ iterates over the input parameters. The observables used in this computation were the CNN-inferred rotation periods, APOGEE calibrated temperatures, metallicities ([M/H]) and $\alpha$-element abundances ([$\alpha$/M]). All MCMC input data are provided with uncertainties in Table~\ref{tab:periods}.

\section{Rotation Periods of TESS Stars} \label{sec:periods}
We estimated periods for TESS-SPOC targets with at least 7 sectors and TASOC targets with at least 4 sectors, the same minimum numbers of sectors as for the training light curves. {To avoid overlap with the noise light curves used in the training sample, we limited the period search to stars with TIC $T_\mathrm{eff} \leq$ 6,000 K}. To determine reliability we followed the method of \citet{Claytor2022} and used a cut in fractional uncertainty to denote reliability. We do not treat the estimated $\sigma_P$ ($= b$ in Eq.~\ref{laplace}) as a formal uncertainty. Rather, the quantity $\sigma_P/P$ serves as a metric of relative credibility of period estimates. $\sigma_P/P \leq 35\%$ translated to $\sim$10\% median percent error in the training set recovery, so we adopt this uncertainty cut as our baseline for reliability. Since there are four neural networks, each with its own period range, we obtained four sets of period candidates for both the TESS-SPOC and TASOC data sets. If two or more neural networks yielded estimates passing our reliability cut for the same star, we averaged the estimates and added their standard deviation in quadrature to the uncertainty. If the newly combined fractional uncertainty was larger than 35\%, we discarded the star. 

{Two failure modes were identified where a rotation signal was present in the light curve, but we failed to identify it using the CNN. An example is TIC 177116487, which exhibits a strong 4.3 day period, but whose predicted uncertainty (1.55 d) excluded it from our credibility cut. These tended to be at the short end of the period range (i.e., 5 days or less), where our choice of loss function biased predictions away from the edges and made predictions less certain. Another case is TIC 220438355, which appears to have rotational modulation around 13 days. The four CNN models predicted very different periods for this star, all with large uncertainties, suggesting that we may miss some stars with periods coincident with TESS systematics.}

\begin{deluxetable}{ll}
    \label{tab:periods}
    \centering
\tablecaption{Properties of \nperiods\ Rotationally Detected TESS SCVZ Stars, MCMC Input \& Fit Parameters}
    \tablehead{\colhead{Label} & \colhead{Description}}
    \startdata
        TIC     & TESS Input Catalog identifier \\
        TOI     & TESS Object of Interest identifier \\
        prot    & CNN-inferred rotation period \\
        e\_prot & rotation period uncertainty \\
        prov    & period provenance: TESS-SPOC or TASOC \\
        spurious    & flag denoting likely spurious period \\
        rper    & photometric activity range $R_\mathrm{per}$ \\
        sph     & photometric activity index $S_\mathrm{ph}$ \\
        cdpp    & combined differential photometric precision \\
        Teff\_TIC   & TIC effective temperature \\
        Tmag    & TESS magnitude \\
        contratio   & TIC flux contamination ratio \\
        parallax    & \textit{Gaia} DR3 parallax \\
        ruwe    & \textit{Gaia} DR3 renormalized unit weight error \\
        phot\_g\_mean\_mag  & \textit{Gaia} DR3 apparent $G$ magnitude \\
        bp\_rp  & \textit{Gaia} DR3 $G_{BP} - G_{RP}$ color index \\
        teff    & APOGEE DR17 effective temperature \\
        teff\_err   & temperature uncertainty \\
        m\_h    & APOGEE DR17 metallicity [M/H] \\
        m\_h\_err   & metallicity uncertainty \\
        alpha\_m    & APOGEE DR17 $\alpha$ enhancement [$\alpha$/M] \\
        alpha\_m\_err   & $\alpha$ enhancement uncertainty \\
        snr\_bad    & APOGEE DR17 spectral signal-to-noise flag \\
        fspot   & spot filling fraction \\
        age     & MCMC gyrochronological age \\
        e\_age+  & 1$\sigma$ age upper credible limit \\
        e\_age-  & 1$\sigma$ age lower credible limit \\
        mass    & MCMC-inferred stellar mass \\
        e\_mass+ & 1$\sigma$ mass upper credible limit \\
        e\_mass- & 1$\sigma$ mass lower credible limit \\
        rad     & MCMC-inferred stellar radius \\
        e\_rad+  & 1$\sigma$ radius upper credible limit \\
        e\_rad-  & 1$\sigma$ radius lower credible limit \\
        Ro      & MCMC-inferred Rossby number \\
        fconv   & MCMC convergence flag \\
    \enddata
    
    \tablecomments{The ``snr\_bad'' flag represents the APOGEE spectral signal-to-noise flag and is set for only 21 stars. This table is available in its entirety in machine-readable format.}
\end{deluxetable}
 
\begin{figure*}
    \centering
    \includegraphics[width=0.8\textwidth]{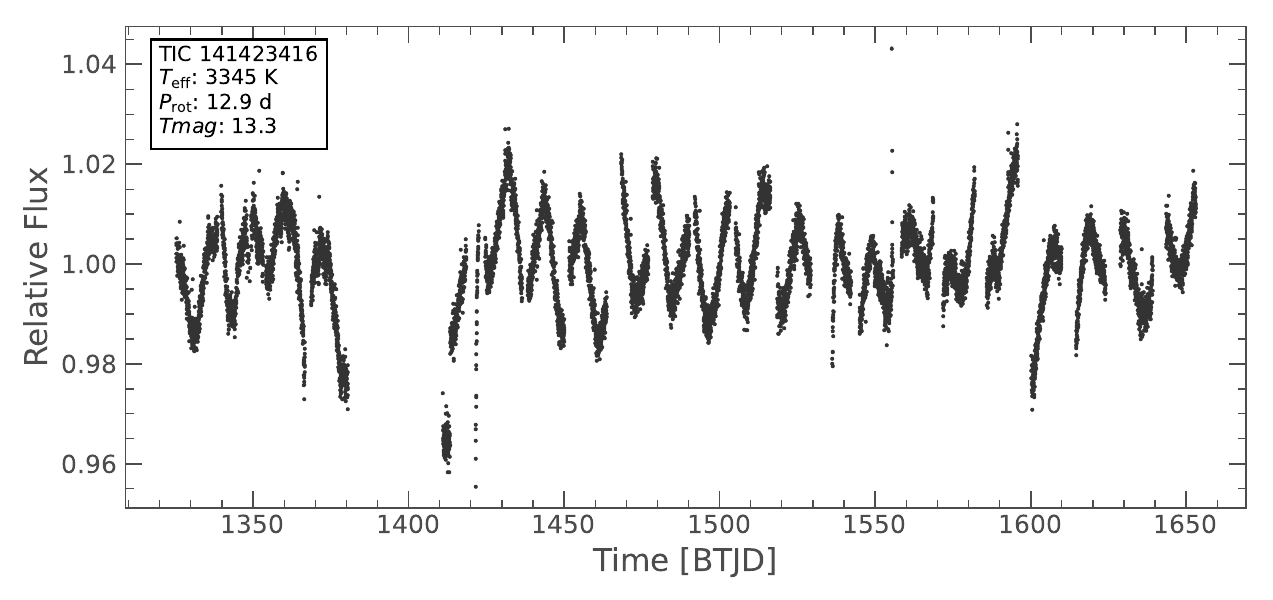}
    \includegraphics[width=0.8\textwidth]{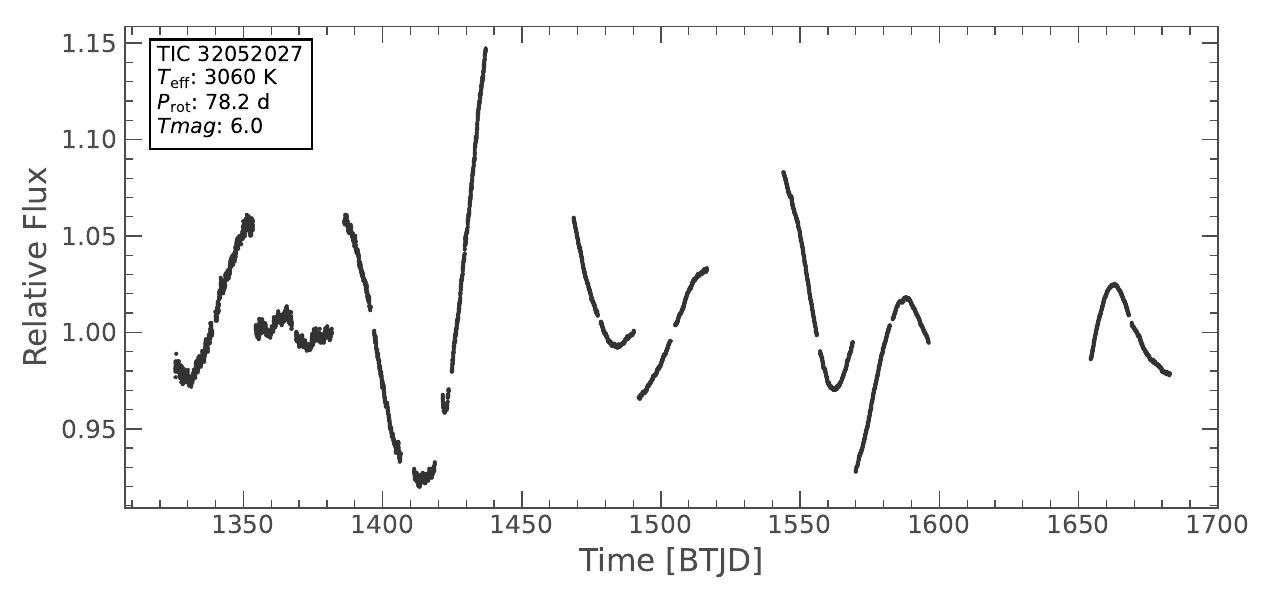}
    \includegraphics[width=0.8\textwidth]{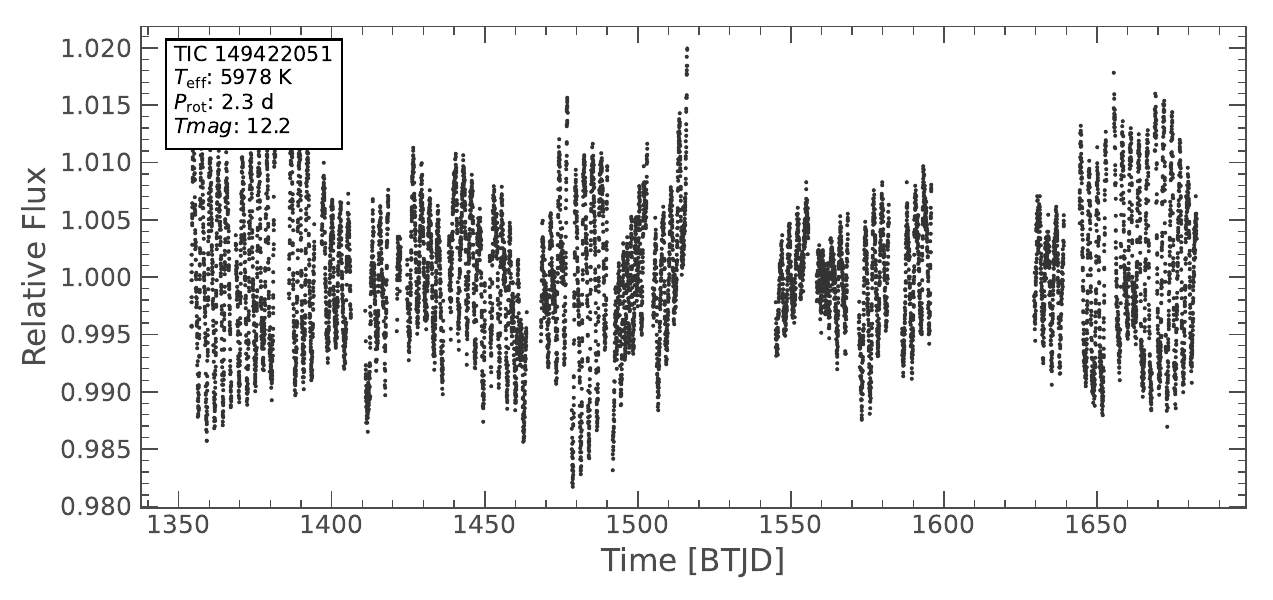}
    \caption{A selection of {TESS-SPOC simple aperture photometry} light curves for which we successfully estimated rotation periods using the neural network. \textit{Top}: an M dwarf with strong, coherent spot modulation. \textit{Middle}: a slowly-rotating M dwarf with several missing sectors. \textit{Bottom}: a rapidly rotating G dwarf.}
    \label{fig:lightcurves}
\end{figure*}

\begin{figure*}[!ht]
    \centering
    \includegraphics[width=0.8\textwidth]{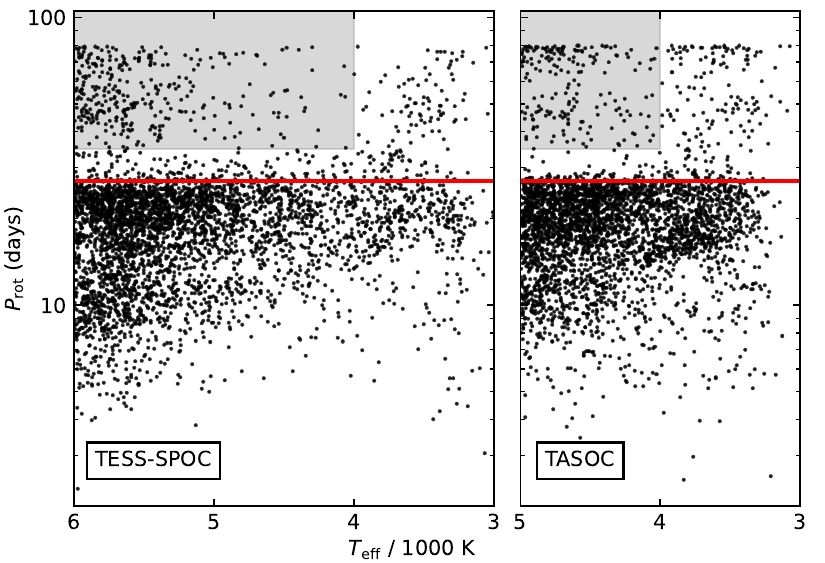}
    \caption{Period--temperature distribution for the TESS-SPOC (left) and TASOC (right) samples. Estimating periods using CNNs, we recover the short-period slope and intermediate-period gap seen in other rotating populations \citep[e.g.,][]{McQuillan2014}. Detection biases create a sharp edge at 27 days {(red line)}, above which the period uncertainties are larger due to the sector-to-sector stitching necessary for long-baseline TESS light curves. Toward $P_\mathrm{rot} = 90$ days, contamination increases because less certain period estimates are biased toward the median period of the training set. The detections above $P_\mathrm{rot} > 35$~d and $T_\mathrm{eff} > 4,000$~K {(gray shaded region)} are likely to be spurious, but most of the M dwarf periods up to 80 days appear to be real.}
    \label{fig:ptgrid}
\end{figure*}

We obtained \nspoc\ TESS-SPOC stars with reliable periods and \ntasoc\ reliable TASOC periods. These combine for a total of \nperiods\ unique targets, \nboth\ of which overlap between the two samples. We discuss the overlap sample in Section~\ref{sec:comparison}. The rotation periods up to 80 days, their photometric amplitudes, selected spectroscopic parameters, and associated flags are presented in Table~\ref{tab:periods}. We also list the stellar parameters for the \nnorot\ rotationally non-detected stars in Table~\ref{tab:norot}. For stars with periods from both TESS-SPOC and TASOC data, we favored the TESS-SPOC period in the final table due to the light curves having twice the duration as the TASOC light curves. We note that while the CNN estimated periods longer than 80 days that passed the uncertainty cut, this regime is highly contaminated by spurious detections. {These stars' light curves comprise a mixture of relatively flat signals, signals in which the systematics still dominate despite our light corrections, and weak or short-lived rotation signals, perhaps from rapidly evolving active regions}. We leave the vetting of periods longer than 80 days to future work, and for now we consider only shorter periods to be reliable. Figure~\ref{fig:lightcurves} shows a small selection of light curves for which we obtained periods. The periods are plotted against TIC effective temperature in Figure~\ref{fig:ptgrid} to illustrate, for the first time, the distribution of main sequence stellar rotation periods longer than 13 days in TESS.

\begin{deluxetable}{ll}
    \label{tab:norot}
    \centering
\tablecaption{Properties of \nnorot\ Rotationally Nondetected TESS SCVZ Stars}
    \tablehead{\colhead{Label} & \colhead{Description}}
    \startdata
        TIC     & TESS Input Catalog identifier \\
        TOI     & TESS Object of Interest identifier \\
        prov    & period provenance: TESS-SPOC or TASOC \\
        rvar    & photometric activity range $R_\mathrm{var}$ \\
        cdpp    & combined differential photometric precision \\
        Teff\_TIC   & TIC effective temperature \\
        Tmag    & TESS magnitude \\
        contratio   & TIC flux contamination ratio \\
        parallax    & \textit{Gaia} DR3 parallax \\
        ruwe    & \textit{Gaia} DR3 renormalized unit weight error \\
        phot\_g\_mean\_mag  & \textit{Gaia} DR3 apparent $G$ magnitude \\
        bp\_rp  & \textit{Gaia} DR3 $G_{BP} - G_{RP}$ color index \\
        teff    & APOGEE DR17 effective temperature \\
        teff\_err   & temperature uncertainty \\
        m\_h    & APOGEE DR17 metallicity [M/H] \\
        m\_h\_err   & metallicity uncertainty \\
        alpha\_m    & APOGEE DR17 $\alpha$ enhancement [$\alpha$/M] \\
        alpha\_m\_err   & $\alpha$ enhancement uncertainty \\
        star\_bad   & APOGEE DR17 stellar parameter fit flag \\
        snr\_bad    & APOGEE DR17 spectral signal-to-noise flag \\
    \enddata
    
    \tablecomments{The ``star\_bad'' flag represents the APOGEE stellar parameter fit flag, set when a best-fit model is close to a grid edge. It is set for 343 stars. The ``snr\_bad'' flag represents the APOGEE spectral signal-to-noise flag and is set for only 119 stars. This table is available in its entirety in machine-readable format.}
\end{deluxetable}
 
\subsection{Features of the Period Distribution: Biases}
Since the TESS-SPOC sample spans a wider range in temperature than our TASOC sample, we will focus our main discussion of the period distribution on the TESS-SPOC sample. {First, we note an apparent edge} in rotation period that occurs at roughly 27 days. While slow rotators tend to be less active than fast rotators at a fixed temperature, the spot modulation amplitudes at which we expect to lose detections vary in period across temperature. In other words, a period detection edge produced by astrophysical variability should not be flat. Rather, the 27-day detection edge is likely {caused by the stitching of the 27-day light curves}. Without a reliable absolute flux calibration in each sector, stitching sector light curves together can destroy coherent signals longer than 27 days in period. While we include sector-to-sector stitching in all our training sets, the 27-day edge suggests that the training sets do not fully capture the sector effects in TESS, or at the very least the sector effects make period estimates much less certain beyond 27 days.

{We had initially included stars hotter than 6,000 K in the period search, where we noticed another detection edge, this one vertical, at 6,000 K.} The underlying sample distribution has no such edge, so it must be produced by the period search. 6,000 K is the lower bound of the noise source sample used for the TESS-SPOC training set, so above this temperature there is overlap between the training light curves and the ``real" data. It is likely that inclusion in the training set as a noise template (multiple instances with varying injected simulated rotation signals) confuses the neural network and causes it to assign a large uncertainty to these targets. Another possibility is that spot modulation amplitudes drop above 6,000 K, where the convective envelope disappears and stars become less active. This drop in amplitude is seen in the \textit{Kepler} stars of Fig.~\ref{fig:period-amplitudes}. The drop in detections above 6,000 K is likely a combination of these effects, {and the apparent edge affirms our choice to exclude the hotter stars from further analysis.}

\subsection{Features of the Period Distribution: Populations}
The period--temperature distribution displays a sloped, short-period edge, similar to what was seen in \textit{Kepler} \citep{McQuillan2014, Santos2019, Santos2021}. This edge represents the point at which field stars converge onto the slowly-rotating sequence \citep{Curtis2020}. {We measured the slope of the short-period edge using a second-order edge-detection algorithm on the smoothed period distribution in 500 K temperature bins (which allowed for the smallest bin to have at least 100 points) and fitting a line to the edge locations in $\log P_\mathrm{rot}$--$T_\mathrm{eff}$ space. To estimate the uncertainties in the slopes, we bootstrap resampled the temperatures and periods with 1,000 iterations, then applied the edge detection and line fit to the resampled data. For the TESS stars, the edge had a slope of $(-2.28 \pm 0.22) \times 10^{-4}$ dex/K., while the same routine for the \textit{Kepler} stars yielded a slope of $(-2.51 \pm 0.08) \times 10^{-4}$ dex/K, agreeing within 1$\sigma$.}

The distribution also displays a gap in rotation period, occurring at roughly 12 days at 5,000 K and increasing to 20 days at 4,000 K. \citet{McQuillan2013} first identified this gap in the \textit{Kepler} field, and it has also been recovered in other field star samples using K2 \citep{Reinhold2020, Gordon2021}. \citet{Lu2022} showed that the gap may close in fully convective star samples. {To validate the detection of the gap in TESS, following \citet{McQuillan2014} we used a Hartigan's dip test for unimodality \citep{Hartigan1985}. Since the period of the gap varies with temperature (or color), we first transformed the periods to a scale $P_\mathrm{rot} \mapsto P'$ such that the period gap in the \textit{Kepler} stars \citep[using the periods of][]{Santos2019, Santos2021} occurs at $P' = 1$ regardless of \textit{Gaia} $G_{BP} - G_{RP}$ color. We fit a line to points redder than $G_{BP} - G_{RP} > 1.2$ chosen along the gap in the \textit{Kepler} $\log P_\mathrm{rot}$ versus $G_{BP} - G_{RP}$ plane, then divided the each TESS rotation period by the value of the fit line at the star's color to compute $P'$. We performed the Hartigan dip test on the $\log P'$ values, obtaining a $p$-value of 0.001, rejecting the null hypothesis of unimodality with 99.9\% confidence. For comparison, the same analysis of the \textit{Kepler} stars yields a $p$-value of less than $10^{-4}$, while \citet{McQuillan2014} claimed a significant detection of the gap in a 60 K-wide temperature bin with a $p$-value of 0.01.}

\begin{figure*}[!t]
    \centering
    \includegraphics[width=\linewidth]{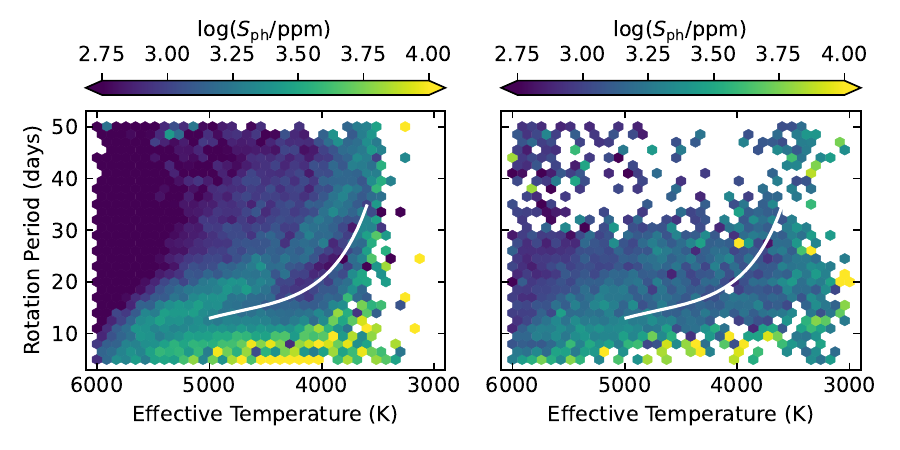}
    \caption{The rotation period distribution versus temperature for both the \textit{Kepler} field \citep[left,][]{Santos2019, Santos2021} and TESS SCVZ (right, this work). The bins are colored by photometric variability index $S_\mathrm{ph}$, {and the white curve represents a 3rd order polynomial fit to the \textit{Kepler} period gap}. As expected, amplitudes generally decrease with increasing period at fixed temperature. The amplitudes near the rotation gap in TESS are slightly smaller than those away from the gap in the same temperature and period ranges, in agreement with \textit{Kepler} as shown left, and with K2 \citep{Reinhold2020}. The TESS bins above a temperature of 4,000 K and above a period of 35 days nearly all have only a single star; this region is sparsely populated, and most periods here are likely to be spurious detections.}
    \label{fig:period-amplitudes}
\end{figure*}

Figure~\ref{fig:period-amplitudes} shows another look at the rotation period distribution, now colored by the photometric variability amplitude $S_\mathrm{ph}$, in comparison with the distribution from the \textit{Kepler} field \citep{Santos2019, Santos2021}. As we expect, stellar variability generally decreases with increasing periods at fixed temperature, since slowly rotating stars are less magnetically active than faster stars. There is a dip in the variability between 3,500 K and 4,500 K, most notably near the location of the rotation period gap, which goes from about (5,000 K, 12 d) and curves upward to (4,000 K, $\sim$20 d) (refer to Figure~\ref{fig:ptgrid}). This is consistent with \citet{Reinhold2020}, who found a similar dip in variability near the period gap in \textit{Kepler} and K2 stars. \citet{Reinhold2019, Reinhold2020} argued that the dip in variability causes the apparent gap in rotation periods, where stars in the gap exhibit modulation too small to be detected in rotation.

\begin{figure*}
    \centering
    \includegraphics[width=0.8\linewidth]{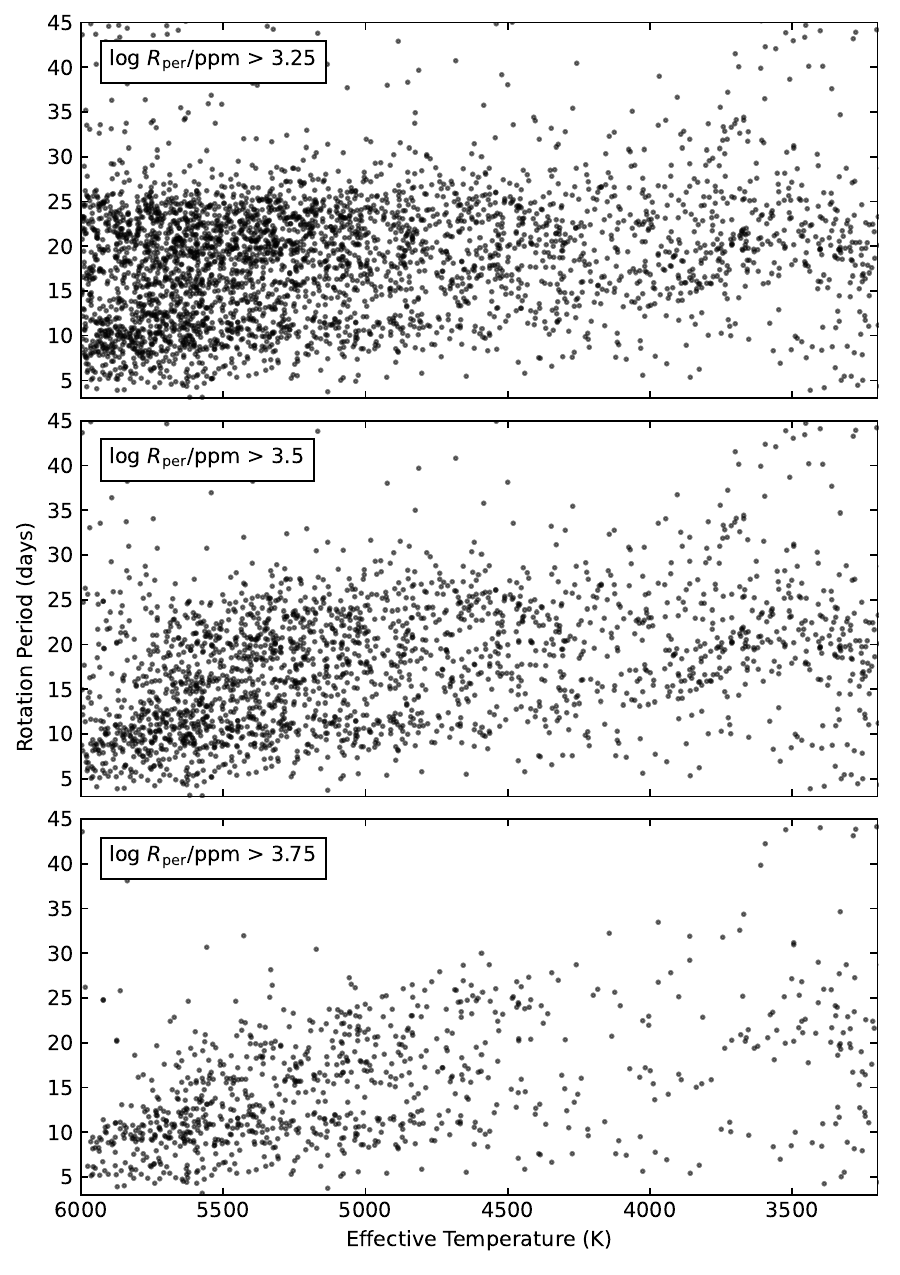}
    \caption{Distribution of periods and TIC temperatures for the TESS-SPOC sample, with varying lower bound on photometric variability range $R_\mathrm{per}$. All stars shown pass our reliability criterion ($\sigma_P/P \leq 35\%$).}
    \label{fig:amplitude_cuts}
\end{figure*}

Figure~\ref{fig:amplitude_cuts} shows the TESS-SPOC period--temperature distribution using different variability range $R_\mathrm{per}$ floor values. Requiring $\log(R_\mathrm{per}/\mathrm{ppm}) > 3.5$ removes many stars from the top-left corner of the diagram, which are hot but have apparently long rotation periods. While we do not expect to find many stars in this regime based on Galactic population synthesis \citep[e.g.,][]{vanSaders2019}, the stars that are here should have low variability because they are hot and therefore have thin-to-nonexistent outer convective envelopes, and because they spin relatively slowly. The stars that are lost from the top panel to the middle panel of Figure~\ref{fig:amplitude_cuts} are likely mostly spurious detections whose measured $R_\mathrm{per}$ is actually the photometric noise, as well as a handful of correctly measured, low-variability stars. As we continue to increase the $R_\mathrm{per}$ floor, we see two effects. First, we lose more low-variability stars on the hot, long-period edge. This is precisely what we expect to see in a period sample: raising the variability floor, we should lose the highest-Rossby number stars first. These are the slowly rotating, hot stars in the top left ``corner" of the distribution. Second, the gap becomes more apparent, consistent with \citet{Reinhold2020}, although stars are not lost from the gap at a significantly higher rate than stars outside the gap.

\subsection{Comparison between TESS-SPOC and TASOC} \label{sec:comparison}
In the TASOC sample (e.g., in the right panel of Figure~\ref{fig:ptgrid}), we again see a weak presence of the period gap as well as the sloped short-period edge. the TASOC sample also show the 27-day horizontal detection edge exhibited by the TESS-SPOC sample, resulting from the increase in uncertainty past 27 days from sector-to-sector stitching.

There are \nboth\ stars in common between the TASOC and TESS-SPOC samples. We estimated two periods for each of these stars using different neural networks fit to different training sets tailored to the different light curve lengths. While the underlying pixel data between the two samples were the same, the apertures used to perform photometry were different, and the TESS-SPOC light curves were more than twice as long (13 sectors) as the TASOC light curves (6 sectors). In addition, the two training sets used different underlying samples of stars for noise and systematics. This gives us a sample to compare period estimates for robustness against photometric aperture, training set, and duration of observation.

\begin{figure}[!t]
    \centering
    \includegraphics[width=\linewidth, trim=0.65cm 0.4cm 0.6cm 0.3cm, clip=true]{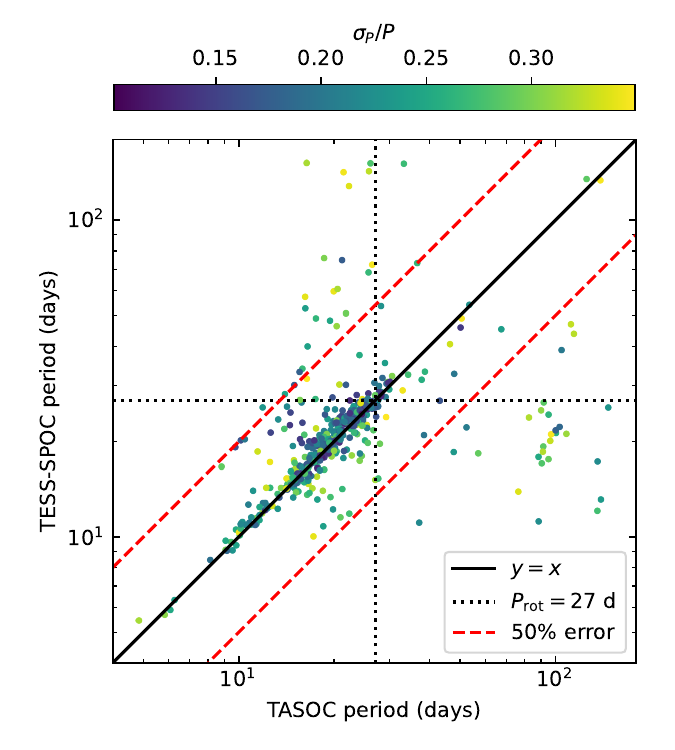}
    \caption{Period comparisons for the \nboth\ stars in both the TESS-SPOC and TASOC samples. The solid black line represents perfect agreement, while the dashed red lines are $\pm 50\%$ error. The black dotted lines are at 27 days on either axis, showing the TESS sector length. There is generally good agreement for most stars, with median percent error of 7\%, and most of the disagreeing estimates have relatively large uncertainty. We note that while we do not include periods greater than 80 days in our analysis or table, we show them here to illustrate at what periods the agreement worsens.}
    \label{fig:spoc-tasoc-compare}
\end{figure}

In Figure~\ref{fig:spoc-tasoc-compare} we compare the period estimates for the overlap sample. They mostly agree, with a median relative error of 7\%. The estimates that disagree have relatively large uncertainties, though the fact that they make our 35\% reliability cut means that there will be some contamination in our period sample. 76\% of stars in the overlap sample have period estimates agreeing to within 20\%. {In absolute units, 50\% of overlap stars agree to within 1.2 d, 75\% agree to within 3.0 d, and 90\% agree to within 9.6 d}. The discrepancies in estimated period likely arise from the different aperture selection, different light curve durations, or differences in the underlying training sets, although here we do not attempt to isolate the main contributor.

\subsection{Comparison with other large field rotation samples}
\begin{figure*}[!t]
    \centering
    \includegraphics[width=0.8\linewidth, trim=0.5in 0.3in 0.8in 0.4in, clip]{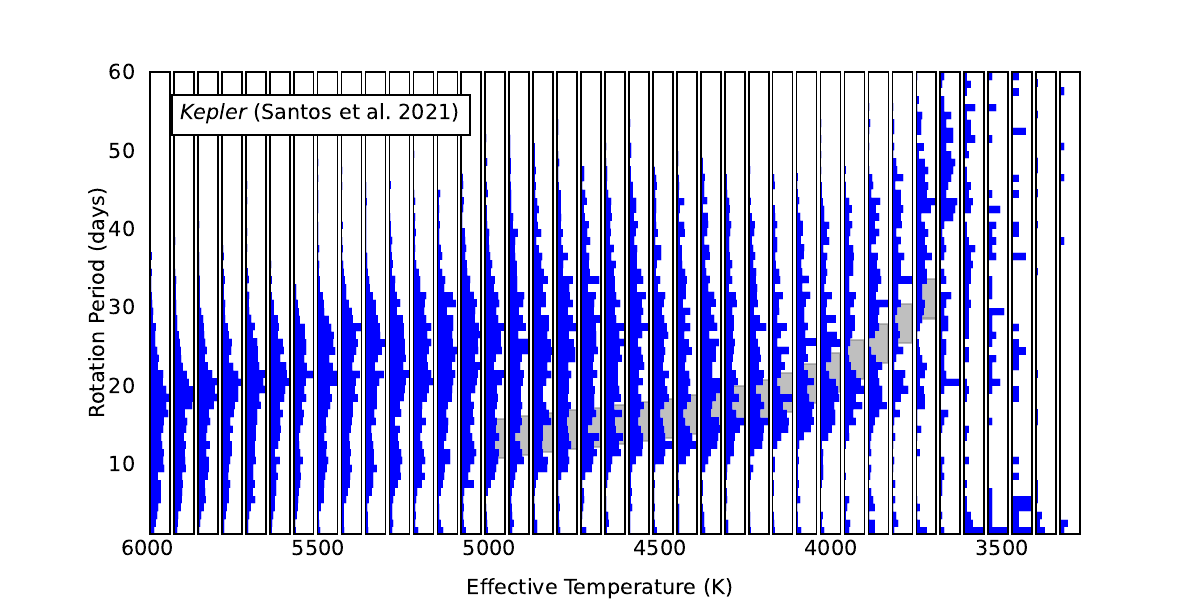}
    \includegraphics[width=0.8\linewidth, trim=0.5in 0in 0.8in 0.4in, clip]{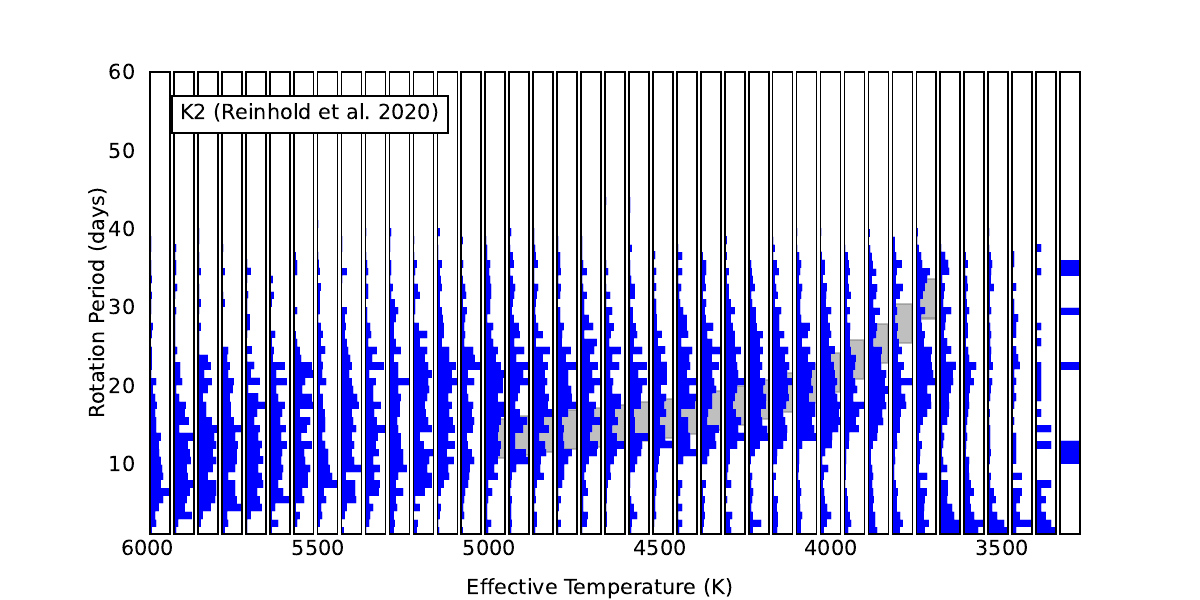}
    \caption{Histogram representations for the period--temperature distribution of \textit{Kepler} \citep[top]{Santos2019, Santos2021} and K2 \citep[bottom]{Reinhold2020}. {Our polynomial fit to the \textit{Kepler} period gap is shown by the gray bands in each histogram, with width of 5 days chosen only for visibility.}}
    \label{fig:histograms1}
\end{figure*}

\begin{figure*}[!t]
    \centering
    \includegraphics[width=0.8\linewidth, trim=0.5in 0.3in 0.8in 0.4in, clip]{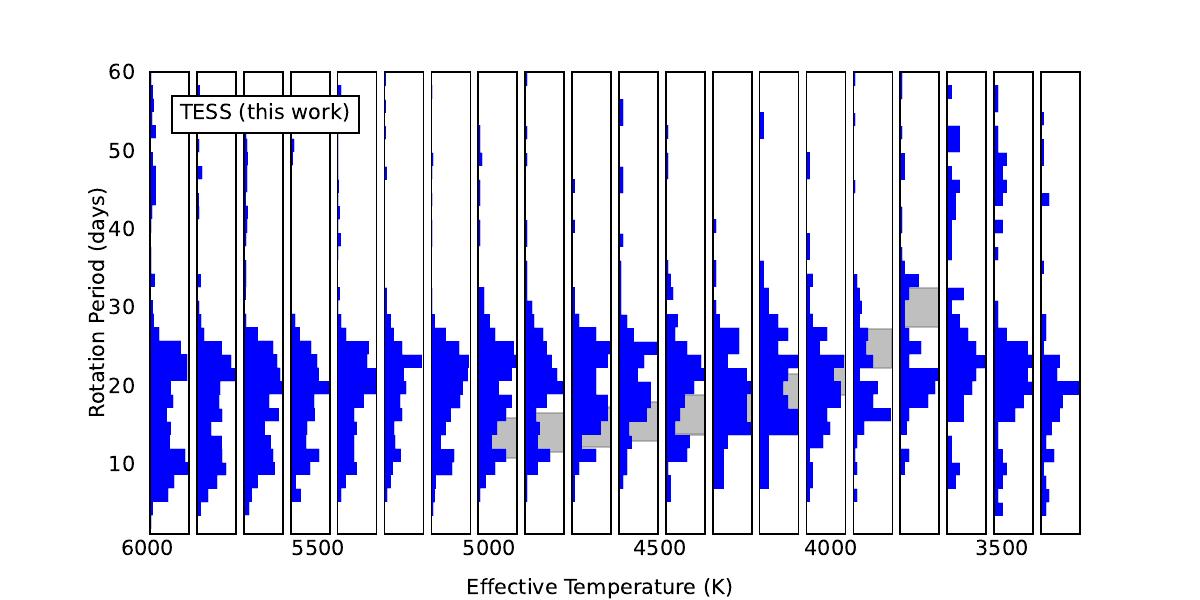}
    \includegraphics[width=0.8\linewidth, trim=0.5in 0in 0.8in 0.4in, clip]{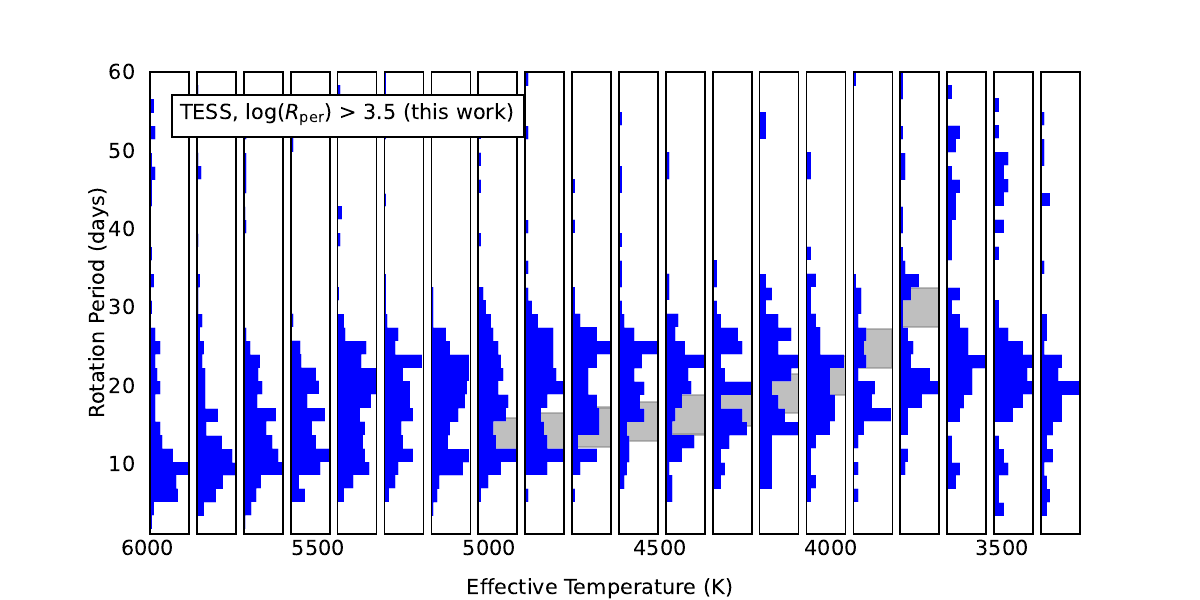}
    \caption{Histogram representations for the period--temperature distribution of TESS, with no photometric variability cut (top), and restricting to $\log (R_\mathrm{per}/\mathrm{ppm}) > 3.5$ (bottom). {Our polynomial fit to the \textit{Kepler} period gap is shown by the gray bands in each histogram, with width of 5 days chosen only for visibility.} Temperatures are from the TIC.}
    \label{fig:histograms2}
\end{figure*}

The TESS rotation period distribution is the product of the underlying distribution of periods, the presence of modulation in the light curve, the availability of data products, and the ability to detect periods across various stellar parameters. To try and understand the relative influence of these effects, we compare the TESS period distribution with other large period data sets, particularly \textit{Kepler} and K2. Figure~\ref{fig:histograms1} shows the period distributions from \textit{Kepler} and K2, while Figure~\ref{fig:histograms2} shows our newly obtained TESS distribution. We represent temperature bins as vertical histograms in the style of \citet{McQuillan2014} to increase the clarity of the period gap in the cool-temperature regime. The number of temperature and period bins is adjusted in each panel to account for the total number of stars in each sample.

The top panel of Figure~\ref{fig:histograms1} displays 42,511 carefully vetted \textit{Kepler} rotation periods from \citet{Santos2021}. The \textit{Kepler} period distribution exhibits a pileup on its upper edge for stars hotter than $\sim 5,500$ K, which is a prediction of the weakened magnetic braking hypothesis \citep{vanSaders2016, vanSaders2019} and has been well-studied in the \textit{Kepler} field \citep{Hall2021, Masuda2022, David2022}. Also present is the rotation period gap, clearly visible at $\sim 15$ days at 5,000 K, $\sim 17$ days at 4,500 K, and continuing to increase and widen at cooler temperatures.

The bottom panel of Figure~\ref{fig:histograms1} shows 13,507 rotation periods from stars in K2 measured by \citet{Reinhold2020}. These represent a high-fidelity subsample with normalized Lomb-Scargle peak height $> 0.5$ and variability range $R_\mathrm{var} > 0.1\%$. Peak heights range from 0 to 1 and quantify how sinusoidal a light curve is, with a perfect sinusoid returning unit peak height, and noisy, non-sinusoidal data returning values close to zero. $R_\mathrm{var}$ is defined similarly to $R_\mathrm{per}$, except that $R_\mathrm{var}$ is the variability range over the entire light curve, rather than a median of ranges per period bin. The K2 distribution shows the period gap most strongly between 5,000 K and about 4,250 K, but it is weakly visible in cooler stars, where it appears to increase in period and widen as in \textit{Kepler}. The hot star pileup is not apparent here. This is likely due to the relatively large temperature uncertainty in the K2 Ecliptic Plane Input Catalog \citep[median 140 K,][]{Huber2016}, which blurs out features in the temperature distribution \citep{vanSaders2019, David2022}. Finally, periods longer than about 35 days are largely absent from the K2 distribution because of K2's 80-day observing campaigns in each field.

The TESS distribution in the top panel of Figure~\ref{fig:histograms2} shows periods for 3,963 TESS-SPOC stars with $\sigma_P/P \leq 35\%$, representing the most credible detections. The period gap is present and is most apparent at temperatures between 4,000 and 5,000 K. It is still visible at temperatures cooler than 4,000 K, but the dearth of reliable detections at periods nearing 35 days makes the gap more difficult to detect. In addition to the gap, we detect a handful of M-dwarfs rotating with periods between 35 and 80 days; similar stars were also observed in the \textit{Kepler} period distribution. We visually inspected the light curves for these stars and confirmed them to be true rotation detections with photometric variability $R_\mathrm{per}$ approaching 1\%. On the hot end, the distribution lacks the long-period edge seen in \textit{Kepler} because of the abundance of hot stars apparently rotating with $\gtrsim$20 d periods. These are likely spurious detections, as their measured amplitudes are close to the noise floor (close to 100 ppm for $Tmag = 8$ and 1\% for $Tmag = 15$). When we raise the variability floor in the bottom panel of Figure~\ref{fig:histograms2}, the hot, slow rotators mostly disappear, but the gap and the slowly rotating M-dwarfs remain.

\begin{figure*}[!t]
    \centering
    \includegraphics[width=0.8\textwidth, trim=0in 0in 0in 0in, clip]{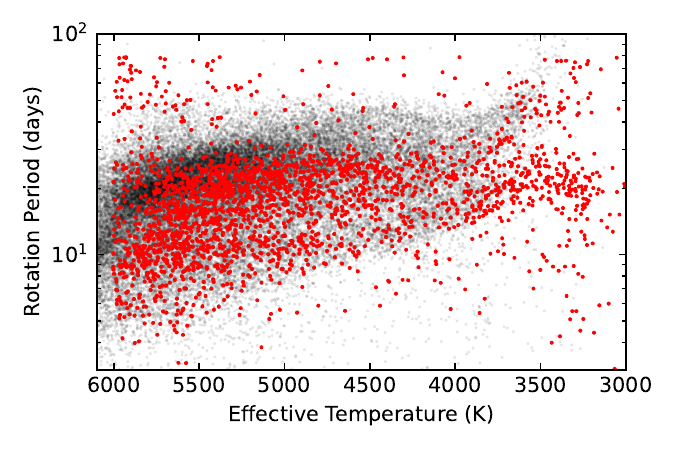}
    \caption{The \textit{Kepler} period--temperature distribution from \citet{Santos2019, Santos2021} (black points) with our new TESS rotation periods overplotted (red points). TESS temperatures are from the TIC.}
    \label{fig:kepler-tess-periods}
\end{figure*}

We offer one final view of the TESS period distribution, now plotted over the \textit{Kepler} distribution of \citet{Santos2019, Santos2021}, in Figure~\ref{fig:kepler-tess-periods}. The short-period edge of the TESS distribution has the same location and slope of \textit{Kepler}'s, suggesting that the edge is a result of rotational evolution, rather than arising from details of the star formation history \citep{Davenport2017, Curtis2020}. The rotation period gap agrees as well, following \textit{Kepler}'s for as long as the TESS gap remains visible into the hot stars. TESS appears to see stars in regions \textit{Kepler} does not: the slowly rotating, hot stars ($T_\mathrm{eff} > 4,000$ K and $P_\mathrm{rot} > 35$ d) have amplitudes close to the noise floor and are likely spurious detections. {Since they are not largely present in the other field star populations, we have flagged these stars as ``spurious" in Table~\ref{tab:periods}, but we note that excluding them from analysis did not affect our results.} On the other hand, the slowly rotating M dwarfs with TESS periods up to 80 days, have been vetted by eye and are mostly real rotation detections. Interestingly, the branch of stars beneath the period gap turns over at temperatures below 3,500 K, which is not seen in \textit{Kepler} but is seen in some younger samples observed by K2 and MEarth \citep{Reinhold2020, Newton2016, Newton2018}.

\vfill
\subsection{Modeling Results} \label{sec:mresults}

Taking the stars with reliable periods from either TESS-SPOC or TASOC, we cross-matched with APOGEE DR17 spectroscopic parameters estimated with ASPCAP \citep{GarciaPerez2016}. To ensure a high quality sample, we removed objects with the ASPCAP STAR\_BAD and SNR\_BAD flags set. We also checked the MULTIPLE\_SUSPECT flag for possible double-lined spectroscopic binaries, but none of our APOGEE rotators were flagged. Some stars in our sample had multiple visits and therefore multiple ASPCAP measurements. For targets with multiple measurements, we averaged the temperatures, metallicities, and $\alpha$ abundances, then added the standard deviation of those measurements in quadrature with the formal ASPCAP uncertainties to obtain an uncertainty for each measurement. This affected 191 targets out of \nspec\ with APOGEE spectroscopy. We then filtered out targets with large Renormalized Unit Weight Error \citep[RUWE $>$ 1.2,][]{Gaia2023} and high flux contamination ratio (TIC \textit{contratio} $>$ 10\%) to clean the sample of potential binary or nearby flux contaminants. This yielded a sample of \ngold\ stars, which we designate as our ``Gold'' sample. We fit models to these stars according to the procedure in Section~\ref{sec:modeling}, taking the posterior medians as the nominal fit parameters. The fit parameters and their uncertainties are presented in Table~\ref{tab:periods}.

\begin{figure}[!t]
    \centering
    \includegraphics[width=\columnwidth]{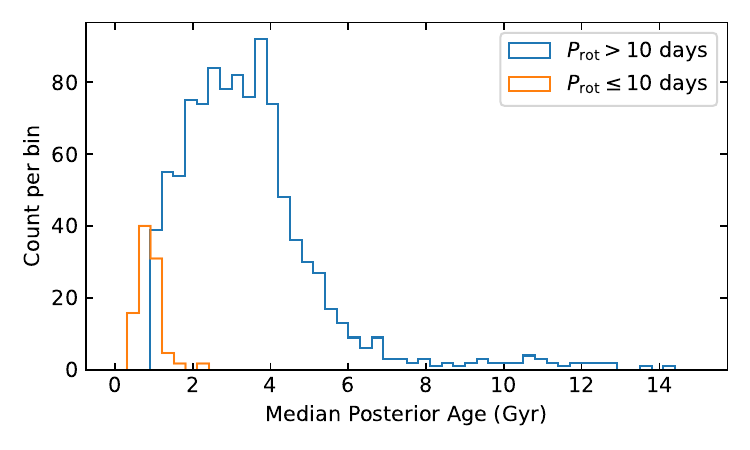}
    \caption{The distribution of rotation-based stellar ages in the TESS SCVZ (the Gold sample in Table~\ref{tab:sample}). The stars shown span an APOGEE temperature range of 3600--6100 K. We separate stars with periods less than 10 days, which are more likely to be tidally synchronized binaries than true rapid rotators \citep{Simonian2019}.}
    \label{fig:ages}
\end{figure}

\subsubsection{The TESS SCVZ age distribution} \label{age_dist}
The ages for our stars, which are estimated using our TESS rotation periods, are shown in Figure~\ref{fig:ages}. We separate stars with rotation periods less than 10 days, which in \textit{Kepler} were more likely to be spun-up by close binary companions than be true rapid rotators \citet{Simonian2019}. The age distribution peaks between 2 and 4 Gyr, which is consistent with other age distributions of Solar neighborhood stars: \citet{Buder2019} used isochrones for GALAH stars, \citet{Claytor2020} used rotation-based ages for \textit{Kepler} dwarfs, \citet{Berger2020} used isochrones for \textit{Kepler} dwarfs, \citet{SilvaAguirre2018} used asteroseismology for \textit{Kepler} giants, and \citet{Mackereth2021} used seismology in the TESS CVZs; all obtained a distribution peaking between 2 and 4 Gyr. We note that our age distribution lacks many of the old ($> 6$ Gyr) stars seen in other samples. This is a consequence of two detection biases: (1) our 27-day detection edge prevents the reliable detection of slowly-rotating stars hotter than 4,000 K, and (2) old stars rotate more slowly and are less active, further complicating their detection in rotation. 

\begin{figure*}[!t]
    \centering
    \includegraphics[width=0.9\textwidth]{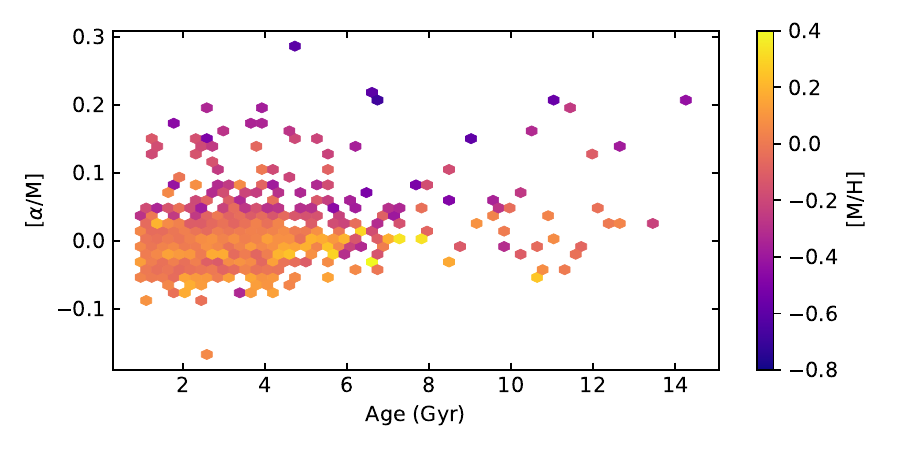}
    \caption{Stellar $\alpha$-element abundances versus age in the TESS SCVZ (the Gold sample in Table~\ref{tab:sample}), tracing Galactic chemical evolution of $\alpha$ enhancement from supernovae. Consistent with trends seen in the \textit{Kepler} field \citep[e.g.,][]{SilvaAguirre2018, Claytor2020}, there is a weak trend of increasing $\alpha$ abundance with age below 8 Gyr associated with the classical ``thin'' Galactic disk. We expect but do not see a strong trend of increasing $\alpha$ abundance with age above 8 Gyr (the classical ``thick'' disk), but this because we detect very few older stars due to (1) our 27-day detection edge, and (2) slow rotation and weak activity making stars more difficult to detect.}
    \label{fig:alpha-age}
\end{figure*}

\subsubsection{Galactic chemical evolution}
With rotation-based ages and high-resolution APOGEE spectroscopic abundances, we can also look for Galactic chemical evolution trends in TESS \citep[e.g.,][]{SilvaAguirre2018, Claytor2020}. Stars' initial composition patterns are set by the compositions of the clouds in which they form, which are in turn enriched by stars that have lived and died before them. Galactic chemical evolution is often traced using the relative abundances of $\alpha$ elements (e.g., O, Mg, Ca) to metals and metals to hydrogen \citep{Bensby2014}. APOGEE provides values for [$\alpha$/M] and [M/H], which we adopt. The background Galactic composition was governed by the dominance of core-collapse supernovae in the early Milky Way, followed by dominance of type Ia supernovae beginning about 8 Gyr ago \citep{Feltzing2017}. Both types of supernovae enriched the interstellar medium with metals, but in different ratios. Consequently, stars display decreasing [$\alpha$/M] and increasing [M/H] with time (reversed as a function of age). We therefore expect old stars to have low metallicity but high $\alpha$ enhancement, while young stars should have higher metallicity and lower $\alpha$-element abundances \citep{Haywood2013, Bensby2014, Martig2016, Feltzing2017, Buder2019}. These young and old populations are representative of the classical Galactic ``thin" and ``thick" disks, respectively.

Figure~\ref{fig:alpha-age} shows stellar $\alpha$-element abundance as a function of rotation-based age in the TESS SCVZ. As expected, young stars are generally $\alpha$-poor and metal-rich. There is a slight increasing trend of $\alpha$ enhancement with age. We detect very few stars in rotation older than 6 Gyr due to the detection biases discussed in Section \ref{age_dist}. Finally, we also detect a few young $\alpha$-rich stars. These are known from other samples \citep[e.g.,][]{Martig2015, SilvaAguirre2018, Claytor2020} and are likely to be the products of stellar mergers. In this scenario, two old, $\alpha$-enhanced stars merge, destroying the stars' rotation histories and yielding a fast-rotating, apparently young, $\alpha$-enhanced product \citep{Zhang2021}.

\subsubsection{Stellar activity}
Finally, with rotationally characterized stars we can begin to investigate trends of photometric activity with model-derived parameters like age and Rossby number. We define the Rossby number as the ratio of the rotation period over the convective overturn timescale $\tau_\mathrm{cz}$, where the convective timescale is computed from our models as the pressure scale height $H_P$ divided by the convective velocity evaluated at a distance $H_P$ above the base of the convection zone. To quantify the photometric activity, we use the photometric activity index $S_\mathrm{ph}$ rather than $R_\mathrm{per}$ so that we can compare the trends in the TESS SCVZ with those in the \textit{Kepler} field observed by \citet{Santos2019, Santos2021}, and \citet{Mathur2023}. The ages and and Rossby numbers for the Mathur et al. sample were computed using the same procedure and models underlying this work, so we can directly compare the \textit{Kepler} and TESS distributions. We start with the Gold sample and discard stars with periods less than 10 days as before, leaving \nrossby\ stars with TESS periods, APOGEE spectroscopic parameters, and well-determined Rossby numbers and ages.

\begin{figure}[!t]
    \centering
    \includegraphics[width=\linewidth]{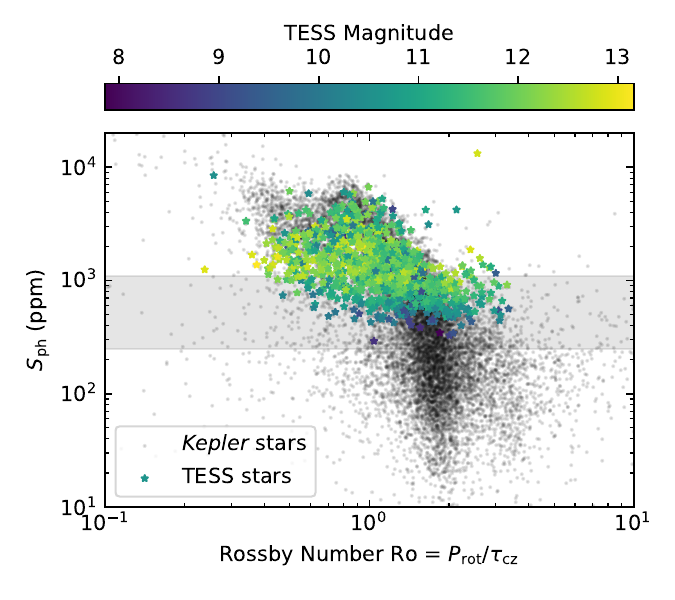}
    \caption{The stellar activity index $S_\mathrm{ph}$ versus Rossby number for \nrossby\ stars in our Gold sample with periods greater than 10 days, plotted over the \textit{Kepler} sample of \citet{Santos2019, Santos2021} and \citet{Mathur2023}. As expected from theory and as seen in the \textit{Kepler} field, photometric activity decreases with increasing Rossby number. The gray region denotes our measured TESS noise floor, ranging from 250 ppm at 8th magnitude to 1,100 ppm at 13th magnitude. We do not detect TESS stars with amplitudes as low as \textit{Kepler} due to TESS's worse photometric precision and therefore higher noise floor.}
    \label{fig:rossby}
\end{figure}

\begin{figure}[!t]
    \centering
    \includegraphics[width=\linewidth]{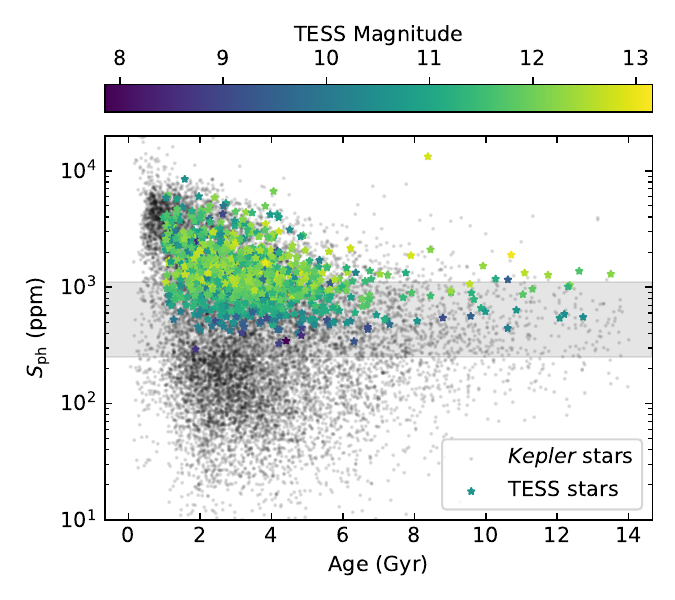}
    \caption{The stellar activity index $S_\mathrm{ph}$ versus rotation-based age for \nrossby\ stars in our Gold sample with periods greater than 10 days, plotted over the \textit{Kepler} sample of \citet{Santos2019, Santos2021} and \citet{Mathur2023}. Photometric activity decreases as stars age, an effect of spin-down seen in the \textit{Kepler} field and predicted by theory.}
    \label{fig:activity}
\end{figure}

Figure~\ref{fig:rossby} shows the photometric activity index $S_\mathrm{ph}$ versus the Rossby number for our binary-cleaned stars, plotted over the distribution of stars from \textit{Kepler}. Activity decreases with increasing Rossby number, as expected. The TESS distribution generally agrees with the \textit{Kepler} distribution. We have a few stars close to the high-activity saturated regime \citep[e.g.,][]{Wright2011}, but most of our stars are magnetically unsaturated. The TESS detection limits are clear here, as our lowest-activity star with a period detection has $S_\mathrm{ph} = 345$ ppm, compared to \textit{Kepler}'s lower limit in the tens of ppm. We have a few hot stars at Ro $\gtrsim 2$ where \textit{Kepler} has almost none. These are the likely spurious period detections from before (e.g., Figure~\ref{fig:ptgrid}).

We show $S_\mathrm{ph}$ as a function of rotation-based age in Figure~\ref{fig:activity}. Photometric activity decreases with age, an effect of stellar spin-down. The TESS distribution follows the range and morphology of the \textit{Kepler} distribution all the way down to the TESS rotation detection limit of $S_\mathrm{ph} \approx 350$ ppm.

\section{Detectability of Rotation} \label{sec:detectability}

Here we consider the detectability of rotation as a function of fundamental stellar parameters. At a fixed rotation period, at lower temperature and higher metallicity, we expect deeper convective envelopes, stronger magnetism, more surface spots, and more easily detectable rotational modulation. Besides changing with static stellar parameters, the strength of a star's magnetism changes as the star ages. Main-sequence stars with outer convective envelopes are born spinning fast and spin down as they age as they lose angular momentum to magnetized stellar winds \citep{Kraft1967, Skumanich1972}. The decrease in rotation speed results in a weakening magnetic field, fewer or smaller spots, and less flux modulation, making rotation more difficult to detect in older stars than in younger stars at fixed mass and composition. 

While we might expect rotation to be harder to detect in lower metallicity stars, an age dependence enters the picture because of the variation of Galactic composition with age. Because the background abundance ratios change with time, any apparent changes in rotation detectability with stellar abundances may actually be caused by the decreasing detectability with age.

To investigate the detectability of rotation with fundamental stellar properties, we consider the fraction of targets for which we detected periods in stellar parameter bins. While the CNN infers a period for each target, we can use the estimated uncertainty to determine whether those periods are reliable. As in \citet{Claytor2022}, we label targets with $\sigma_P/P < 0.35$ (corresponding to $\sim$10\% median percent error) as successful detections, and anything else as a nondetection.

\begin{figure*}[!t]
    \centering
    \includegraphics[width=0.7\textwidth]{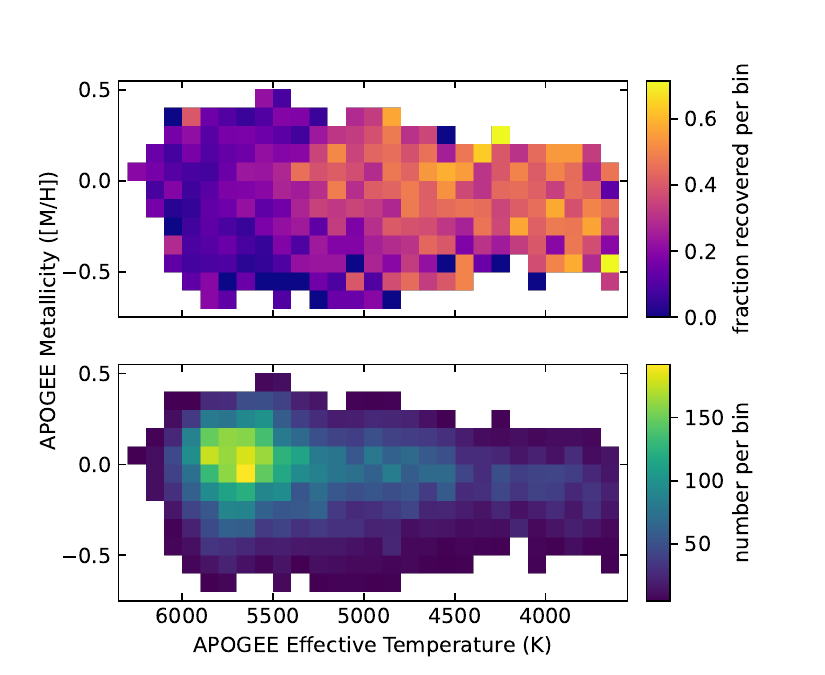}
    \caption{The detectability of rotation across APOGEE temperature and metallicity. {The color bar corresponding to the detection fraction has a maximum value of 70\%}. We preferentially detect rotation in cool stars, which have deeper convective envelopes and are more active. When the detectability drops off as a function of temperature, we see a weak trend of increasing detectability with increasing metallicity. This may be due to metallicity increasing the convective depth at fixed temperature. It may also be an age effect, as young, active stars tend to be more metal-rich.}
    \label{fig:teff_recovery}
\end{figure*}

Figure~\ref{fig:teff_recovery} shows the rotation detection fraction versus temperature and metallicity for all our rotationally-searched stars with APOGEE spectroscopy. Only bins with at least five targets are shown so that the diagram is not muddled by small number fluctuations. As expected, cooler stars, especially cooler than 5,000 K, are detected in period significantly more often than hotter stars. In the range $5,000~\mathrm{K} < T_\mathrm{eff} \lesssim 6,000~\mathrm{K}$, where the detections begin to decrease as a function of temperature, there appears to be a weak trend in metallicity, with higher-metallicity ([M/H] $\gtrsim -0.1$) stars being detected in period more frequently than lower-metallicity stars. This is consistent with \citet{Claytor2020, Amard2020, See2021}, who found that rotation was more easily detected in \textit{Kepler} stars with higher metallicity at fixed mass. We see the same bias toward higher metallicity among our rotators, which may be due either to the deeper convective envelope resulting from enhanced opacity, or to more rapid rotation (and therefore higher activity) from increased moment of inertia and slower spin-down.

\begin{figure*}[!t]
    \centering
    \includegraphics[width=0.8\textwidth]{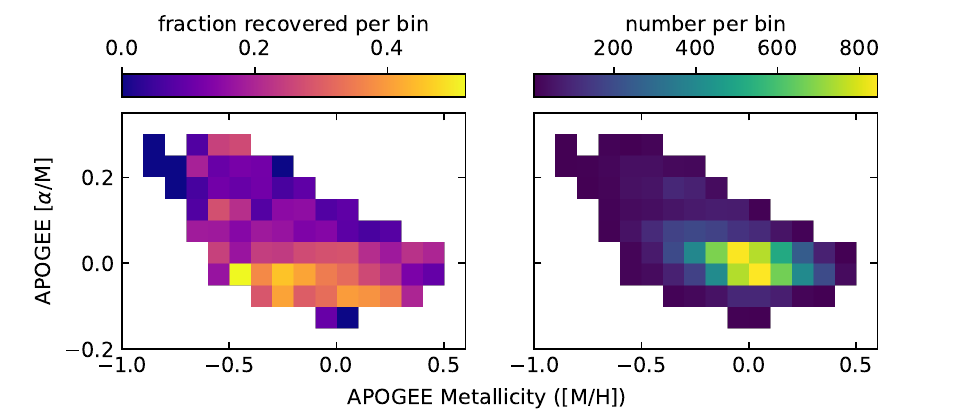}
    \caption{The detectability of rotation across APOGEE metallicity and $\alpha$-element abundace. {The color bar corresponding to the detection fraction has a maximum value of 50\%}. We detect fewer stars in rotation at high [$\alpha$/M] because these stars tend to be old, slowly rotating, and less active. At fixed [$\alpha$/M], there is scatter in the trend of detectability with metallicity from bin to bin.}
    \label{fig:met_recovery}
\end{figure*}

Another view of the detection fraction is shown in Figure~\ref{fig:met_recovery}, this time as a function of metallicity and $\alpha$-element enhancement. At fixed metallicity, we detect fewer stars in rotation at high [$\alpha$/M] due to the underlying relationship between age and $\alpha$ enhancement. High-$\alpha$ stars tend to be older, spin more slowly, and are less active, so we expect them to be more difficult to detect in rotation. This view also allows us to inspect the period detection fraction across metallicity at fixed $\alpha$ enhancement. At fixed [$\alpha$/M], there is significant scatter in the detection fraction across metallicity. Some bins (e.g., 0 $<$ [$\alpha$/M] $<$ 0.05) show gradually increasing detectability with increasing metallicity, while others (e.g., -0.05 $<$ [$\alpha$/M] $<$ 0) worsen in detection at higher metallicity. Due to the amount of noise in the bins, it is difficult to conclude whether the apparently enhanced detection fraction at higher metallicity is caused by higher activity from deeper convection zones or by the underlying age distribution.

\section{Spot Filling Fraction} \label{sec:spots}

The links between temperature, metallicity, age, convection, rotation, and photometric variability shed light on the generation of magnetism in cool, main-sequence stars. The strength of rotational modulation in the light curve, and therefore the detectability of rotation, hint at the presence of cool spots created by magnetic fields concentrated near the stellar surface. Because spots are created by the same dynamo that rotation and convection drive, we can use the prevalence of spots in different temperature and rotation ranges to infer dynamo properties in those regimes.

\citet{Cao2022} found that temperature-sensitive spectral features include contributions from the quiet photosphere and cooler spots. Thus, fitting APOGEE spectra with two temperature components, they inferred the surface spot filling fractions and the temperature contrasts of a sample of stars. They used a modified version of the FERRE code \citep{AllendePrieto2006}, the spectral fitting code used by the ASPCAP pipeline, to infer spot filling fractions for all stars in APOGEE DR17. Following this method, we obtained spot filling fractions and updated effective temperatures for the stars in our sample with APOGEE spectra.

We began with the \ngold\ stars in our Gold sample (described in Section \ref{sec:mresults}). We made cuts in \textit{Gaia} DR3 magnitudes and colors using $M_G > 4.4$ and $G_{BP} - G_{RP} > 1$ to target below the field main-sequence turnoff and ensure all our stars are securely on the main sequence. This yielded \ncool\ cool, main-sequence stars, but a few (less than 20) showed an excess in $M_G$, indicating that they were likely leftover binary systems \citep[e.g.,][]{Berger2018}. To remove these, we fit a line to the main sequence and computed the magnitude excess as $\Delta M_G = M_G - \langle M_G \rangle$, where $\langle M_G \rangle$ was the fit main-sequence magnitude. The distribution of magnitude excesses had two clear peaks, with a trough we visually identified at $\Delta M_G = -0.4$. We removed stars with $|\Delta M_G| < 0.4$, leaving \nplatinum\ stars. We designate these as our ``Platinum" sample, a pure, cool, main-sequence sample robustly free from binary contamination. 

\begin{figure}[!t]
    \centering
    \includegraphics[width=\linewidth]{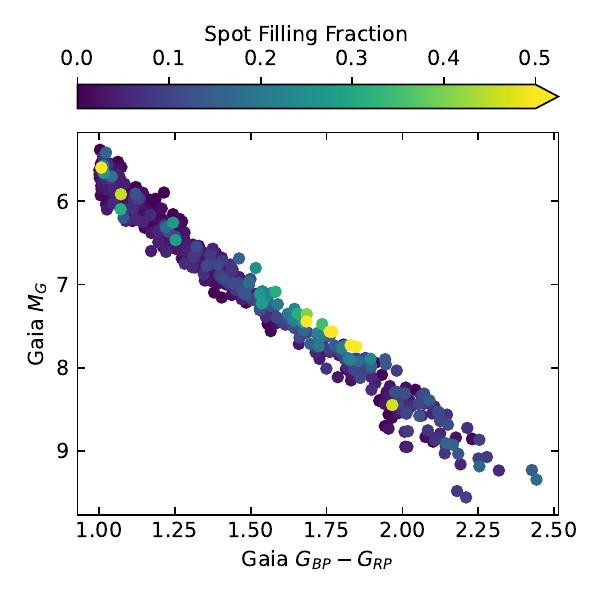}
    \caption{\textit{Gaia} DR3 color-magnitude diagram of our Platinum cool main-sequence sample, colored by surface spot filling fraction. The sample is carefully cleaned of potential binary systems, which can interlope as falsely high spot filling fractions.}
    \label{fig:platinum}
\end{figure}

We show the Platinum sample on a \textit{Gaia} color-magnitude diagram in Figure~\ref{fig:platinum}, with points colored by spot filling fraction. 
While most stars in our sample have spot filling fractions less than 10\%, the mid-K range ($1.5 < G_{BP} - G_{RP} < 2$) exhibits elevated fractions. Here, filling fractions reach $\approx$ 0.3--0.5, behavior that was first observed by \citet{Cao2023}, which they attributed to internal differential rotation. {In this color range, there is a weak but nonzero correlation (Spearman $r = -0.47$ with $p \sim 10^{-8}$) of increasing filling fraction with decreasing $M_G$}. This may represent an increase of spot coverage with increasing metallicity in this temperature regime; a correlation between spot filling fraction and metallicity {(Spearman $r = 0.65$ with $p < 10^{-16}$)} is still present after strong binary rejection, and the trend disappears outside this temperature range. {Six stars in this sample have spot filling fractions above 50\%: TIC IDs 149449414, 270429868, 293935949, 309791920, 364326409, and 370114295.}

\begin{figure*}[!t]
    \centering
    \includegraphics[width=0.8\textwidth]{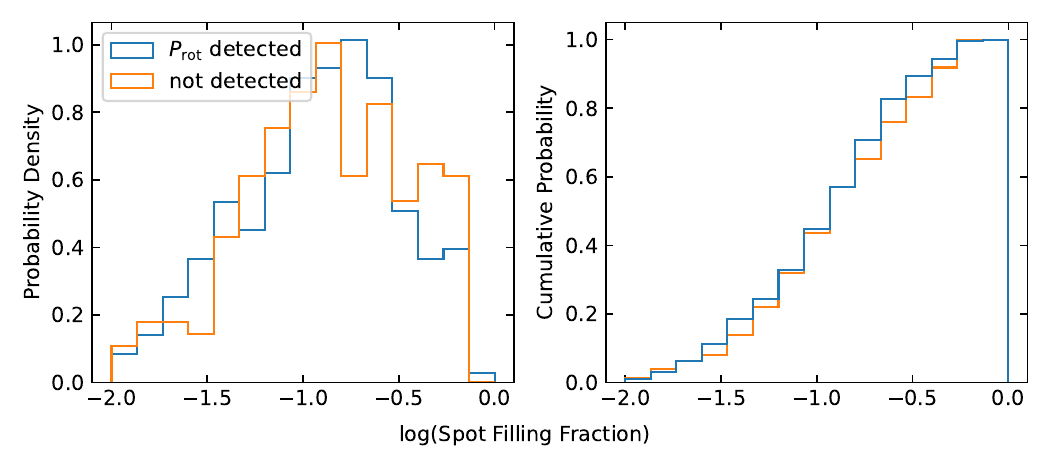}
    \caption{The distribution of spot filling fractions separated by whether rotation was detected in the \textit{Gaia} color range $1.5 < G_{BP} - G_{RP} < 2$. While we might expect more spotted stars to be easier to detect in rotation, there are too few stars to draw any statistically significant conclusions.}
    \label{fig:fspot_detection}
\end{figure*}

With spot filling fractions, we can now investigate the detectability of rotation as a function of surface spot coverage. Figure~\ref{fig:fspot_detection} shows the \nkdetected\ Platinum sample K-dwarfs with $1.5 < G_{BP} - G_{RP} < 2$, along with the \nknondet\ stars in the same regime but with no rotation detection. The left panel shows the subsamples' distributions of spot filling fractions, while the right panel shows the cumulative frequency distributions. A Kolmogorov–Smirnov test returns a $p$-value of 0.3, rejecting the null hypothesis that the two samples are drawn from the same underlying distribution with only 70\% (i.e., just over 1$\sigma$) significance. There are too few stars in this regime to confirm any difference in spot filling fraction between the period detection and nondetection samples.

\begin{figure*}[!t]
    \centering
    \includegraphics[width=0.8\textwidth]{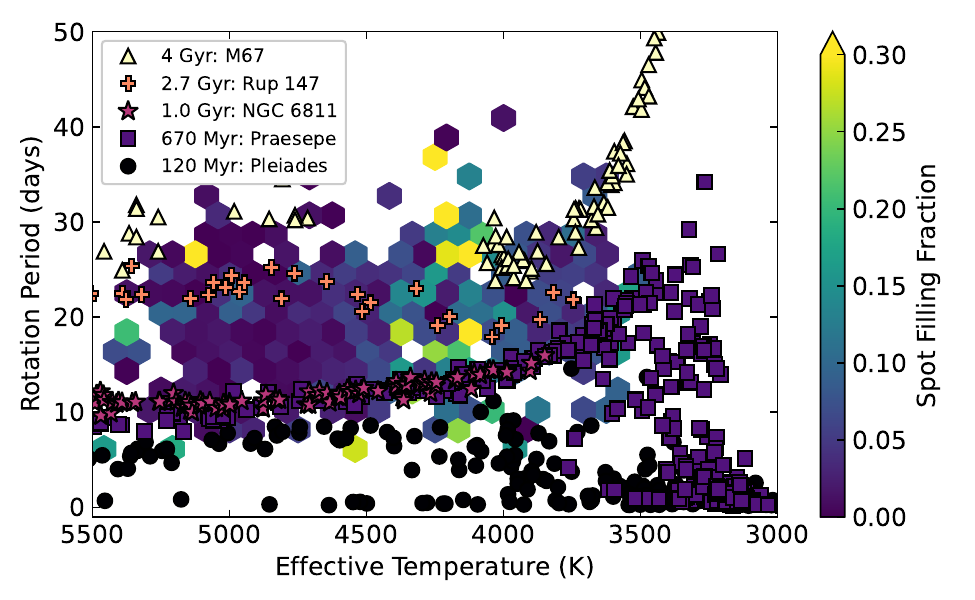}
    \caption{The period distribution of our Platinum sample, colored by spot filling fraction, and plotted with rotation sequences of benchmark open clusters. In the K temperature range (between 4,000 and 4,500 K), NGC 6811 and Rup 147 exhibit stalled rotational braking, a departure from current gyrochronological models. The spot filling fractions are elevated here as well.}
    \label{fig:clusters}
\end{figure*}

\citet{Cao2023} suggested that core-envelope decoupling gives rise to anomalous rotation behavior in cool stars, evidenced by elevated spot filling fractions in cluster K-dwarfs between 4,000 and 4,500 K. The process of decoupling and recoupling drives a radial shear layer and enhanced surface magnetism. With field star rotation periods up to 27 days, we can investigate the behavior of rotation and spottedness in the TESS SCVZ. Figure~\ref{fig:clusters} shows the period--temperature distribution of our Platinum sample, again colored by spot filling fraction, with the rotation sequences from benchmark open clusters Pleiades \citep{Rebull2016}, Praesepe \citep{Douglas2017, Douglas2019}, NGC 6811 \citep{Curtis2019}, Ruprecht 147 \citep{Curtis2020}, and M67 \citep{Barnes2016, Dungee2022}. Here we use the two-component fit effective temperature, rather than the TIC or ASPCAP values, for consistency with the spot filling fractions. As a function of temperature, spot filling fractions increase in the mid-K range---the same behavior \citet{Cao2023} identified in Praesepe. At fixed temperature, we might expect filling fractions to be higher at shorter periods, where stars rotate faster and are more magnetically active. Instead, spot filling fractions in the mid-K range appear to be elevated across the entire span of recovered periods ($\sim$10--35 d). \citet{Cao2023} predict shear-enhanced magnetism to persist in this temperature range until ages of a few Gyr (temperature-dependent), so it is likely we do not reach long enough periods for the spot filling fraction to decrease.

The increase in spot filling fraction occurs in the color and temperature range where open clusters NGC 6811 and Rup 147 were shown to exhibit an unexpected epoch of stalled rotational braking \citep{Curtis2019, Curtis2020}. NGC 6811 is 1 Gyr old \citep{Curtis2019}, but for temperatures cooler than 5,000 K its rotation sequence rests upon that of the 670-Myr-old Praesepe \citep{Douglas2017, Douglas2019}. Somewhere between the ages of these clusters, stellar spin-down departs from the classical picture of gyrochronology. By 2.7 Gyr \citep[Rup 147,][]{Curtis2020}, stars at the hot end have resumed braking, but the cooler stars lag behind, suggesting that the epoch of stalled braking lasts longer for lower-mass stars \citep[e.g.,][]{Gallet2015, Lanzafame2015, Somers2016, Spada2020, Cao2023}.

\citet{Spada2020} showed that a two-zone interior model, which allows the core and envelope to rotate at different rates, can nearly reproduce the stalled spin-down behavior exhibited by these clusters. In these models the core and envelope decouple, and the envelope continues to spin down from magnetic braking while the core maintains its rotation speed. During recoupling, angular momentum is transferred from the core to the envelope, and the apparent spin-down is temporarily slowed or halted. After recoupling, the star again behaves as a solid body and undergoes classical \citep[i.e., power law,][]{Skumanich1972} braking. While \citet{Curtis2020} argued in favor of the two-zone model, they could not rule out a temporary reduction in the braking torque, either from reduced wind or weakening of the magnetic field, as a possible cause. We suggest that the coincidence of elevated spot filling fractions in field stars with the stalled braking seen in open clusters supports the shear-driven dynamo hypothesis argued by \citet{Cao2023}. 

\section{Summary \& Conclusion} \label{sec:conclusion}
We used deep learning to infer reliable periods for \nperiods\ main sequence stars near the southern ecliptic pole from year-long TESS full-frame image light curves. Our periods represent the first large-scale recovery and measurement of rotation in TESS stars rotating more slowly than 13.7 days, the limit previously imposed by TESS's complicated systematics. We fit stellar evolutionary models to the stars using rotation and high-resolution spectroscopic parameters to determine stellar ages, masses, convection timescales, Rossby numbers, and more. We investigated the detectability of rotation as a function of fundamental stellar parameters as well as new spot filling fractions inferred from spectroscopy. Our key results and conclusions are as follows:

\begin{itemize}
    \item We find evidence for the intermediate rotation period gap, first discovered in the \textit{Kepler} field and seen in K2 field star samples across the ecliptic plane, the first such detection from TESS stars. The period gap in TESS closely aligns with the gaps from previous missions, cementing the conclusion that the gap is a product of stellar structure and evolution and not star formation history.
    \item The rotation period gap coincides with a dip in photometric variability, consistent with the findings of \citet{Reinhold2019, Reinhold2020} in other field star populations. 
    \item The distribution of rotation periods in TESS closely resembles the distributions seen by \textit{Kepler} and K2. Its lower edge features a slope of increasing period with decreasing temperature, similar to the distributions from previous missions, and we detect slowly rotating M-dwarfs with a similar location and distribution as in \textit{Kepler}.
    \item We detect a higher fraction of stars in rotation at cooler effective temperatures, where stars rotate faster at fixed age and have deeper convective envelopes resulting in higher activity amplitudes. We also preferentially detect rotation in stars at higher metallicities at fixed temperature. This may owe to deepening convective envelopes with increasing metallicity, or to increased moment of inertia with increasing metallicity resulting in slower spin down, and faster period (and therefore higher activity) at fixed age.
    \item In \textit{Gaia} color regimes with a range of spot filling fractions, stars detected in rotation showed no significant difference in spot filling fraction compared to stars with no period detection.
    \item Field stars with elevated spot filling fractions coincide with open cluster stars that exhibit a temporary stall in magnetic braking. These coincide at least partly with the period gap and its variability depression, suggesting a common cause.
\end{itemize}

While TESS systematics have presented unique challenges that remain difficult to solve with conventional period-finding techniques, deep learning presents a way to circumvent instrument systematics without having to solve systematics removal for every individual case. Since first observing the southern hemisphere in 2018, TESS has also observed the North, revisited both hemispheres, and continues to observe the entire sky in its search for transiting exoplanets. As it does, it continues to build a vast trove of stellar light curves to search for rotation in stars across the entire sky.

Our simulation-driven CNN approach enables the inference of more than just rotation. The existing training sets include activity level, latitudinal differential rotation, spot lifetimes, and activity cycles. These quantities can be probed with minimal modification to our CNN framework and would provide new avenues of investigation of stellar rotational, activity, and magnetic evolution.

Understanding the complicated rotational evolution of low-mass stars and the related anomalies in activity and spot coverage will require more rotation periods for more diverse populations of stars. As we grow the number of rotation periods obtained with TESS, precise and homogeneously derived temperatures and metallicities will be imperative to pinpoint the regimes where stellar rotation and activity processes change. The Milky Way Mapper (MWM) of the Sloan Digital Sky Survey V \citep{Kollmeier2017} is obtaining APOGEE spectroscopy for 6 million stars across the whole, including 300,000 observed with TESS two-minute cadence in the SCVZ. MWM will provide homogeneous temperatures, metallicities, and detailed chemical abundances for all these stars, offering unprecedented precision on the fundamental parameters of a large rotation sample.

Upcoming space missions will provide crucial avenues to rotation periods as well. The methods in this work will be applicable to photometry obtained by the \textit{Nancy Grace Roman Space Telescope} \citep{Spergel2015}. \textit{Roman} will perform a Galactic Bulge Time Domain Survey \citep{Penny2019, Johnson2020} with cadence similar to TESS with the addition of lower cadence photometry in at least one secondary band. Not only will a rotation be made accessible in a relatively unprobed population of stars toward the Galactic bulge, but the multi-band coverage will provide access to time-domain temperature resolution, enabling the study of stellar spot and facula distributions for hundreds of thousands of stars. Furthermore, the potential to observe two globular clusters near the Galactic center with \textit{Roman} \citep{Grunblatt2023} would provide the first large gyrochronology anchors at both old ages and sub-Solar metallicities.

\section{Acknowledgments}
We thank the anonymous reviewer for insightful comments that improved the quality of this manuscript.

We gratefully acknowledge Gagandeep Anand, Ashley Chontos, Monique Chyba, Ryan Dungee, Rafael Garcia, Daniel Huber, Corin Marasco, Savita Mathur, Peter Sadowski, {\^A}ngela Santos, Jessica Schonhut-Stasik, Benjamin Shappee, Xudong Sun, and Jamie Tayar for helpful conversations that improved the quality of this manuscript.

The technical support and advanced computing resources from the University of Hawai‘i Information Technology Services | Cyberinfrastructure are gratefully acknowledged.

Z.R.C. and J.v.S. acknowledge support from the National Aeronautics and Space Administration (80NSSC21K0246, 80NSSC18K18584)

This paper includes data collected by the TESS mission. Funding for the TESS mission is provided by the NASA's Science Mission Directorate.

Funding for the Sloan Digital Sky Survey IV has been provided by the Alfred P. Sloan Foundation, the U.S. Department of Energy Office of Science, and the Participating Institutions. 
SDSS-IV acknowledges support and resources from the Center for High Performance Computing  at the University of Utah. 
The SDSS website is www.sdss4.org. 
SDSS-IV is managed by the Astrophysical Research Consortium for the Participating Institutions of the SDSS Collaboration including the Brazilian Participation Group, the Carnegie Institution for Science, Carnegie Mellon University, Center for Astrophysics | Harvard \& Smithsonian, the Chilean Participation Group, the French Participation Group, Instituto de Astrof\'isica de Canarias, The Johns Hopkins University, Kavli Institute for the Physics and Mathematics of the Universe (IPMU) / University of Tokyo, the Korean Participation Group, Lawrence Berkeley National Laboratory, Leibniz Institut f\"ur Astrophysik Potsdam (AIP),  Max-Planck-Institut f\"ur Astronomie (MPIA Heidelberg), Max-Planck-Institut f\"ur Astrophysik (MPA Garching), Max-Planck-Institut f\"ur Extraterrestrische Physik (MPE), National Astronomical Observatories of China, New Mexico State University, New York University, University of Notre Dame, Observat\'ario Nacional / MCTI, The Ohio State University, Pennsylvania State University, Shanghai Astronomical Observatory, United Kingdom Participation Group, Universidad Nacional Aut\'onoma de M\'exico, University of Arizona, University of Colorado Boulder, University of Oxford, University of Portsmouth, University of Utah, University of Virginia, University of Washington, University of Wisconsin, Vanderbilt University, and Yale University.

This work has made use of data from the European Space Agency (ESA) mission {\it Gaia} (\url{https://www.cosmos.esa.int/gaia}), processed by the {\it Gaia} Data Processing and Analysis Consortium (DPAC, \url{https://www.cosmos.esa.int/web/gaia/dpac/consortium}).
Funding for the DPAC has been provided by national institutions, in particular the institutions participating in the {\it Gaia} Multilateral Agreement.

\software{\texttt{butterpy} \citep{butterpy, Claytor2022}, \texttt{kiauhoku} \citep{Claytor2020, kiauhoku}, \texttt{NumPy} \citep{Numpy2020}, \texttt{Pandas} \citep{Pandas2010}, \texttt{Matplotlib} \citep{Matplotlib2007}, \texttt{AstroPy} \citep{Astropy2013, Astropy2018, Astropy2022}, \texttt{SciPy} \citep{Scipy2020}, \texttt{PyTorch} \citep{Pytorch2019}, \texttt{Lightkurve} \citep{Lightkurve2018}, \texttt{TESScut} \citep{Brasseur2019}, \texttt{iPython} \citep{iPython2007}, \texttt{starspot} \citep{Starspot2021}, Astroquery, \citep{Ginsburg2019}}

\appendix
\section{Public TESS Photometry and Tools}
\label{app:lightcurves}
There are several publicly available light curve sets, pipelines, and tools designed and optimized for TESS data. We list some of the most widely used in Table~\ref{tab:lightcurves}. Tools like \texttt{Lightkurve} \citep{Lightkurve2018} and \texttt{eleanor} \citep{Feinstein2019} are general tools to download, process, and analyze TESS data. \texttt{eleanor} is a flexible tool that allows for several different systematics correction routines to be used on the same light curves. However, it requires large downloads, making it somewhat inconvenient for working with large data. \texttt{Unpopular} \citep{Hattori2022} is a light curve processing pipeline optimized for systematics removal while preserving multi-sector astrophysical signals. It may be ideal for the problem of rotation, but it requires downloading large FFI cutouts, or the entire set of FFIs, for it to work optimally. \texttt{Lightkurve} does no automatic processing and provides simple tools for downloading and interacting with image and light curve data. We use \texttt{Lightkurve} for all our photometry and light curve processing.

Among the many public light curve datasets, the TESS Quick-Look Pipeline \citep[QLP, e.g.,][]{Huang2020a} and DIAmante \citep{Montalto2020} are designed for planet searches, so their light curve processing is aggressive and can remove the stellar signals we are interested in. The difference imaging analysis (DIA) light curves of \citet{Oelkers2018b} are for general use, but only sectors 1--5 of the first year are available. The GSFC-ELEANOR-LITE light curves \citep{Powell2022} are a brand new data set using \texttt{eleanor} to create general-use light curves for all TESS stars brighter than 16th magnitude in the TESS band pass. They will be worth considering for large scale investigations in TESS, but currently only four sectors are publicly available. The TESS Science Processing Operations Center \citep[TESS-SPOC,][]{Caldwell2020} has FFI light curves for nearly 40,000 bright SCVZ targets, with background subtraction and systematics correction, as well as underlying pixel data and apertures, available. They are suitable for general use and are easily downloaded from MAST. Finally, the TESS Asteroseismic Science Operations Center \citep[TASOC, e.g.,][]{Handberg2021} is producing data products for all targets brighter than 15th TESS magnitude. They provide two different light curve products optimized for signals at different timescales with varying levels of systematics correction. 

\begin{deluxetable*}{lcc}
    \centerwidetable
    \tablecaption{TESS Full Frame Image Light Curves, Pipelines, and Tools}
    \tablehead{\colhead{Name} & \colhead{Reference(s)} & \colhead{Science Use}}
    \startdata
        \texttt{Lightkurve}* & \citet{Lightkurve2018} & general \\
        \texttt{eleanor} & \citet{Feinstein2019} & general \\
        \texttt{Unpopular} & \citet{Hattori2022} & general \\
        \hline
        DIA & \citet{Oelkers2018b} & general \\
        QLP & \citet{Huang2020a, Huang2020b, Kunimoto2021} & exoplanet detection \\
        TESS-SPOC* & \citet{Caldwell2020} & general \\
        DIAmante & \citet{Montalto2020} & exoplanet detection \\
        T'DA/TASOC* & \citet{Handberg2021, Lund2021} & asteroseismology \\
        GSFC-ELEANOR-LITE & \citet{Powell2022} & general \\
    \enddata
    \tablecomments{Software tools are listed first, followed by public light curve data sets. Tools and data sets used in this work are marked with an asterisk. All light curve data sets are documented and publicly available as MAST High Level Science Products at \url{https://archive.stsci.edu/hlsp}, except for DIA \citep{Oelkers2018b}, which is available at \url{https://filtergraph.com/tess_ffi}.}
    \label{tab:lightcurves}    
\end{deluxetable*}

\onecolumngrid
\section{Training Set Noise Properties}
\label{app:noise}
In Section~\ref{sec:templates}, we describe the real, quiescent light curves used for our ``pure noise'' templates. The pure noise light curves are taken from hot stars (6,000 K -- 8,000 K for TESS-SPOC; 5,800 K -- 6,000 K for TASOC), which should be relatively quiet in the TESS band pass. Here we examine the noise properties of subsets of our TESS-SPOC sample to demonstrate that the noise light curves are representative of the target light curves.

\begin{figure}
    \centering
    \includegraphics[width=0.6\textwidth]{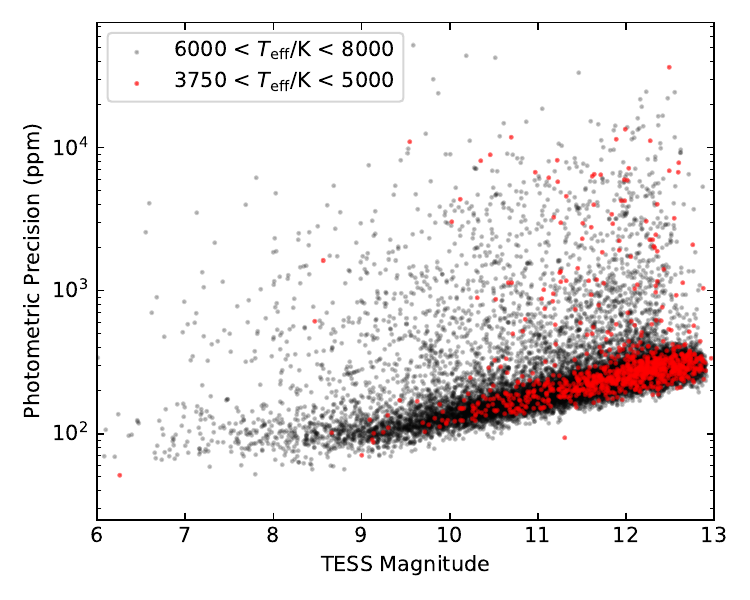}
    \caption{The 6-hour CDPP of our TESS-SPOC ``pure-noise'' light curves (black points), which are taken from hot stars observed in the SCVZ during TESS Cycle 1, and a subsample of cool star target light curves (red points). The photometric noise properties are virtually identical for the two samples.}
    \label{fig:noise-comparison}
\end{figure}

Figure~\ref{fig:noise-comparison} shows the 6-hour CDPP versus TESS magnitude for our TESS-SPOC noise light curves (black points) and for a subset of cool star light curves (red points). Regardless of temperature, the photometric noise follows a clear function of brightness, and the distributions of noise levels are the same at fixed TESS magnitude. Thus, the noise properties of our training light curves are representative of those of our ``real'' stellar light curves.

\begin{figure*}
    \centering
    \includegraphics[width=\textwidth]{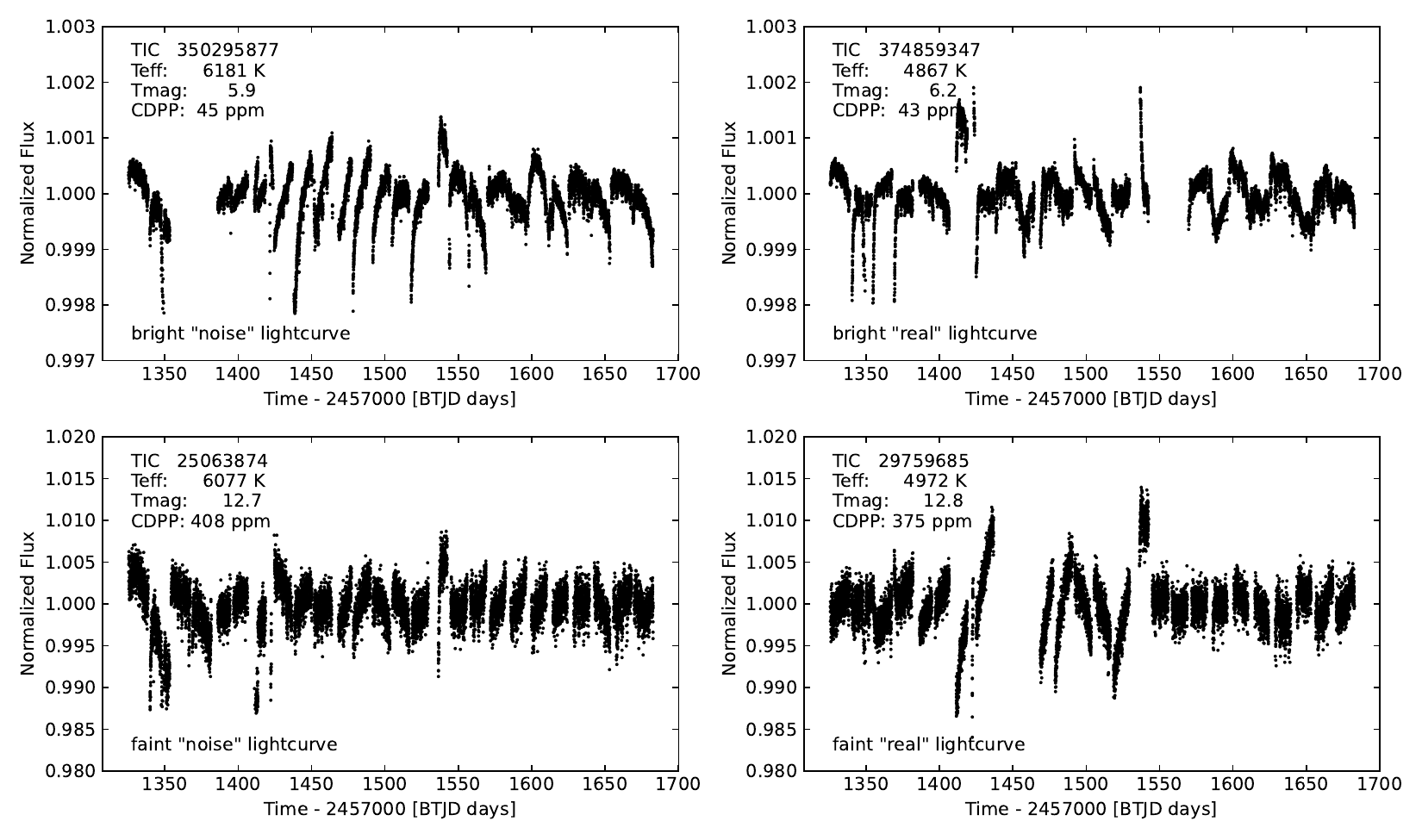}
    \caption{Four example light curves comparing the noise and systematics for our quiescent training``noise'' sample  (left) and our ``real'' data (right). The photometric noise, quantified by the 6-hour CDPP, is a function of brightness and the same for both the training and real data. The TESS systematics are generally independent of the target and are shared by all light curves. They are shown here as observing gaps, steep jumps (e.g., at BTJD of 1425 and 1550 days), and ramps of sector-dependent increasing or decreasing flux with time.}
    \label{fig:lightcurve-comparison}
\end{figure*}

We also compare the TESS systematics in four example light curves in Figure~\ref{fig:lightcurve-comparison}. The left panels show examples of noise light curves taken from a bright (top) and a faint (bottom) star, while the right panels show light curves from cool stars in our period search sample. Photometric noise is again quantified by the 6-hour CDPP and is the same for both the hot and cool stars at fixed brightness. Systematic effects are present in all light curves as observing gaps, steep jumps at specific dates, and increasing or decreasing flux with time from unremoved (or overcorrected) background light. Since these effects are independent of stellar properties, the systematics in the noise light curves are also representative of those present in the period search light curves.

\onecolumngrid
\section{Optimizing the Neural Network Architecture}
\label{app:architecture}
In Section~\ref{sec:network} we lay out the various convolutional neural network (CNN) architectures that we trained and assessed to optimize our network's performance. Here we discuss the details of that optimization and the justification for our choices of architecture.

For both the TESS-SPOC and TASOC data products, we trained four different CNNs, each with 3 convolution layers, but each CNN had different numbers of convolution kernels to give the networks different flexibility in learning features. The architectures were (A) 8, 16, and 32 kernels; (B) 6, 32, and 64 kernels; (C) 32, 64, and 128; and (D) 64, 128, and 256. We also used four different training sets for both TESS-SPOC and TASOC, each with a different upper limit on rotation period. The period upper limits were 30, 60, 90, and 180 days, intended to optimize different networks for different period ranges. We trained all four architectures on each period range, compared performance metrics, and chose the architecture that had the best performance on average across all four training sets. For performance metrics, we considered, (1) average test loss, (2) median relative uncertainty, (3) percentage of test targets recovered to within 10\% and 20\% accuracy, and (4) the 1st and 99th percentiles of the filtered period estimates. To illustrate the meaning of these values, we will use the 180-day TESS-SPOC training set as an example.

\begin{figure*}
    \centering
    \includegraphics[width=0.6\textwidth]{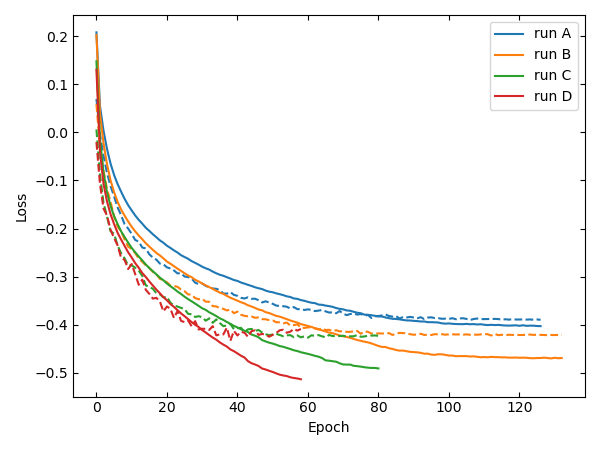}
    \caption{Learning curves of all four CNN architectures for the 180-day training set. The solid lines track the training loss, while the dashed lines show the test loss, which was used to assess performance of the networks once trained. {The number of training epochs is determined by the loss evaluated on the validation set: training stops when the validation loss stops decreasing.}}
    \label{fig:learning}
\end{figure*}

During training, each training set is partitioned into a training, validation, and test set. The training set is used to fit the network parameters, the validation set is used to determine when to stop training to avoid overfitting, and the test set is used to assess performance. We monitored the average loss for all three partitions during training so that we can construct learning curves, which show the loss values versus training epoch. Figure~\ref{fig:learning} shows the learning curves for all four architectures on the 180-day training set. The solid lines represent the training loss, while the dashed lines represent the test loss. Left unchecked, training loss will continue to decrease, but the loss on a held-out validation set will plateau or begin to increase once the network begins overfitting, which we use as our stopping criterion. The test loss is highest for run A, the simplest architecture we used. This indicates that run A is not complex enough to fully learn the features in the data, or at least that it begins overfitting before it can fully learn the features. Run B performs better, but is comparable to runs C and D, which fully train in fewer epochs. We can rule out run A for this case, but more metrics are needed to properly assess which run performs best.

\begin{figure*}
    \centering
    \includegraphics[width=0.8\textwidth]{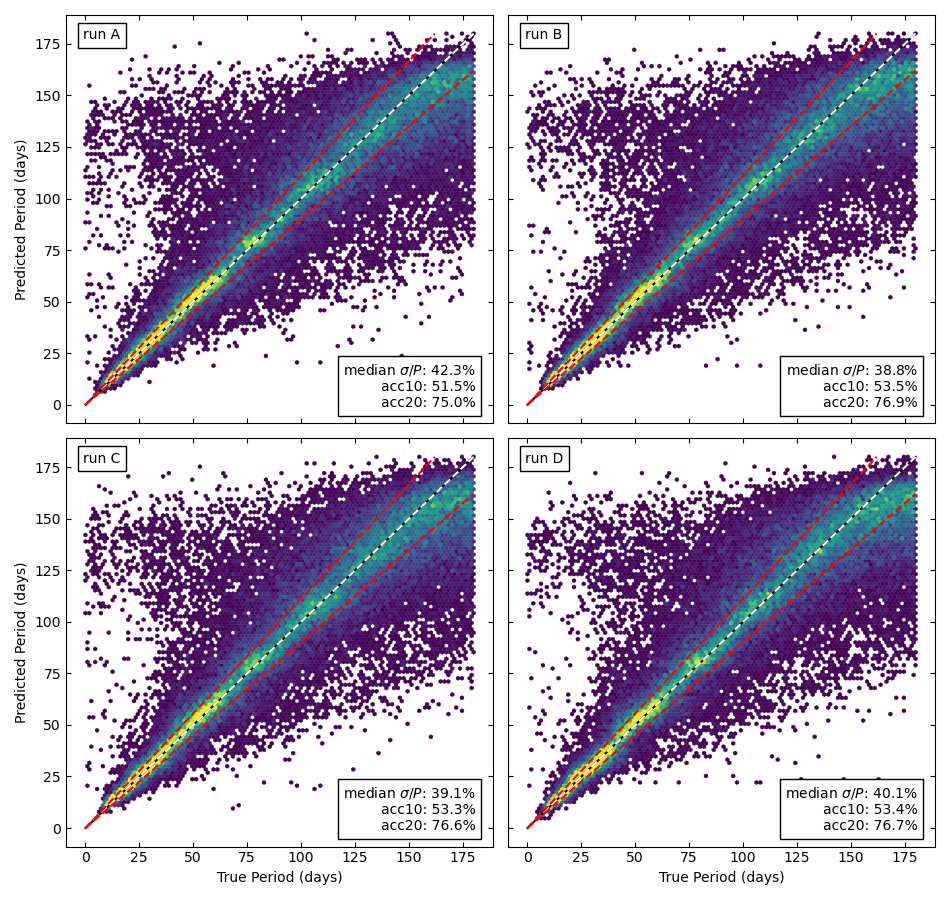}
    \caption{Period detections for different CNN architectures, filtered by relative uncertainty. Architectures are increasingly complex from A to D, and recovery statistics are shown in the legend of each panel.}
    \label{fig:recovery}
\end{figure*}

One of the strengths of our method is the ability to estimate an uncertainty, which we can use as a metric of predicted reliability \citep{Claytor2022}. Specifically, we use the fractional uncertainty $\sigma_P/P$ to normalize for period dependence. A better-trained network should have lower values of $\sigma_P/P$, indicating more reliable estimates. We use the median $\sigma_P/P$ as an additional metric of performance in addition to using it to filter out bad estimates. Figure~\ref{fig:recovery} shows the \emph{filtered} period estimates for each run, but note that the median fractional uncertainty listed in each panel is computed over the \emph{unfiltered} periods. Run B has the lowest estimated uncertainty, so by this metric it performs the best and has the most reliable estimates. 

We also use accuracy metrics to assess performance. The ``acc10" and ``acc20" metrics quantify what fraction of test targets are recovered to within 10\% and 20\% accuracy after filtering by uncertainty. The  ``acc10" metrics for each run are near 50\%, which also means that the median relative error on the period estimates is near 10\% for all runs. Run B has the highest accuracy metrics, so it once again performs best.

\begin{figure*}
    \centering
    \includegraphics[width=0.8\textwidth]{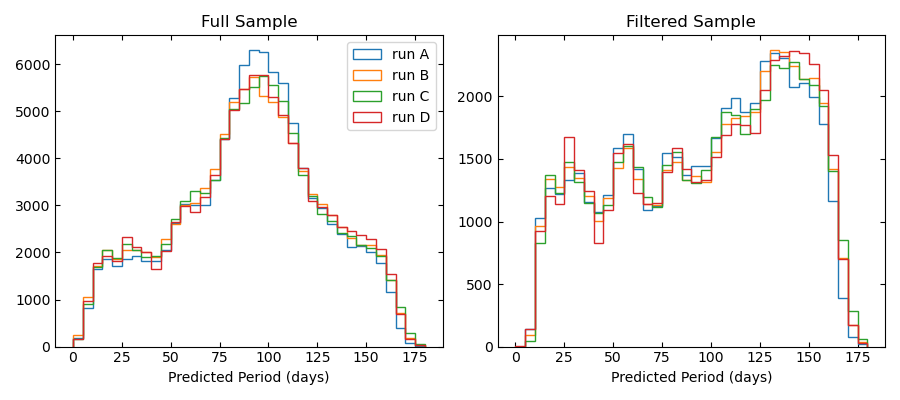}
    \caption{Distributions of rotation period estimtaes for different CNN architectures, for both the unfiltered (left) and filtered (right) samples.}
    \label{fig:recovery-histograms}
\end{figure*}

Estimating uncertainty biases estimates toward the median of the distribution, making period inference near the edges of the training set period range more difficult \citep{Claytor2022}. We attempt to mitigate this by tabulating the 1st and 99th percentiles of each (unfiltered and filtered) inferred period range. Figure~\ref{fig:recovery-histograms} shows the distribution of periods for both the unfiltered (left) and filtered (right) estimates. Though it is difficult to assess by eye, run A has the lowest 1st percentile (12.1 d) in the filtered sample, although all runs have first percentiles in the 12--13 day range. This also gives us a lower limit for where we can expect successful period estimates from this training set: networks trained on the 180-day set struggle to infer periods less than 12 days, motivating the need for training sets with smaller period ranges. 

We prioritized metrics as follows: we considered the average test loss to rule out runs that failed to compete in loss value (e.g., runs B, C, and D achieved comparable loss values, but run A fell short). We then prioritized the accuracy metrics and uncertainty together, then if those were comparable we used the 1st and 99th percentile values to break ties. 

When considering all our metrics for the 180-day TESS-SPOC training set, run B performs the best overall. We then repeated this process for each training set and chose the architecture that performed best over all training sets. Following this procedure, we chose architecture C for the TESS-SPOC data and architecture A for TASOC. We note that it may be optimal to use the optimal architecture for each training set, rather than adopt one architecture for all sets. We will consider before publication and release of the final period catalog.

\clearpage
\bibliography{references}

\begin{thebibliography}{}
\expandafter\ifx\csname natexlab\endcsname\relax\def\natexlab#1{#1}\fi
\providecommand{\url}[1]{\href{#1}{#1}}
\providecommand{\dodoi}[1]{doi:~\href{http://doi.org/#1}{\nolinkurl{#1}}}
\providecommand{\doeprint}[1]{\href{http://ascl.net/#1}{\nolinkurl{http://ascl.net/#1}}}
\providecommand{\doarXiv}[1]{\href{https://arxiv.org/abs/#1}{\nolinkurl{https://arxiv.org/abs/#1}}}

\bibitem[{{Abdurro'uf} {et~al.}(2022){Abdurro'uf}, {Accetta}, {Aerts}, {Silva Aguirre}, {Ahumada}, {Ajgaonkar}, {Filiz Ak}, {Alam}, {Allende Prieto}, {Almeida}, \& et~al.}]{Abdurro'uf2022}
{Abdurro'uf}, {Accetta}, K., {Aerts}, C., {et~al.} 2022, \apjs, 259, 35, \dodoi{10.3847/1538-4365/ac4414}

\bibitem[{{Adelberger} {et~al.}(2011){Adelberger}, {Garc{\'\i}a}, {Robertson}, {Snover}, {Balantekin}, {Heeger}, {Ramsey-Musolf}, {Bemmerer}, {Junghans}, {Bertulani}, {Chen}, {Costantini}, {Prati}, {Couder}, {Uberseder}, {Wiescher}, {Cyburt}, {Davids}, {Freedman}, {Gai}, {Gazit}, {Gialanella}, {Imbriani}, {Greife}, {Hass}, {Haxton}, {Itahashi}, {Kubodera}, {Langanke}, {Leitner}, {Leitner}, {Vetter}, {Winslow}, {Marcucci}, {Motobayashi}, {Mukhamedzhanov}, {Tribble}, {Nollett}, {Nunes}, {Park}, {Parker}, {Schiavilla}, {Simpson}, {Spitaleri}, {Strieder}, {Trautvetter}, {Suemmerer}, \& {Typel}}]{Adelberger2011}
{Adelberger}, E.~G., {Garc{\'\i}a}, A., {Robertson}, R.~G.~H., {et~al.} 2011, Reviews of Modern Physics, 83, 195, \dodoi{10.1103/RevModPhys.83.195}

\bibitem[{{Allende Prieto} {et~al.}(2006){Allende Prieto}, {Beers}, {Wilhelm}, {Newberg}, {Rockosi}, {Yanny}, \& {Lee}}]{AllendePrieto2006}
{Allende Prieto}, C., {Beers}, T.~C., {Wilhelm}, R., {et~al.} 2006, \apj, 636, 804, \dodoi{10.1086/498131}

\bibitem[{{Amard} {et~al.}(2020){Amard}, {Roquette}, \& {Matt}}]{Amard2020}
{Amard}, L., {Roquette}, J., \& {Matt}, S.~P. 2020, \mnras, 499, 3481, \dodoi{10.1093/mnras/staa3038}

\bibitem[{{Angus}(2021)}]{Starspot2021}
{Angus}, R. 2021, {starspot: code for measuring stellar rotation periods}, v0.2,  Zenodo, \dodoi{10.5281/zenodo.4613887}.
\newblock \url{https://doi.org/10.5281/zenodo.4613887}

\bibitem[{{Angus} {et~al.}(2015){Angus}, {Aigrain}, {Foreman-Mackey}, \& {McQuillan}}]{Angus2015}
{Angus}, R., {Aigrain}, S., {Foreman-Mackey}, D., \& {McQuillan}, A. 2015, \mnras, 450, 1787, \dodoi{10.1093/mnras/stv423}

\bibitem[{{Astropy Collaboration} {et~al.}(2013){Astropy Collaboration}, {Robitaille}, {Tollerud}, {Greenfield}, {Droettboom}, {Bray}, {Aldcroft}, {Davis}, {Ginsburg}, {Price-Whelan}, {Kerzendorf}, {Conley}, {Crighton}, {Barbary}, {Muna}, {Ferguson}, {Grollier}, {Parikh}, {Nair}, {Unther}, {Deil}, {Woillez}, {Conseil}, {Kramer}, {Turner}, {Singer}, {Fox}, {Weaver}, {Zabalza}, {Edwards}, {Azalee Bostroem}, {Burke}, {Casey}, {Crawford}, {Dencheva}, {Ely}, {Jenness}, {Labrie}, {Lim}, {Pierfederici}, {Pontzen}, {Ptak}, {Refsdal}, {Servillat}, \& {Streicher}}]{Astropy2013}
{Astropy Collaboration}, {Robitaille}, T.~P., {Tollerud}, E.~J., {et~al.} 2013, \aap, 558, A33, \dodoi{10.1051/0004-6361/201322068}

\bibitem[{{Astropy Collaboration} {et~al.}(2018){Astropy Collaboration}, {Price-Whelan}, {Sip{\H{o}}cz}, {G{\"u}nther}, {Lim}, {Crawford}, {Conseil}, {Shupe}, {Craig}, {Dencheva}, {Ginsburg}, {Vand erPlas}, {Bradley}, {P{\'e}rez-Su{\'a}rez}, {de Val-Borro}, {Aldcroft}, {Cruz}, {Robitaille}, {Tollerud}, {Ardelean}, {Babej}, {Bach}, {Bachetti}, {Bakanov}, {Bamford}, {Barentsen}, {Barmby}, {Baumbach}, {Berry}, {Biscani}, {Boquien}, {Bostroem}, {Bouma}, {Brammer}, {Bray}, {Breytenbach}, {Buddelmeijer}, {Burke}, {Calderone}, {Cano Rodr{\'\i}guez}, {Cara}, {Cardoso}, {Cheedella}, {Copin}, {Corrales}, {Crichton}, {D'Avella}, {Deil}, {Depagne}, {Dietrich}, {Donath}, {Droettboom}, {Earl}, {Erben}, {Fabbro}, {Ferreira}, {Finethy}, {Fox}, {Garrison}, {Gibbons}, {Goldstein}, {Gommers}, {Greco}, {Greenfield}, {Groener}, {Grollier}, {Hagen}, {Hirst}, {Homeier}, {Horton}, {Hosseinzadeh}, {Hu}, {Hunkeler}, {Ivezi{\'c}}, {Jain}, {Jenness}, {Kanarek}, {Kendrew}, {Kern}, {Kerzendorf}, {Khvalko}, {King}, {Kirkby}, {Kulkarni},
  {Kumar}, {Lee}, {Lenz}, {Littlefair}, {Ma}, {Macleod}, {Mastropietro}, {McCully}, {Montagnac}, {Morris}, {Mueller}, {Mumford}, {Muna}, {Murphy}, {Nelson}, {Nguyen}, {Ninan}, {N{\"o}the}, {Ogaz}, {Oh}, {Parejko}, {Parley}, {Pascual}, {Patil}, {Patil}, {Plunkett}, {Prochaska}, {Rastogi}, {Reddy Janga}, {Sabater}, {Sakurikar}, {Seifert}, {Sherbert}, {Sherwood-Taylor}, {Shih}, {Sick}, {Silbiger}, {Singanamalla}, {Singer}, {Sladen}, {Sooley}, {Sornarajah}, {Streicher}, {Teuben}, {Thomas}, {Tremblay}, {Turner}, {Terr{\'o}n}, {van Kerkwijk}, {de la Vega}, {Watkins}, {Weaver}, {Whitmore}, {Woillez}, {Zabalza}, \& {Astropy Contributors}}]{Astropy2018}
{Astropy Collaboration}, {Price-Whelan}, A.~M., {Sip{\H{o}}cz}, B.~M., {et~al.} 2018, \aj, 156, 123, \dodoi{10.3847/1538-3881/aabc4f}

\bibitem[{{Astropy Collaboration} {et~al.}(2022){Astropy Collaboration}, {Price-Whelan}, {Lim}, {Earl}, {Starkman}, {Bradley}, {Shupe}, {Patil}, {Corrales}, {Brasseur}, {N{\"o}the}, {Donath}, {Tollerud}, {Morris}, {Ginsburg}, {Vaher}, {Weaver}, {Tocknell}, {Jamieson}, {van Kerkwijk}, {Robitaille}, {Merry}, {Bachetti}, {G{\"u}nther}, {Aldcroft}, {Alvarado-Montes}, {Archibald}, {B{\'o}di}, {Bapat}, {Barentsen}, {Baz{\'a}n}, {Biswas}, {Boquien}, {Burke}, {Cara}, {Cara}, {Conroy}, {Conseil}, {Craig}, {Cross}, {Cruz}, {D'Eugenio}, {Dencheva}, {Devillepoix}, {Dietrich}, {Eigenbrot}, {Erben}, {Ferreira}, {Foreman-Mackey}, {Fox}, {Freij}, {Garg}, {Geda}, {Glattly}, {Gondhalekar}, {Gordon}, {Grant}, {Greenfield}, {Groener}, {Guest}, {Gurovich}, {Handberg}, {Hart}, {Hatfield-Dodds}, {Homeier}, {Hosseinzadeh}, {Jenness}, {Jones}, {Joseph}, {Kalmbach}, {Karamehmetoglu}, {Ka{\l}uszy{\'n}ski}, {Kelley}, {Kern}, {Kerzendorf}, {Koch}, {Kulumani}, {Lee}, {Ly}, {Ma}, {MacBride}, {Maljaars}, {Muna}, {Murphy}, {Norman},
  {O'Steen}, {Oman}, {Pacifici}, {Pascual}, {Pascual-Granado}, {Patil}, {Perren}, {Pickering}, {Rastogi}, {Roulston}, {Ryan}, {Rykoff}, {Sabater}, {Sakurikar}, {Salgado}, {Sanghi}, {Saunders}, {Savchenko}, {Schwardt}, {Seifert-Eckert}, {Shih}, {Jain}, {Shukla}, {Sick}, {Simpson}, {Singanamalla}, {Singer}, {Singhal}, {Sinha}, {Sip{\H{o}}cz}, {Spitler}, {Stansby}, {Streicher}, {{\v{S}}umak}, {Swinbank}, {Taranu}, {Tewary}, {Tremblay}, {de Val-Borro}, {Van Kooten}, {Vasovi{\'c}}, {Verma}, {de Miranda Cardoso}, {Williams}, {Wilson}, {Winkel}, {Wood-Vasey}, {Xue}, {Yoachim}, {Zhang}, {Zonca}, \& {Astropy Project Contributors}}]{Astropy2022}
{Astropy Collaboration}, {Price-Whelan}, A.~M., {Lim}, P.~L., {et~al.} 2022, \apj, 935, 167, \dodoi{10.3847/1538-4357/ac7c74}

\bibitem[{{Avallone} {et~al.}(2022){Avallone}, {Tayar}, {van Saders}, {Berger}, {Claytor}, {Beaton}, {Teske}, {Godoy-Rivera}, \& {Pan}}]{Avallone2022}
{Avallone}, E.~A., {Tayar}, J.~N., {van Saders}, J.~L., {et~al.} 2022, \apj, 930, 7, \dodoi{10.3847/1538-4357/ac60a1}

\bibitem[{Balona {et~al.}(2011)Balona, Guzik, Uytterhoeven, Smith, Tenenbaum, \& Twicken}]{Balona2011}
Balona, L.~A., Guzik, J.~A., Uytterhoeven, K., {et~al.} 2011, Monthly Notices of the Royal Astronomical Society, 415, 3531, \dodoi{10.1111/j.1365-2966.2011.18973.x}

\bibitem[{{Barnes}(2003)}]{Barnes2003}
{Barnes}, S.~A. 2003, \apj, 586, 464, \dodoi{10.1086/367639}

\bibitem[{{Barnes} {et~al.}(2016){Barnes}, {Weingrill}, {Fritzewski}, {Strassmeier}, \& {Platais}}]{Barnes2016}
{Barnes}, S.~A., {Weingrill}, J., {Fritzewski}, D., {Strassmeier}, K.~G., \& {Platais}, I. 2016, \apj, 823, 16, \dodoi{10.3847/0004-637X/823/1/16}

\bibitem[{{Bensby} {et~al.}(2014){Bensby}, {Feltzing}, \& {Oey}}]{Bensby2014}
{Bensby}, T., {Feltzing}, S., \& {Oey}, M.~S. 2014, \aap, 562, A71, \dodoi{10.1051/0004-6361/201322631}

\bibitem[{{Berger} {et~al.}(2018){Berger}, {Huber}, {Gaidos}, \& {van Saders}}]{Berger2018}
{Berger}, T.~A., {Huber}, D., {Gaidos}, E., \& {van Saders}, J.~L. 2018, \apj, 866, 99, \dodoi{10.3847/1538-4357/aada83}

\bibitem[{{Berger} {et~al.}(2020){Berger}, {Huber}, {van Saders}, {Gaidos}, {Tayar}, \& {Kraus}}]{Berger2020}
{Berger}, T.~A., {Huber}, D., {van Saders}, J.~L., {et~al.} 2020, \aj, 159, 280, \dodoi{10.3847/1538-3881/159/6/280}

\bibitem[{{Birky} {et~al.}(2020){Birky}, {Hogg}, {Mann}, \& {Burgasser}}]{Birky2020}
{Birky}, J., {Hogg}, D.~W., {Mann}, A.~W., \& {Burgasser}, A. 2020, \apj, 892, 31, \dodoi{10.3847/1538-4357/ab7004}

\bibitem[{{Borucki} {et~al.}(2010){Borucki}, {Koch}, {Basri}, {Batalha}, {Brown}, {Caldwell}, {Caldwell}, {Christensen-Dalsgaard}, {Cochran}, {DeVore}, {Dunham}, {Dupree}, {Gautier}, {Geary}, {Gilliland}, {Gould}, {Howell}, {Jenkins}, {Kondo}, {Latham}, {Marcy}, {Meibom}, {Kjeldsen}, {Lissauer}, {Monet}, {Morrison}, {Sasselov}, {Tarter}, {Boss}, {Brownlee}, {Owen}, {Buzasi}, {Charbonneau}, {Doyle}, {Fortney}, {Ford}, {Holman}, {Seager}, {Steffen}, {Welsh}, {Rowe}, {Anderson}, {Buchhave}, {Ciardi}, {Walkowicz}, {Sherry}, {Horch}, {Isaacson}, {Everett}, {Fischer}, {Torres}, {Johnson}, {Endl}, {MacQueen}, {Bryson}, {Dotson}, {Haas}, {Kolodziejczak}, {Van Cleve}, {Chandrasekaran}, {Twicken}, {Quintana}, {Clarke}, {Allen}, {Li}, {Wu}, {Tenenbaum}, {Verner}, {Bruhweiler}, {Barnes}, \& {Prsa}}]{Borucki2010}
{Borucki}, W.~J., {Koch}, D., {Basri}, G., {et~al.} 2010, Science, 327, 977, \dodoi{10.1126/science.1185402}

\bibitem[{Bradley {et~al.}(2015)Bradley, Guzik, Miles, Uytterhoeven, Jackiewicz, \& Kinemuchi}]{Bradley2015}
Bradley, P.~A., Guzik, J.~A., Miles, L.~F., {et~al.} 2015, The Astronomical Journal, 149, 68, \dodoi{10.1088/0004-6256/149/2/68}

\bibitem[{{Brasseur} {et~al.}(2019){Brasseur}, {Phillip}, {Fleming}, {Mullally}, \& {White}}]{Brasseur2019}
{Brasseur}, C.~E., {Phillip}, C., {Fleming}, S.~W., {Mullally}, S.~E., \& {White}, R.~L. 2019, {Astrocut: Tools for creating cutouts of TESS images}.
\newblock \doeprint{1905.007}

\bibitem[{{Breton} {et~al.}(2021){Breton}, {Santos}, {Bugnet}, {Mathur}, {Garc{\'\i}a}, \& {Pall{\'e}}}]{Breton2021}
{Breton}, S.~N., {Santos}, A.~R.~G., {Bugnet}, L., {et~al.} 2021, \aap, 647, A125, \dodoi{10.1051/0004-6361/202039947}

\bibitem[{{Buder} {et~al.}(2019){Buder}, {Lind}, {Ness}, {Asplund}, {Duong}, {Lin}, {Kos}, {Casagrande}, {Casey}, {Bland-Hawthorn}, {de Silva}, {D'Orazi}, {Freeman}, {Martell}, {Schlesinger}, {Sharma}, {Simpson}, {Zucker}, {Zwitter}, {{\v{C}}otar}, {Dotter}, {Hayden}, {Hyde}, {Kafle}, {Lewis}, {Nataf}, {Nordlander}, {Reid}, {Rix}, {Sk{\'u}lad{\'o}ttir}, {Stello}, {Ting}, {Traven}, {Wyse}, \& {Galah Collaboration}}]{Buder2019}
{Buder}, S., {Lind}, K., {Ness}, M.~K., {et~al.} 2019, \aap, 624, A19, \dodoi{10.1051/0004-6361/201833218}

\bibitem[{Caldwell {et~al.}(2020)Caldwell, Jenkins, \& Ting}]{10.17909/t9-wpz1-8s54}
Caldwell, D.~A., Jenkins, J.~M., \& Ting, E.~B. 2020, TESS Light Curves From Full Frame Images ("TESS-SPOC"),  STScI/MAST, \dodoi{10.17909/T9-WPZ1-8S54}.
\newblock \url{http://archive.stsci.edu/doi/resolve/resolve.html?doi=10.17909/t9-wpz1-8s54}

\bibitem[{{Caldwell} {et~al.}(2020){Caldwell}, {Tenenbaum}, {Twicken}, {Jenkins}, {Ting}, {Smith}, {Hedges}, {Fausnaugh}, {Rose}, \& {Burke}}]{Caldwell2020}
{Caldwell}, D.~A., {Tenenbaum}, P., {Twicken}, J.~D., {et~al.} 2020, Research Notes of the American Astronomical Society, 4, 201, \dodoi{10.3847/2515-5172/abc9b3}

\bibitem[{{Canto Martins} {et~al.}(2020){Canto Martins}, {Gomes}, {Messias}, {de Lira}, {Le{\~a}o}, {Almeida}, {Teixeira}, {das Chagas}, {Bravo}, {Bewketu Belete}, \& {De Medeiros}}]{CantoMartins2020}
{Canto Martins}, B.~L., {Gomes}, R.~L., {Messias}, Y.~S., {et~al.} 2020, \apjs, 250, 20, \dodoi{10.3847/1538-4365/aba73f}

\bibitem[{{Cao} \& {Pinsonneault}(2022)}]{Cao2022}
{Cao}, L., \& {Pinsonneault}, M.~H. 2022, \mnras, 517, 2165, \dodoi{10.1093/mnras/stac2706}

\bibitem[{{Cao} {et~al.}(2023){Cao}, {Pinsonneault}, \& {van Saders}}]{Cao2023}
{Cao}, L., {Pinsonneault}, M.~H., \& {van Saders}, J.~L. 2023, \apjl, 951, L49, \dodoi{10.3847/2041-8213/acd780}

\bibitem[{{Carlberg} {et~al.}(2011){Carlberg}, {Majewski}, {Patterson}, {Bizyaev}, {Smith}, \& {Cunha}}]{Carlberg2011}
{Carlberg}, J.~K., {Majewski}, S.~R., {Patterson}, R.~J., {et~al.} 2011, \apj, 732, 39, \dodoi{10.1088/0004-637X/732/1/39}

\bibitem[{Castelli \& Kurucz(2004)}]{Castelli2004}
Castelli, F., \& Kurucz, R.~L. 2004, arXiv e-prints, \dodoi{10.48550/ARXIV.ASTRO-PH/0405087}

\bibitem[{{Ceillier} {et~al.}(2017){Ceillier}, {Tayar}, {Mathur}, {Salabert}, {Garc{\'\i}a}, {Stello}, {Pinsonneault}, {van Saders}, {Beck}, \& {Bloemen}}]{Ceillier2017}
{Ceillier}, T., {Tayar}, J., {Mathur}, S., {et~al.} 2017, \aap, 605, A111, \dodoi{10.1051/0004-6361/201629884}

\bibitem[{{Chontos} {et~al.}(2021){Chontos}, {Huber}, {Berger}, {Kjeldsen}, {Serenelli}, {Silva Aguirre}, {Ball}, {Basu}, {Bedding}, {Chaplin}, {Claytor}, {Corsaro}, {Garcia}, {Howell}, {Lundkvist}, {Mathur}, {Metcalfe}, {Nielsen}, {Mian Joel Ong}, {{\c{C}}elik Orhan}, {{\"O}rtel}, {Salama}, {Stassun}, {Townsend}, {van Saders}, {Winther}, {Yildiz}, {Butler}, {Tinney}, \& {Wittenmyer}}]{Chontos2021}
{Chontos}, A., {Huber}, D., {Berger}, T.~A., {et~al.} 2021, \apj, 922, 229, \dodoi{10.3847/1538-4357/ac1269}

\bibitem[{{Christiansen} {et~al.}(2012){Christiansen}, {Jenkins}, {Caldwell}, {Burke}, {Tenenbaum}, {Seader}, {Thompson}, {Barclay}, {Clarke}, {Li}, {Smith}, {Stumpe}, {Twicken}, \& {Van Cleve}}]{Christiansen2012}
{Christiansen}, J.~L., {Jenkins}, J.~M., {Caldwell}, D.~A., {et~al.} 2012, \pasp, 124, 1279, \dodoi{10.1086/668847}

\bibitem[{Claytor {et~al.}(2021)Claytor, Lucas, \& Llama}]{butterpy}
Claytor, Z.~R., Lucas, M., \& Llama, J. 2021, {Butterpy: realistic star spot evolution and light curves in Python}, 0.1.0,  Zenodo, \dodoi{10.5281/zenodo.4722052}.
\newblock \url{https://doi.org/10.5281/zenodo.4722052}

\bibitem[{{Claytor} {et~al.}(2022){Claytor}, {van Saders}, {Llama}, {Sadowski}, {Quach}, \& {Avallone}}]{Claytor2022}
{Claytor}, Z.~R., {van Saders}, J.~L., {Llama}, J., {et~al.} 2022, \apj, 927, 219, \dodoi{10.3847/1538-4357/ac498f}

\bibitem[{{Claytor} {et~al.}(2020{\natexlab{a}}){Claytor}, {van Saders}, {Santos}, {Garc{\'\i}a}, {Mathur}, {Tayar}, {Pinsonneault}, \& {Shetrone}}]{Claytor2020}
{Claytor}, Z.~R., {van Saders}, J.~L., {Santos}, {\^A}. R.~G., {et~al.} 2020{\natexlab{a}}, \apj, 888, 43, \dodoi{10.3847/1538-4357/ab5c24}

\bibitem[{{Claytor} {et~al.}(2020{\natexlab{b}}){Claytor}, {van Saders}, {Santos}, {Garc{\'\i}a}, {Mathur}, {Tayar}, {Pinsonneault}, \& {Shetrone}}]{kiauhoku}
---. 2020{\natexlab{b}}, {kiauhoku: Stellar model grid interpolation}, Astrophysics Source Code Library, record ascl:2011.027.
\newblock \doeprint{2011.027}

\bibitem[{{Claytor, Zachary R.} {et~al.}(2022){Claytor, Zachary R.}, {van Saders, Jennifer L.}, {Llama, Joe}, {Sadowski, Peter}, {Quach, Brandon}, \& {Avallone, Ellis}}]{10.17909/davg-m919}
{Claytor, Zachary R.}, {van Saders, Jennifer L.}, {Llama, Joe}, {et~al.} 2022, Simulated TESS Light Curves for Measuring Rotation with Deep Learning ("SMARTS"),  STScI/MAST, \dodoi{10.17909/DAVG-M919}.
\newblock \url{http://archive.stsci.edu/doi/resolve/resolve.html?doi=10.17909/davg-m919}

\bibitem[{{Curtis} {et~al.}(2019){Curtis}, {Ag{\"u}eros}, {Douglas}, \& {Meibom}}]{Curtis2019}
{Curtis}, J.~L., {Ag{\"u}eros}, M.~A., {Douglas}, S.~T., \& {Meibom}, S. 2019, \apj, 879, 49, \dodoi{10.3847/1538-4357/ab2393}

\bibitem[{{Curtis} {et~al.}(2020){Curtis}, {Ag{\"u}eros}, {Matt}, {Covey}, {Douglas}, {Angus}, {Saar}, {Cody}, {Vanderburg}, {Law}, {Kraus}, {Latham}, {Baranec}, {Riddle}, {Ziegler}, {Lund}, {Torres}, {Meibom}, {Aguirre}, \& {Wright}}]{Curtis2020}
{Curtis}, J.~L., {Ag{\"u}eros}, M.~A., {Matt}, S.~P., {et~al.} 2020, \apj, 904, 140, \dodoi{10.3847/1538-4357/abbf58}

\bibitem[{{Davenport}(2017)}]{Davenport2017}
{Davenport}, J. R.~A. 2017, \apj, 835, 16, \dodoi{10.3847/1538-4357/835/1/16}

\bibitem[{{David} {et~al.}(2022){David}, {Angus}, {Curtis}, {van Saders}, {Colman}, {Contardo}, {Lu}, \& {Zinn}}]{David2022}
{David}, T.~J., {Angus}, R., {Curtis}, J.~L., {et~al.} 2022, \apj, 933, 114, \dodoi{10.3847/1538-4357/ac6dd3}

\bibitem[{{Demarque} {et~al.}(2008){Demarque}, {Guenther}, {Li}, {Mazumdar}, \& {Straka}}]{Demarque2008}
{Demarque}, P., {Guenther}, D.~B., {Li}, L.~H., {Mazumdar}, A., \& {Straka}, C.~W. 2008, \apss, 316, 31, \dodoi{10.1007/s10509-007-9698-y}

\bibitem[{{Douglas} {et~al.}(2017){Douglas}, {Ag{\"u}eros}, {Covey}, \& {Kraus}}]{Douglas2017}
{Douglas}, S.~T., {Ag{\"u}eros}, M.~A., {Covey}, K.~R., \& {Kraus}, A. 2017, \apj, 842, 83, \dodoi{10.3847/1538-4357/aa6e52}

\bibitem[{{Douglas} {et~al.}(2019){Douglas}, {Curtis}, {Ag{\"u}eros}, {Cargile}, {Brewer}, {Meibom}, \& {Jansen}}]{Douglas2019}
{Douglas}, S.~T., {Curtis}, J.~L., {Ag{\"u}eros}, M.~A., {et~al.} 2019, \apj, 879, 100, \dodoi{10.3847/1538-4357/ab2468}

\bibitem[{{Dungee} {et~al.}(2022){Dungee}, {van Saders}, {Gaidos}, {Chun}, {Garc{\'\i}a}, {Magnier}, {Mathur}, \& {Santos}}]{Dungee2022}
{Dungee}, R., {van Saders}, J., {Gaidos}, E., {et~al.} 2022, \apj, 938, 118, \dodoi{10.3847/1538-4357/ac90be}

\bibitem[{Ester {et~al.}(1996)Ester, Kriegel, Sander, \& Xu}]{Ester1996}
Ester, M., Kriegel, H.-P., Sander, J., \& Xu, X. 1996, in Proceedings of the Second International Conference on Knowledge Discovery and Data Mining, KDD'96 (AAAI Press), 226–231

\bibitem[{{Feinstein} {et~al.}(2020){Feinstein}, {Montet}, {Ansdell}, {Nord}, {Bean}, {G{\"u}nther}, {Gully-Santiago}, \& {Schlieder}}]{Feinstein2020}
{Feinstein}, A.~D., {Montet}, B.~T., {Ansdell}, M., {et~al.} 2020, \aj, 160, 219, \dodoi{10.3847/1538-3881/abac0a}

\bibitem[{{Feinstein} {et~al.}(2019){Feinstein}, {Montet}, {Foreman-Mackey}, {Bedell}, {Saunders}, {Bean}, {Christiansen}, {Hedges}, {Luger}, {Scolnic}, \& {Cardoso}}]{Feinstein2019}
{Feinstein}, A.~D., {Montet}, B.~T., {Foreman-Mackey}, D., {et~al.} 2019, \pasp, 131, 094502, \dodoi{10.1088/1538-3873/ab291c}

\bibitem[{{Feltzing} {et~al.}(2017){Feltzing}, {Howes}, {McMillan}, \& {Stonkut{\.{e}}}}]{Feltzing2017}
{Feltzing}, S., {Howes}, L.~M., {McMillan}, P.~J., \& {Stonkut{\.{e}}}, E. 2017, \mnras, 465, L109, \dodoi{10.1093/mnrasl/slw209}

\bibitem[{{Ferguson} {et~al.}(2005){Ferguson}, {Alexander}, {Allard}, {Barman}, {Bodnarik}, {Hauschildt}, {Heffner-Wong}, \& {Tamanai}}]{Ferguson2005}
{Ferguson}, J.~W., {Alexander}, D.~R., {Allard}, F., {et~al.} 2005, \apj, 623, 585, \dodoi{10.1086/428642}

\bibitem[{{Fetherolf} {et~al.}(2023){Fetherolf}, {Pepper}, {Simpson}, {Kane}, {Mo{\v{c}}nik}, {English}, {Antoci}, {Huber}, {Jenkins}, {Stassun}, {Twicken}, {Vanderspek}, \& {Winn}}]{Fetherolf2023}
{Fetherolf}, T., {Pepper}, J., {Simpson}, E., {et~al.} 2023, \apjs, 268, 4, \dodoi{10.3847/1538-4365/acdee5}

\bibitem[{{Gaia Collaboration} {et~al.}(2023){Gaia Collaboration}, {Vallenari}, {Brown}, {Prusti}, {de Bruijne}, {Arenou}, {Babusiaux}, {Biermann}, {Creevey}, {Ducourant}, \& et~al.}]{Gaia2023}
{Gaia Collaboration}, {Vallenari}, A., {Brown}, A.~G.~A., {et~al.} 2023, \aap, 674, A1, \dodoi{10.1051/0004-6361/202243940}

\bibitem[{{Gallet} \& {Bouvier}(2015)}]{Gallet2015}
{Gallet}, F., \& {Bouvier}, J. 2015, \aap, 577, A98, \dodoi{10.1051/0004-6361/201525660}

\bibitem[{{Garc{\'\i}a P{\'e}rez} {et~al.}(2016){Garc{\'\i}a P{\'e}rez}, {Allende Prieto}, {Holtzman}, {Shetrone}, {M{\'e}sz{\'a}ros}, {Bizyaev}, {Carrera}, {Cunha}, {Garc{\'\i}a-Hern{\'a}ndez}, {Johnson}, {Majewski}, {Nidever}, {Schiavon}, {Shane}, {Smith}, {Sobeck}, {Troup}, {Zamora}, {Weinberg}, {Bovy}, {Eisenstein}, {Feuillet}, {Frinchaboy}, {Hayden}, {Hearty}, {Nguyen}, {O'Connell}, {Pinsonneault}, {Wilson}, \& {Zasowski}}]{GarciaPerez2016}
{Garc{\'\i}a P{\'e}rez}, A.~E., {Allende Prieto}, C., {Holtzman}, J.~A., {et~al.} 2016, \aj, 151, 144, \dodoi{10.3847/0004-6256/151/6/144}

\bibitem[{{Ginsburg} {et~al.}(2019){Ginsburg}, {Sip{\H{o}}cz}, {Brasseur}, {Cowperthwaite}, {Craig}, {Deil}, {Guillochon}, {Guzman}, {Liedtke}, {Lian Lim}, {Lockhart}, {Mommert}, {Morris}, {Norman}, {Parikh}, {Persson}, {Robitaille}, {Segovia}, {Singer}, {Tollerud}, {de Val-Borro}, {Valtchanov}, {Woillez}, {Astroquery Collaboration}, \& {a subset of astropy Collaboration}}]{Ginsburg2019}
{Ginsburg}, A., {Sip{\H{o}}cz}, B.~M., {Brasseur}, C.~E., {et~al.} 2019, \aj, 157, 98, \dodoi{10.3847/1538-3881/aafc33}

\bibitem[{{Gordon} {et~al.}(2021){Gordon}, {Davenport}, {Angus}, {Foreman-Mackey}, {Agol}, {Covey}, {Ag{\"u}eros}, \& {Kipping}}]{Gordon2021}
{Gordon}, T.~A., {Davenport}, J. R.~A., {Angus}, R., {et~al.} 2021, \apj, 913, 70, \dodoi{10.3847/1538-4357/abf63e}

\bibitem[{{Grevesse} \& {Sauval}(1998)}]{Grevesse1998}
{Grevesse}, N., \& {Sauval}, A.~J. 1998, \ssr, 85, 161, \dodoi{10.1023/A:1005161325181}

\bibitem[{{Grunblatt} {et~al.}(2023){Grunblatt}, {Wilson}, {Winter}, {Gaudi}, {Huber}, {Yahalomi}, {Bellini}, {Claytor}, {Martinez Palomera}, {Barclay}, {Fu}, \& {Price-Whelan}}]{Grunblatt2023}
{Grunblatt}, S.~K., {Wilson}, R.~F., {Winter}, A., {et~al.} 2023, arXiv e-prints, arXiv:2306.10647, \dodoi{10.48550/arXiv.2306.10647}

\bibitem[{{Hall} {et~al.}(2021){Hall}, {Davies}, {van Saders}, {Nielsen}, {Lund}, {Chaplin}, {Garc{\'\i}a}, {Amard}, {Breimann}, {Khan}, {See}, \& {Tayar}}]{Hall2021}
{Hall}, O.~J., {Davies}, G.~R., {van Saders}, J., {et~al.} 2021, Nature Astronomy, 5, 707, \dodoi{10.1038/s41550-021-01335-x}

\bibitem[{{Handberg} {et~al.}(2021){Handberg}, {Lund}, {White}, {Hall}, {Buzasi}, {Pope}, {Hansen}, {von Essen}, {Carboneau}, {Huber}, {Vanderspek}, {Fausnaugh}, {Tenenbaum}, {Jenkins}, \& {T'DA Collaboration}}]{Handberg2021}
{Handberg}, R., {Lund}, M.~N., {White}, T.~R., {et~al.} 2021, \aj, 162, 170, \dodoi{10.3847/1538-3881/ac09f1}

\bibitem[{{Handberg, Rasmus} {et~al.}(2019){Handberg, Rasmus}, {Lund, Mikkel}, {Huber, Daniel}, \& {Buzasi, Derek}}]{10.17909/t9-4smn-dx89}
{Handberg, Rasmus}, {Lund, Mikkel}, {Huber, Daniel}, \& {Buzasi, Derek}. 2019, TESS Data For Asteroseismology Lightcurves ("TASOC"),  STScI/MAST, \dodoi{10.17909/T9-4SMN-DX89}.
\newblock \url{http://archive.stsci.edu/doi/resolve/resolve.html?doi=10.17909/t9-4smn-dx89}

\bibitem[{Harris {et~al.}(2020)Harris, Millman, van~der Walt, Gommers, Virtanen, Cournapeau, Wieser, Taylor, Berg, Smith, Kern, Picus, Hoyer, van Kerkwijk, Brett, Haldane, del R{'{\i}}o, Wiebe, Peterson, G{'{e}}rard-Marchant, Sheppard, Reddy, Weckesser, Abbasi, Gohlke, \& Oliphant}]{Numpy2020}
Harris, C.~R., Millman, K.~J., van~der Walt, S.~J., {et~al.} 2020, Nature, 585, 357, \dodoi{10.1038/s41586-020-2649-2}

\bibitem[{Hartigan \& Hartigan(1985)}]{Hartigan1985}
Hartigan, J.~A., \& Hartigan, P.~M. 1985, The Annals of Statistics, 13, 70 , \dodoi{10.1214/aos/1176346577}

\bibitem[{{Hattori} {et~al.}(2022){Hattori}, {Foreman-Mackey}, {Hogg}, {Montet}, {Angus}, {Pritchard}, {Curtis}, \& {Sch{\"o}lkopf}}]{Hattori2022}
{Hattori}, S., {Foreman-Mackey}, D., {Hogg}, D.~W., {et~al.} 2022, \aj, 163, 284, \dodoi{10.3847/1538-3881/ac625a}

\bibitem[{{Haywood} {et~al.}(2013){Haywood}, {Di Matteo}, {Lehnert}, {Katz}, \& {G{\'o}mez}}]{Haywood2013}
{Haywood}, M., {Di Matteo}, P., {Lehnert}, M.~D., {Katz}, D., \& {G{\'o}mez}, A. 2013, \aap, 560, A109, \dodoi{10.1051/0004-6361/201321397}

\bibitem[{{Hedges} {et~al.}(2020){Hedges}, {Angus}, {Barentsen}, {Saunders}, {Montet}, \& {Gully-Santiago}}]{Hedges2020}
{Hedges}, C., {Angus}, R., {Barentsen}, G., {et~al.} 2020, Research Notes of the American Astronomical Society, 4, 220, \dodoi{10.3847/2515-5172/abd106}

\bibitem[{{Holcomb} {et~al.}(2022){Holcomb}, {Robertson}, {Hartigan}, {Oelkers}, \& {Robinson}}]{Holcomb2022}
{Holcomb}, R.~J., {Robertson}, P., {Hartigan}, P., {Oelkers}, R.~J., \& {Robinson}, C. 2022, \apj, 936, 138, \dodoi{10.3847/1538-4357/ac8990}

\bibitem[{{Hon} {et~al.}(2022){Hon}, {Kuszlewicz}, {Huber}, {Stello}, \& {Reyes}}]{Hon2022}
{Hon}, M., {Kuszlewicz}, J.~S., {Huber}, D., {Stello}, D., \& {Reyes}, C. 2022, \aj, 164, 135, \dodoi{10.3847/1538-3881/ac8931}

\bibitem[{{Hon} {et~al.}(2021){Hon}, {Huber}, {Kuszlewicz}, {Stello}, {Sharma}, {Tayar}, {Zinn}, {Vrard}, \& {Pinsonneault}}]{Hon2021}
{Hon}, M., {Huber}, D., {Kuszlewicz}, J.~S., {et~al.} 2021, \apj, 919, 131, \dodoi{10.3847/1538-4357/ac14b1}

\bibitem[{Howell {et~al.}(2014)Howell, Sobeck, Haas, Still, Barclay, Mullally, Troeltzsch, Aigrain, Bryson, Caldwell, Chaplin, Cochran, Huber, Marcy, Miglio, Najita, Smith, Twicken, \& Fortney}]{Howell2014}
Howell, S.~B., Sobeck, C., Haas, M., {et~al.} 2014, Publications of the Astronomical Society of the Pacific, 126, 398, \dodoi{10.1086/676406}

\bibitem[{{Huang} {et~al.}(2020{\natexlab{a}}){Huang}, {Vanderburg}, {P{\'a}l}, {Sha}, {Yu}, {Fong}, {Fausnaugh}, {Shporer}, {Guerrero}, {Vanderspek}, \& {Ricker}}]{Huang2020a}
{Huang}, C.~X., {Vanderburg}, A., {P{\'a}l}, A., {et~al.} 2020{\natexlab{a}}, Research Notes of the American Astronomical Society, 4, 204, \dodoi{10.3847/2515-5172/abca2e}

\bibitem[{{Huang} {et~al.}(2020{\natexlab{b}}){Huang}, {Vanderburg}, {P{\'a}l}, {Sha}, {Yu}, {Fong}, {Fausnaugh}, {Shporer}, {Guerrero}, {Vanderspek}, \& {Ricker}}]{Huang2020b}
---. 2020{\natexlab{b}}, Research Notes of the American Astronomical Society, 4, 206, \dodoi{10.3847/2515-5172/abca2d}

\bibitem[{Huber {et~al.}(2016)Huber, Bryson, Haas, Barclay, Barentsen, Howell, Sharma, Stello, \& Thompson}]{Huber2016}
Huber, D., Bryson, S.~T., Haas, M.~R., {et~al.} 2016, The Astrophysical Journal Supplement Series, 224, 2, \dodoi{10.3847/0067-0049/224/1/2}

\bibitem[{{Hunter}(2007)}]{Matplotlib2007}
{Hunter}, J.~D. 2007, Computing in Science Engineering, 9, 90, \dodoi{10.1109/MCSE.2007.55}

\bibitem[{{IRSA}(2022)}]{10.26131/IRSA544}
{IRSA}. 2022, Gaia Source Catalogue DR3,  IPAC, \dodoi{10.26131/IRSA544}.
\newblock \url{https://catcopy.ipac.caltech.edu/dois/doi.php?id=10.26131/IRSA544}

\bibitem[{{Jenkins} {et~al.}(2016){Jenkins}, {Twicken}, {McCauliff}, {Campbell}, {Sanderfer}, {Lung}, {Mansouri-Samani}, {Girouard}, {Tenenbaum}, {Klaus}, {Smith}, {Caldwell}, {Chacon}, {Henze}, {Heiges}, {Latham}, {Morgan}, {Swade}, {Rinehart}, \& {Vanderspek}}]{Jenkins2016}
{Jenkins}, J.~M., {Twicken}, J.~D., {McCauliff}, S., {et~al.} 2016, in Society of Photo-Optical Instrumentation Engineers (SPIE) Conference Series, Vol. 9913, Software and Cyberinfrastructure for Astronomy IV, ed. G.~{Chiozzi} \& J.~C. {Guzman}, 99133E

\bibitem[{{Johnson} {et~al.}(2020){Johnson}, {Penny}, {Gaudi}, {Kerins}, {Rattenbury}, {Robin}, {Calchi Novati}, \& {Henderson}}]{Johnson2020}
{Johnson}, S.~A., {Penny}, M., {Gaudi}, B.~S., {et~al.} 2020, \aj, 160, 123, \dodoi{10.3847/1538-3881/aba75b}

\bibitem[{{Kingma} \& {Ba}(2014)}]{Kingma2014}
{Kingma}, D.~P., \& {Ba}, J. 2014, arXiv e-prints, arXiv:1412.6980.
\newblock \doarXiv{1412.6980}

\bibitem[{{Kollmeier} {et~al.}(2017){Kollmeier}, {Zasowski}, {Rix}, {Johns}, {Anderson}, {Drory}, {Johnson}, {Pogge}, {Bird}, {Blanc}, {Brownstein}, {Crane}, {De Lee}, {Klaene}, {Kreckel}, {MacDonald}, {Merloni}, {Ness}, {O'Brien}, {Sanchez-Gallego}, {Sayres}, {Shen}, {Thakar}, {Tkachenko}, {Aerts}, {Blanton}, {Eisenstein}, {Holtzman}, {Maoz}, {Nandra}, {Rockosi}, {Weinberg}, {Bovy}, {Casey}, {Chaname}, {Clerc}, {Conroy}, {Eracleous}, {G{\"a}nsicke}, {Hekker}, {Horne}, {Kauffmann}, {McQuinn}, {Pellegrini}, {Schinnerer}, {Schlafly}, {Schwope}, {Seibert}, {Teske}, \& {van Saders}}]{Kollmeier2017}
{Kollmeier}, J.~A., {Zasowski}, G., {Rix}, H.-W., {et~al.} 2017, arXiv e-prints, arXiv:1711.03234, \dodoi{10.48550/arXiv.1711.03234}

\bibitem[{{Kounkel} {et~al.}(2022){Kounkel}, {Stassun}, {Bouma}, {Covey}, {Hillenbrand}, \& {Curtis}}]{Kounkel2022}
{Kounkel}, M., {Stassun}, K.~G., {Bouma}, L.~G., {et~al.} 2022, \aj, 164, 137, \dodoi{10.3847/1538-3881/ac866d}

\bibitem[{{Kraft}(1967)}]{Kraft1967}
{Kraft}, R.~P. 1967, \apj, 150, 551, \dodoi{10.1086/149359}

\bibitem[{{Kunimoto} {et~al.}(2021){Kunimoto}, {Huang}, {Tey}, {Fong}, {Hesse}, {Shporer}, {Guerrero}, {Fausnaugh}, {Vanderspek}, \& {Ricker}}]{Kunimoto2021}
{Kunimoto}, M., {Huang}, C., {Tey}, E., {et~al.} 2021, Research Notes of the American Astronomical Society, 5, 234, \dodoi{10.3847/2515-5172/ac2ef0}

\bibitem[{{Lanzafame} \& {Spada}(2015)}]{Lanzafame2015}
{Lanzafame}, A.~C., \& {Spada}, F. 2015, \aap, 584, A30, \dodoi{10.1051/0004-6361/201526770}

\bibitem[{{Lightkurve Collaboration} {et~al.}(2018){Lightkurve Collaboration}, {Cardoso}, {Hedges}, {Gully-Santiago}, {Saunders}, {Cody}, {Barclay}, {Hall}, {Sagear}, {Turtelboom}, {Zhang}, {Tzanidakis}, {Mighell}, {Coughlin}, {Bell}, {Berta-Thompson}, {Williams}, {Dotson}, \& {Barentsen}}]{Lightkurve2018}
{Lightkurve Collaboration}, {Cardoso}, J.~V.~d.~M., {Hedges}, C., {et~al.} 2018, {Lightkurve: Kepler and TESS time series analysis in Python}, Astrophysics Source Code Library.
\newblock \doeprint{1812.013}

\bibitem[{{Lu} {et~al.}(2020){Lu}, {Angus}, {Ag{\"u}eros}, {Blancato}, {Ness}, {Rowland}, {Curtis}, \& {Grunblatt}}]{Lu2020}
{Lu}, Y.~L., {Angus}, R., {Ag{\"u}eros}, M.~A., {et~al.} 2020, \aj, 160, 168, \dodoi{10.3847/1538-3881/abada4}

\bibitem[{{Lu} {et~al.}(2021){Lu}, {Angus}, {Curtis}, {David}, \& {Kiman}}]{Lu2021}
{Lu}, Y.~L., {Angus}, R., {Curtis}, J.~L., {David}, T.~J., \& {Kiman}, R. 2021, \aj, 161, 189, \dodoi{10.3847/1538-3881/abe4d6}

\bibitem[{{Lu} {et~al.}(2022){Lu}, {Curtis}, {Angus}, {David}, \& {Hattori}}]{Lu2022}
{Lu}, Y.~L., {Curtis}, J.~L., {Angus}, R., {David}, T.~J., \& {Hattori}, S. 2022, \aj, 164, 251, \dodoi{10.3847/1538-3881/ac9bee}

\bibitem[{{Luger} {et~al.}(2019){Luger}, {Bedell}, {Vanderspek}, \& {Burke}}]{Luger2019}
{Luger}, R., {Bedell}, M., {Vanderspek}, R., \& {Burke}, C.~J. 2019, arXiv e-prints, arXiv:1903.12182, \dodoi{10.48550/arXiv.1903.12182}

\bibitem[{{Lund} {et~al.}(2021){Lund}, {Handberg}, {Buzasi}, {Carboneau}, {Hall}, {Pereira}, {Huber}, {Hey}, {Van Reeth}, \& {T'DA Collaboration}}]{Lund2021}
{Lund}, M.~N., {Handberg}, R., {Buzasi}, D.~L., {et~al.} 2021, \apjs, 257, 53, \dodoi{10.3847/1538-4365/ac214a}

\bibitem[{{Mackereth} {et~al.}(2021){Mackereth}, {Miglio}, {Elsworth}, {Mosser}, {Mathur}, {Garcia}, {Nardiello}, {Hall}, {Vrard}, {Ball}, {Basu}, {Beaton}, {Beck}, {Bergemann}, {Bossini}, {Casagrande}, {Campante}, {Chaplin}, {Chiappini}, {Girardi}, {J{\o}rgensen}, {Khan}, {Montalb{\'a}n}, {Nielsen}, {Pinsonneault}, {Rodrigues}, {Serenelli}, {Silva Aguirre}, {Stello}, {Tayar}, {Teske}, {van Saders}, \& {Willett}}]{Mackereth2021}
{Mackereth}, J.~T., {Miglio}, A., {Elsworth}, Y., {et~al.} 2021, \mnras, 502, 1947, \dodoi{10.1093/mnras/stab098}

\bibitem[{{Majewski} {et~al.}(2017){Majewski}, {Schiavon}, {Frinchaboy}, {Allende Prieto}, {Barkhouser}, {Bizyaev}, {Blank}, {Brunner}, {Burton}, {Carrera}, {Chojnowski}, {Cunha}, {Epstein}, {Fitzgerald}, {Garc{\'\i}a P{\'e}rez}, {Hearty}, {Henderson}, {Holtzman}, {Johnson}, {Lam}, {Lawler}, {Maseman}, {M{\'e}sz{\'a}ros}, {Nelson}, {Nguyen}, {Nidever}, {Pinsonneault}, {Shetrone}, {Smee}, {Smith}, {Stolberg}, {Skrutskie}, {Walker}, {Wilson}, {Zasowski}, {Anders}, {Basu}, {Beland}, {Blanton}, {Bovy}, {Brownstein}, {Carlberg}, {Chaplin}, {Chiappini}, {Eisenstein}, {Elsworth}, {Feuillet}, {Fleming}, {Galbraith-Frew}, {Garc{\'\i}a}, {Garc{\'\i}a-Hern{\'a}ndez}, {Gillespie}, {Girardi}, {Gunn}, {Hasselquist}, {Hayden}, {Hekker}, {Ivans}, {Kinemuchi}, {Klaene}, {Mahadevan}, {Mathur}, {Mosser}, {Muna}, {Munn}, {Nichol}, {O'Connell}, {Parejko}, {Robin}, {Rocha-Pinto}, {Schultheis}, {Serenelli}, {Shane}, {Silva Aguirre}, {Sobeck}, {Thompson}, {Troup}, {Weinberg}, \& {Zamora}}]{Majewski2017}
{Majewski}, S.~R., {Schiavon}, R.~P., {Frinchaboy}, P.~M., {et~al.} 2017, \aj, 154, 94, \dodoi{10.3847/1538-3881/aa784d}

\bibitem[{{Martig} {et~al.}(2015){Martig}, {Rix}, {Silva Aguirre}, {Hekker}, {Mosser}, {Elsworth}, {Bovy}, {Stello}, {Anders}, {Garc{\'\i}a}, {Tayar}, {Rodrigues}, {Basu}, {Carrera}, {Ceillier}, {Chaplin}, {Chiappini}, {Frinchaboy}, {Garc{\'\i}a-Hern{\'a}ndez}, {Hearty}, {Holtzman}, {Johnson}, {Majewski}, {Mathur}, {M{\'e}sz{\'a}ros}, {Miglio}, {Nidever}, {Pan}, {Pinsonneault}, {Schiavon}, {Schneider}, {Serenelli}, {Shetrone}, \& {Zamora}}]{Martig2015}
{Martig}, M., {Rix}, H.-W., {Silva Aguirre}, V., {et~al.} 2015, \mnras, 451, 2230, \dodoi{10.1093/mnras/stv1071}

\bibitem[{{Martig} {et~al.}(2016){Martig}, {Fouesneau}, {Rix}, {Ness}, {M{\'e}sz{\'a}ros}, {Garc{\'\i}a-Hern{\'a}ndez}, {Pinsonneault}, {Serenelli}, {Silva Aguirre}, \& {Zamora}}]{Martig2016}
{Martig}, M., {Fouesneau}, M., {Rix}, H.-W., {et~al.} 2016, \mnras, 456, 3655, \dodoi{10.1093/mnras/stv2830}

\bibitem[{{Masuda} {et~al.}(2022){Masuda}, {Petigura}, \& {Hall}}]{Masuda2022}
{Masuda}, K., {Petigura}, E.~A., \& {Hall}, O.~J. 2022, \mnras, 510, 5623, \dodoi{10.1093/mnras/stab3650}

\bibitem[{{Mathur} {et~al.}(2023){Mathur}, {Claytor}, {Santos}, {Garc{\'\i}a}, {Amard}, {Bugnet}, {Corsaro}, {Bonanno}, {Breton}, {Godoy-Rivera}, {Pinsonneault}, \& {van Saders}}]{Mathur2023}
{Mathur}, S., {Claytor}, Z.~R., {Santos}, {\^A}. R.~G., {et~al.} 2023, \apj, 952, 131, \dodoi{10.3847/1538-4357/acd118}

\bibitem[{{McQuillan} {et~al.}(2013){McQuillan}, {Aigrain}, \& {Mazeh}}]{McQuillan2013}
{McQuillan}, A., {Aigrain}, S., \& {Mazeh}, T. 2013, \mnras, 432, 1203, \dodoi{10.1093/mnras/stt536}

\bibitem[{{McQuillan} {et~al.}(2014){McQuillan}, {Mazeh}, \& {Aigrain}}]{McQuillan2014}
{McQuillan}, A., {Mazeh}, T., \& {Aigrain}, S. 2014, \apjs, 211, 24, \dodoi{10.1088/0067-0049/211/2/24}

\bibitem[{{Mendoza} {et~al.}(2007){Mendoza}, {Seaton}, {Buerger}, {Bellor{\'\i}n}, {Mel{\'e}ndez}, {Gonz{\'a}lez}, {Rodr{\'\i}guez}, {Delahaye}, {Palacios}, {Pradhan}, \& {Zeippen}}]{Mendoza2007}
{Mendoza}, C., {Seaton}, M.~J., {Buerger}, P., {et~al.} 2007, \mnras, 378, 1031, \dodoi{10.1111/j.1365-2966.2007.11837.x}

\bibitem[{{Montalto} {et~al.}(2020){Montalto}, {Borsato}, {Granata}, {Lacedelli}, {Malavolta}, {Manthopoulou}, {Nardiello}, {Nascimbeni}, \& {Piotto}}]{Montalto2020}
{Montalto}, M., {Borsato}, L., {Granata}, V., {et~al.} 2020, \mnras, 498, 1726, \dodoi{10.1093/mnras/staa2438}

\bibitem[{{Murphy} {et~al.}(2019){Murphy}, {Hey}, {Van Reeth}, \& {Bedding}}]{Murphy2019}
{Murphy}, S.~J., {Hey}, D., {Van Reeth}, T., \& {Bedding}, T.~R. 2019, \mnras, 485, 2380, \dodoi{10.1093/mnras/stz590}

\bibitem[{{Newton} {et~al.}(2016){Newton}, {Irwin}, {Charbonneau}, {Berta-Thompson}, {Dittmann}, \& {West}}]{Newton2016}
{Newton}, E.~R., {Irwin}, J., {Charbonneau}, D., {et~al.} 2016, \apj, 821, 93, \dodoi{10.3847/0004-637X/821/2/93}

\bibitem[{{Newton} {et~al.}(2018){Newton}, {Mondrik}, {Irwin}, {Winters}, \& {Charbonneau}}]{Newton2018}
{Newton}, E.~R., {Mondrik}, N., {Irwin}, J., {Winters}, J.~G., \& {Charbonneau}, D. 2018, \aj, 156, 217, \dodoi{10.3847/1538-3881/aad73b}

\bibitem[{{Noyes} {et~al.}(1984){Noyes}, {Hartmann}, {Baliunas}, {Duncan}, \& {Vaughan}}]{Noyes1984}
{Noyes}, R.~W., {Hartmann}, L.~W., {Baliunas}, S.~L., {Duncan}, D.~K., \& {Vaughan}, A.~H. 1984, \apj, 279, 763, \dodoi{10.1086/161945}

\bibitem[{{Oelkers} \& {Stassun}(2018)}]{Oelkers2018b}
{Oelkers}, R.~J., \& {Stassun}, K.~G. 2018, \aj, 156, 132, \dodoi{10.3847/1538-3881/aad68e}

\bibitem[{{Paegert} {et~al.}(2021){Paegert}, {Stassun}, {Collins}, {Pepper}, {Torres}, {Jenkins}, {Twicken}, \& {Latham}}]{Paegert2021}
{Paegert}, M., {Stassun}, K.~G., {Collins}, K.~A., {et~al.} 2021, arXiv e-prints, arXiv:2108.04778, \dodoi{10.48550/arXiv.2108.04778}

\bibitem[{{Parker}(1955)}]{Parker1955}
{Parker}, E.~N. 1955, \apj, 122, 293, \dodoi{10.1086/146087}

\bibitem[{Paszke {et~al.}(2019)Paszke, Gross, Massa, Lerer, Bradbury, Chanan, Killeen, Lin, Gimelshein, Antiga, Desmaison, Kopf, Yang, DeVito, Raison, Tejani, Chilamkurthy, Steiner, Fang, Bai, \& Chintala}]{Pytorch2019}
Paszke, A., Gross, S., Massa, F., {et~al.} 2019, in Advances in Neural Information Processing Systems 32, ed. H.~Wallach, H.~Larochelle, A.~Beygelzimer, F.~d'Alch\'{e} Buc, E.~Fox, \& R.~Garnett (Curran Associates, Inc.), 8024--8035.
\newblock \url{http://papers.neurips.cc/paper/9015-pytorch-an-imperative-style-high-performance-deep-learning-library.pdf}

\bibitem[{{Penny} {et~al.}(2019){Penny}, {Gaudi}, {Kerins}, {Rattenbury}, {Mao}, {Robin}, \& {Calchi Novati}}]{Penny2019}
{Penny}, M.~T., {Gaudi}, B.~S., {Kerins}, E., {et~al.} 2019, \apjs, 241, 3, \dodoi{10.3847/1538-4365/aafb69}

\bibitem[{{Perez} \& {Granger}(2007)}]{iPython2007}
{Perez}, F., \& {Granger}, B.~E. 2007, Computing in Science Engineering, 9, 21, \dodoi{10.1109/MCSE.2007.53}

\bibitem[{{Powell} {et~al.}(2022){Powell}, {Kruse}, {Montet}, {Feinstein}, {Lewis}, {Foreman-Mackey}, {Barclay}, {Quintana}, {Col{\'o}n}, {Kostov}, {Boyd}, {Smale}, {Mullally}, {Schlieder}, {Schnittman}, {Carroll}, {Carriere}, {Salmon}, {Strong}, {Acks}, {Pfaff}, {Gerner}, \& {Burch}}]{Powell2022}
{Powell}, B.~P., {Kruse}, E., {Montet}, B.~T., {et~al.} 2022, Research Notes of the American Astronomical Society, 6, 111, \dodoi{10.3847/2515-5172/ac74c4}

\bibitem[{{Rebull} {et~al.}(2016){Rebull}, {Stauffer}, {Bouvier}, {Cody}, {Hillenbrand}, {Soderblom}, {Valenti}, {Barrado}, {Bouy}, {Ciardi}, {Pinsonneault}, {Stassun}, {Micela}, {Aigrain}, {Vrba}, {Somers}, {Christiansen}, {Gillen}, \& {Collier Cameron}}]{Rebull2016}
{Rebull}, L.~M., {Stauffer}, J.~R., {Bouvier}, J., {et~al.} 2016, \aj, 152, 113, \dodoi{10.3847/0004-6256/152/5/113}

\bibitem[{{Reinhold} {et~al.}(2019){Reinhold}, {Bell}, {Kuszlewicz}, {Hekker}, \& {Shapiro}}]{Reinhold2019}
{Reinhold}, T., {Bell}, K.~J., {Kuszlewicz}, J., {Hekker}, S., \& {Shapiro}, A.~I. 2019, \aap, 621, A21, \dodoi{10.1051/0004-6361/201833754}

\bibitem[{{Reinhold} \& {Hekker}(2020)}]{Reinhold2020}
{Reinhold}, T., \& {Hekker}, S. 2020, \aap, 635, A43, \dodoi{10.1051/0004-6361/201936887}

\bibitem[{{Ricker} {et~al.}(2015){Ricker}, {Winn}, {Vanderspek}, {Latham}, {Bakos}, {Bean}, {Berta-Thompson}, {Brown}, {Buchhave}, {Butler}, {Butler}, {Chaplin}, {Charbonneau}, {Christensen-Dalsgaard}, {Clampin}, {Deming}, {Doty}, {De Lee}, {Dressing}, {Dunham}, {Endl}, {Fressin}, {Ge}, {Henning}, {Holman}, {Howard}, {Ida}, {Jenkins}, {Jernigan}, {Johnson}, {Kaltenegger}, {Kawai}, {Kjeldsen}, {Laughlin}, {Levine}, {Lin}, {Lissauer}, {MacQueen}, {Marcy}, {McCullough}, {Morton}, {Narita}, {Paegert}, {Palle}, {Pepe}, {Pepper}, {Quirrenbach}, {Rinehart}, {Sasselov}, {Sato}, {Seager}, {Sozzetti}, {Stassun}, {Sullivan}, {Szentgyorgyi}, {Torres}, {Udry}, \& {Villasenor}}]{Ricker2015}
{Ricker}, G.~R., {Winn}, J.~N., {Vanderspek}, R., {et~al.} 2015, Journal of Astronomical Telescopes, Instruments, and Systems, 1, 014003, \dodoi{10.1117/1.JATIS.1.1.014003}

\bibitem[{{Rogers} \& {Nayfonov}(2002)}]{Rogers2002}
{Rogers}, F.~J., \& {Nayfonov}, A. 2002, \apj, 576, 1064, \dodoi{10.1086/341894}

\bibitem[{{Santana} {et~al.}(2021){Santana}, {Beaton}, {Covey}, {O'Connell}, {Longa-Pe{\~n}a}, {Cohen}, {Fern{\'a}ndez-Trincado}, {Hayes}, {Zasowski}, {Sobeck}, {Majewski}, {Chojnowski}, {De Lee}, {Oelkers}, {Stringfellow}, {Almeida}, {Anguiano}, {Donor}, {Frinchaboy}, {Hasselquist}, {Johnson}, {Kollmeier}, {Nidever}, {Price-Whelan}, {Rojas-Arriagada}, {Schultheis}, {Shetrone}, {Simon}, {Aerts}, {Borissova}, {Drout}, {Geisler}, {Law}, {Medina}, {Minniti}, {Monachesi}, {Mu{\~n}oz}, {Poleski}, {Roman-Lopes}, {Schlaufman}, {Stutz}, {Teske}, {Tkachenko}, {Van Saders}, {Weinberger}, \& {Zoccali}}]{Santana2021}
{Santana}, F.~A., {Beaton}, R.~L., {Covey}, K.~R., {et~al.} 2021, \aj, 162, 303, \dodoi{10.3847/1538-3881/ac2cbc}

\bibitem[{{Santos} {et~al.}(2021){Santos}, {Breton}, {Mathur}, \& {Garc{\'\i}a}}]{Santos2021}
{Santos}, A.~R.~G., {Breton}, S.~N., {Mathur}, S., \& {Garc{\'\i}a}, R.~A. 2021, \apjs, 255, 17, \dodoi{10.3847/1538-4365/ac033f}

\bibitem[{{Santos} {et~al.}(2019){Santos}, {Garc{\'\i}a}, {Mathur}, {Bugnet}, {van Saders}, {Metcalfe}, {Simonian}, \& {Pinsonneault}}]{Santos2019}
{Santos}, A.~R.~G., {Garc{\'\i}a}, R.~A., {Mathur}, S., {et~al.} 2019, \apjs, 244, 21, \dodoi{10.3847/1538-4365/ab3b56}

\bibitem[{{Sarmento} {et~al.}(2021){Sarmento}, {Rojas-Ayala}, {Delgado Mena}, \& {Blanco-Cuaresma}}]{Sarmento2021}
{Sarmento}, P., {Rojas-Ayala}, B., {Delgado Mena}, E., \& {Blanco-Cuaresma}, S. 2021, \aap, 649, A147, \dodoi{10.1051/0004-6361/202039703}

\bibitem[{{See} {et~al.}(2021){See}, {Roquette}, {Amard}, \& {Matt}}]{See2021}
{See}, V., {Roquette}, J., {Amard}, L., \& {Matt}, S.~P. 2021, \apj, 912, 127, \dodoi{10.3847/1538-4357/abed47}

\bibitem[{{Serenelli} {et~al.}(2017){Serenelli}, {Johnson}, {Huber}, {Pinsonneault}, {Ball}, {Tayar}, {Silva Aguirre}, {Basu}, {Troup}, {Hekker}, {Kallinger}, {Stello}, {Davies}, {Lund}, {Mathur}, {Mosser}, {Stassun}, {Chaplin}, {Elsworth}, {Garc{\'\i}a}, {Handberg}, {Holtzman}, {Hearty}, {Garc{\'\i}a-Hern{\'a}ndez}, {Gaulme}, \& {Zamora}}]{Serenelli2017}
{Serenelli}, A., {Johnson}, J., {Huber}, D., {et~al.} 2017, \apjs, 233, 23, \dodoi{10.3847/1538-4365/aa97df}

\bibitem[{{Silva Aguirre} {et~al.}(2018){Silva Aguirre}, {Bojsen-Hansen}, {Slumstrup}, {Casagrande}, {Kawata}, {Ciuc{\v{a}}}, {Handberg}, {Lund}, {Mosumgaard}, {Huber}, {Johnson}, {Pinsonneault}, {Serenelli}, {Stello}, {Tayar}, {Bird}, {Cassisi}, {Hon}, {Martig}, {Nissen}, {Rix}, {Sch{\"o}nrich}, {Sahlholdt}, {Trick}, \& {Yu}}]{SilvaAguirre2018}
{Silva Aguirre}, V., {Bojsen-Hansen}, M., {Slumstrup}, D., {et~al.} 2018, \mnras, 475, 5487, \dodoi{10.1093/mnras/sty150}

\bibitem[{{Silva Aguirre} {et~al.}(2020){Silva Aguirre}, {Stello}, {Stokholm}, {Mosumgaard}, {Ball}, {Basu}, {Bossini}, {Bugnet}, {Buzasi}, {Campante}, {Carboneau}, {Chaplin}, {Corsaro}, {Davies}, {Elsworth}, {Garc{\'\i}a}, {Gaulme}, {Hall}, {Handberg}, {Hon}, {Kallinger}, {Kang}, {Lund}, {Mathur}, {Mints}, {Mosser}, {{\c{C}}elik Orhan}, {Rodrigues}, {Vrard}, {Y{\i}ld{\i}z}, {Zinn}, {{\"O}rtel}, {Beck}, {Bell}, {Guo}, {Jiang}, {Kuszlewicz}, {Kuehn}, {Li}, {Lundkvist}, {Pinsonneault}, {Tayar}, {Cunha}, {Hekker}, {Huber}, {Miglio}, {F.~G. Monteiro}, {Slumstrup}, {Winther}, {Angelou}, {Benomar}, {B{\'o}di}, {De Moura}, {Deheuvels}, {Derekas}, {Di Mauro}, {Dupret}, {Jim{\'e}nez}, {Lebreton}, {Matthews}, {Nardetto}, {do Nascimento}, {Pereira}, {Rodr{\'\i}guez D{\'\i}az}, {Serenelli}, {Spitoni}, {Stonkut{\.{e}}}, {Su{\'a}rez}, {Szab{\'o}}, {Van Eylen}, {Ventura}, {Verma}, {Weiss}, {Wu}, {Barclay}, {Christensen-Dalsgaard}, {Jenkins}, {Kjeldsen}, {Ricker}, {Seager}, \& {Vanderspek}}]{SilvaAguirre2020}
{Silva Aguirre}, V., {Stello}, D., {Stokholm}, A., {et~al.} 2020, \apjl, 889, L34, \dodoi{10.3847/2041-8213/ab6443}

\bibitem[{{Simonian} {et~al.}(2019){Simonian}, {Pinsonneault}, \& {Terndrup}}]{Simonian2019}
{Simonian}, G. V.~A., {Pinsonneault}, M.~H., \& {Terndrup}, D.~M. 2019, \apj, 871, 174, \dodoi{10.3847/1538-4357/aaf97c}

\bibitem[{{Skumanich}(1972)}]{Skumanich1972}
{Skumanich}, A. 1972, \apj, 171, 565, \dodoi{10.1086/151310}

\bibitem[{Smith {et~al.}(2016)Smith, Morris, Jenkins, Bryson, Caldwell, \& Girouard}]{Smith2016}
Smith, J.~C., Morris, R.~L., Jenkins, J.~M., {et~al.} 2016, Publications of the Astronomical Society of the Pacific, 128, 124501, \dodoi{10.1088/1538-3873/128/970/124501}

\bibitem[{{Somers} \& {Pinsonneault}(2016)}]{Somers2016}
{Somers}, G., \& {Pinsonneault}, M.~H. 2016, \apj, 829, 32, \dodoi{10.3847/0004-637X/829/1/32}

\bibitem[{{Spada} \& {Lanzafame}(2020)}]{Spada2020}
{Spada}, F., \& {Lanzafame}, A.~C. 2020, \aap, 636, A76, \dodoi{10.1051/0004-6361/201936384}

\bibitem[{{Spergel} {et~al.}(2015){Spergel}, {Gehrels}, {Baltay}, {Bennett}, {Breckinridge}, {Donahue}, {Dressler}, {Gaudi}, {Greene}, {Guyon}, {Hirata}, {Kalirai}, {Kasdin}, {Macintosh}, {Moos}, {Perlmutter}, {Postman}, {Rauscher}, {Rhodes}, {Wang}, {Weinberg}, {Benford}, {Hudson}, {Jeong}, {Mellier}, {Traub}, {Yamada}, {Capak}, {Colbert}, {Masters}, {Penny}, {Savransky}, {Stern}, {Zimmerman}, {Barry}, {Bartusek}, {Carpenter}, {Cheng}, {Content}, {Dekens}, {Demers}, {Grady}, {Jackson}, {Kuan}, {Kruk}, {Melton}, {Nemati}, {Parvin}, {Poberezhskiy}, {Peddie}, {Ruffa}, {Wallace}, {Whipple}, {Wollack}, \& {Zhao}}]{Spergel2015}
{Spergel}, D., {Gehrels}, N., {Baltay}, C., {et~al.} 2015, arXiv e-prints, arXiv:1503.03757, \dodoi{10.48550/arXiv.1503.03757}

\bibitem[{{Spiegel} \& {Zahn}(1992)}]{Spiegel1992}
{Spiegel}, E.~A., \& {Zahn}, J.~P. 1992, \aap, 265, 106

\bibitem[{{Stassun} {et~al.}(2019){Stassun}, {Oelkers}, {Paegert}, {Torres}, {Pepper}, {De Lee}, {Collins}, {Latham}, {Muirhead}, {Chittidi}, {Rojas-Ayala}, {Fleming}, {Rose}, {Tenenbaum}, {Ting}, {Kane}, {Barclay}, {Bean}, {Brassuer}, {Charbonneau}, {Ge}, {Lissauer}, {Mann}, {McLean}, {Mullally}, {Narita}, {Plavchan}, {Ricker}, {Sasselov}, {Seager}, {Sharma}, {Shiao}, {Sozzetti}, {Stello}, {Vanderspek}, {Wallace}, \& {Winn}}]{Stassun2019}
{Stassun}, K.~G., {Oelkers}, R.~J., {Paegert}, M., {et~al.} 2019, \aj, 158, 138, \dodoi{10.3847/1538-3881/ab3467}

\bibitem[{{Stello} {et~al.}(2022){Stello}, {Saunders}, {Grunblatt}, {Hon}, {Reyes}, {Huber}, {Bedding}, {Elsworth}, {Garc{\'\i}a}, {Hekker}, {Kallinger}, {Mathur}, {Mosser}, \& {Pinsonneault}}]{Stello2022}
{Stello}, D., {Saunders}, N., {Grunblatt}, S., {et~al.} 2022, \mnras, 512, 1677, \dodoi{10.1093/mnras/stac414}

\bibitem[{{van der Walt} {et~al.}(2014){van der Walt}, {Sch{\"o}nberger}, {Nunez-Iglesias}, {Boulogne}, {Warner}, {Yager}, {Gouillart}, {Yu}, \& {scikit-image Contributors}}]{vanderWalt2014}
{van der Walt}, S., {Sch{\"o}nberger}, J.~L., {Nunez-Iglesias}, J., {et~al.} 2014, PeerJ, 2, e453, \dodoi{10.7717/peerj.453}

\bibitem[{{van Saders} {et~al.}(2016){van Saders}, {Ceillier}, {Metcalfe}, {Silva Aguirre}, {Pinsonneault}, {Garc{\'\i}a}, {Mathur}, \& {Davies}}]{vanSaders2016}
{van Saders}, J.~L., {Ceillier}, T., {Metcalfe}, T.~S., {et~al.} 2016, \nat, 529, 181, \dodoi{10.1038/nature16168}

\bibitem[{{van Saders} {et~al.}(2019){van Saders}, {Pinsonneault}, \& {Barbieri}}]{vanSaders2019}
{van Saders}, J.~L., {Pinsonneault}, M.~H., \& {Barbieri}, M. 2019, \apj, 872, 128, \dodoi{10.3847/1538-4357/aafafe}

\bibitem[{{Vanderspek} {et~al.}(2018){Vanderspek}, {Doty}, {Fausnaugh}, {Villase{\~n}or}, {Jenkins}, {Berta-Thompson}, {Burke}, \& {Ricker}}]{TESSHandbook}
{Vanderspek}, R., {Doty}, J.~P., {Fausnaugh}, M., {et~al.} 2018, TESS Instrument Handbook, \url{https://archive.stsci.edu/missions/tess/doc/TESS_Instrument_Handbook_v0.1.pdf}

\bibitem[{{Virtanen} {et~al.}(2020){Virtanen}, {Gommers}, {Oliphant}, {Haberland}, {Reddy}, {Cournapeau}, {Burovski}, {Peterson}, {Weckesser}, {Bright}, {van der Walt}, {Brett}, {Wilson}, {Jarrod Millman}, {Mayorov}, {Nelson}, {Jones}, {Kern}, {Larson}, {Carey}, {Polat}, {Feng}, {Moore}, {Vand erPlas}, {Laxalde}, {Perktold}, {Cimrman}, {Henriksen}, {Quintero}, {Harris}, {Archibald}, {Ribeiro}, {Pedregosa}, {van Mulbregt}, \& {Contributors}}]{Scipy2020}
{Virtanen}, P., {Gommers}, R., {Oliphant}, T.~E., {et~al.} 2020, Nature Methods, 17, 261, \dodoi{https://doi.org/10.1038/s41592-019-0686-2}

\bibitem[{{Weber} \& {Davis}(1967)}]{Weber1967}
{Weber}, E.~J., \& {Davis}, Leverett, J. 1967, \apj, 148, 217, \dodoi{10.1086/149138}

\bibitem[{{W}es {M}c{K}inney(2010)}]{Pandas2010}
{W}es {M}c{K}inney. 2010, in {P}roceedings of the 9th {P}ython in {S}cience {C}onference, ed. {S}t\'efan van~der {W}alt \& {J}arrod {M}illman, 56 -- 61

\bibitem[{{Wolniewicz} {et~al.}(2021){Wolniewicz}, {Berger}, \& {Huber}}]{Wolniewicz2021}
{Wolniewicz}, L.~M., {Berger}, T.~A., \& {Huber}, D. 2021, \aj, 161, 231, \dodoi{10.3847/1538-3881/abee1d}

\bibitem[{{Wright} {et~al.}(2011){Wright}, {Drake}, {Mamajek}, \& {Henry}}]{Wright2011}
{Wright}, N.~J., {Drake}, J.~J., {Mamajek}, E.~E., \& {Henry}, G.~W. 2011, \apj, 743, 48, \dodoi{10.1088/0004-637X/743/1/48}

\bibitem[{{Zhang} {et~al.}(2021){Zhang}, {Xiang}, {Zhang}, {Ting}, {Rix}, {Wu}, {Huang}, {Sun}, {Tian}, {Wang}, \& {Liu}}]{Zhang2021}
{Zhang}, M., {Xiang}, M., {Zhang}, H.-W., {et~al.} 2021, \apj, 922, 145, \dodoi{10.3847/1538-4357/ac22a5}

\end{thebibliography}
\bibliographystyle{aasjournal}

\end{document}